\documentclass[acmtog]{acmart}
\AtBeginDocument{%
  \providecommand\BibTeX{{%
    \normalfont B\kern-0.5em{\scshape i\kern-0.25em b}\kern-0.8em\TeX}}}

\usepackage[ruled]{algorithm2e} 

\SetAlFnt{\small}
\SetAlCapFnt{\small}
\SetAlCapNameFnt{\small}
\SetAlCapHSkip{0pt}

\usepackage{diagbox}
\usepackage{algorithmic}
\usepackage{tabularx}
\usepackage{caption}
\usepackage{subfigure}
\usepackage{siunitx}
\usepackage{multirow}
\usepackage{makecell}
\usepackage{titletoc}
\usepackage{soul}
\usepackage{comment}

\usepackage{colortbl}

\usepackage{amsmath,amsfonts,bm}

\def\eqref#1{equation~\ref{#1}}

\def\1{\bm{1}}

\def\rva{{\mathbf{a}}}

\def\rvd{{\mathbf{d}}}

\def\rvp{{\mathbf{p}}}
\def\rvq{{\mathbf{q}}}

\def\rvs{{\mathbf{s}}}
\def\rvt{{\mathbf{t}}}

\def\rvv{{\mathbf{v}}}

\def\rvx{{\mathbf{x}}}

\def\rvz{{\mathbf{z}}}
\def\rvmu{{\mathbf{\mu}}}

\DeclareMathAlphabet{\mathsfit}{\encodingdefault}{\sfdefault}{m}{sl}
\SetMathAlphabet{\mathsfit}{bold}{\encodingdefault}{\sfdefault}{bx}{n}

\definecolor{my_magenta}{HTML}{D41159}
\definecolor{my_blue}{HTML}{1A85FF}
\definecolor{my_ref_blue}{HTML}{34A7F1}

\setcopyright{acmcopyright}
\setcopyright{acmlicensed}
\setcopyright{rightsretained}

\acmJournal{TOG}

\copyrightyear{2025}
\acmYear{2025}
\acmConference[SA Conference Papers '25]{SIGGRAPH Asia 2025 Conference
Papers}{December 15--18, 2025}{Hong Kong, Hong Kong}
\acmBooktitle{SIGGRAPH Asia 2025 Conference Papers (SA Conference Papers
'25), December 15--18, 2025, Hong Kong, Hong
Kong}\acmDOI{10.1145/3757377.3763819}
\acmISBN{979-8-4007-2137-3/2025/12}

\begin{document}

\title{Physics-Based Motion Imitation with Adversarial Differential Discriminators}

\author{Ziyu Zhang}
\authornote{Joint First Authors.}
\email{zza333@sfu.ca}
\affiliation{%
  \institution{Simon Fraser University}
  \country{Canada}
}

\author{Sergey Bashkirov}
\authornotemark[1]
\email{sergey.bashkirov@sony.com}
\affiliation{%
  \institution{Sony Playstation}
  \country{USA}
}
\author{Dun Yang}
\email{dya66@sfu.ca}
\affiliation{%
  \institution{Simon Fraser University}
  \country{Canada}
}
\author{Yi Shi}
\email{ysa273@sfu.ca}
\affiliation{%
  \institution{Simon Fraser University}
  \country{Canada}
}
\author{Michael Taylor}
\email{mike.taylor@sony.com}
\affiliation{%
  \institution{Sony Playstation}
  \country{USA}
}
\author{Xue Bin Peng}
\email{xbpeng@sfu.ca}
\affiliation{%
  \institution{Simon Fraser University}
  \country{Canada}
}
\affiliation{%
  \institution{Nvidia}
  \country{Canada}
}

\begin{abstract}
Multi-objective optimization problems, which require the simultaneous optimization of multiple objectives, are prevalent across numerous applications. Existing multi-objective optimization methods often rely on manually-tuned aggregation functions to formulate a joint optimization objective. The performance of such hand-tuned methods is heavily dependent on careful weight selection, a time-consuming and laborious process. These limitations also arise in the setting of reinforcement-learning-based motion tracking methods for physically simulated characters, where intricately crafted reward functions are typically used to achieve high-fidelity results. Such solutions not only require domain expertise and significant manual tuning, but also limit the applicability of the resulting reward function across diverse skills. To bridge this gap, we present a novel adversarial multi-objective optimization technique that is broadly applicable to a range of multi-objective reinforcement-learning tasks, including motion tracking. Our proposed Adversarial Differential Discriminator (ADD) receives a single positive sample, yet is still effective at guiding the optimization process. We demonstrate that our technique can enable characters to closely replicate a variety of acrobatic and agile behaviors, achieving comparable quality to state-of-the-art motion-tracking methods, without relying on manually-designed reward functions.
\end{abstract}

%
%



\begin{teaserfigure}
    \centering
    \includegraphics[width=\textwidth]{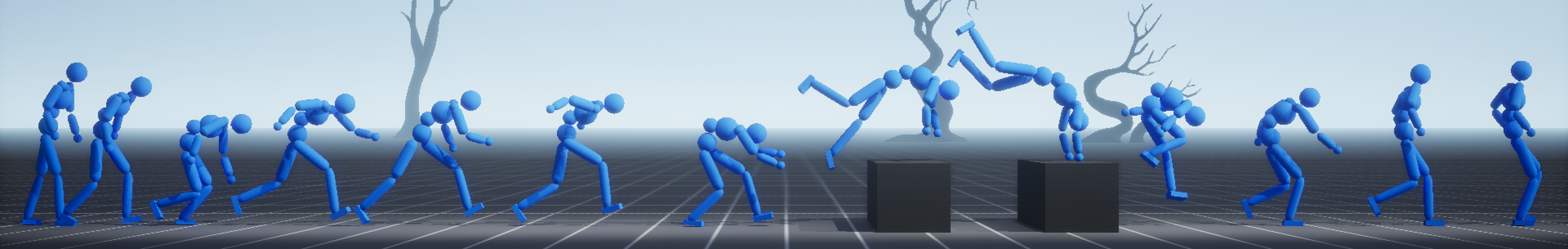}\\
    \caption{We propose an adversarial multi-objective optimization technique that enables physically simulated characters to closely imitate a broad range of highly agile and athletic skills, without requiring manual reward engineering. Here, a physically simulated character learns to jump over obstacles by imitating a \textit{double kong} reference motion.}
    \label{fig:teaser}
\end{teaserfigure}

\maketitle
\section{Introduction}

Physics-based character animation has seen rapid progress over the past several years. Data-driven reinforcement learning methods have enabled the field to grow from synthesizing controllers for relatively simple and common behaviors, such as locomotion, to controllers that can replicate a wide range of highly dynamic and complex skills. The effectiveness of RL methods is heavily dependent on the reward function, which tends to require significant manual effort to design and tune. As an alternative to manual reward engineering, adversarial imitation learning techniques automatically learn reward functions from data, in the form of an adversarial discriminator. 
Despite these impressive advances, a gap in quality has persisted between physically simulated animation and human motion capture. 
Prior adversarial imitation learning methods, though effective at producing natural and life-like behaviors, often diverge from the exact reference motions they aim to reproduce. Rather than focusing on precise replication of target motions, these methods capture the overall \textit{style} of the motions via a distribution matching objective. However, closely replicating a target motion can be vital for many animation applications.
In this work, we present a novel adversarial multi-objective optimization (MOO) technique, which can be broadly applied to a wide range of MOO problems. When applied to motion tracking for character animation, our technique enables physics-based controllers to accurately mimic challenging reference motions recorded from human actors, without relying on manually engineered reward functions.

A commonly used technique for solving MOO problems is the loss balancing method, which combines multiple objectives into a scalar function via a weighted sum. 
Although simple, this approach heavily relies on careful weight selection, a process that can be time-consuming and labor-intensive when done manually. 
To address this limitation, our technique automatically aggregates multiple objectives by replacing the traditional weighted sum of objective functions with an \textit{adversarial differential discriminator} (ADD).
The discriminator takes as input a vector of objective values, herein referred to as a \emph{differential} vector, since the objective values represent the difference between the performance of a model and the ideal performance for each objective. During training, the discriminator learns to classify whether a given differential vector corresponds to an ideal solution or not. This design enables the discriminator to automatically and dynamically determine how to balance the various objectives over the course of training, automatically honing in on more challenging objectives as the model's performance improves. Furthermore, with our formulation, the only positive sample provided to the discriminator is a zero vector representing the differential vector of an ideal ``zero-error'' solution. A key finding of this work is that a discriminator trained with only a single positive sample remains effective in guiding the optimization process.

The central contribution of this paper is a novel GAN-based framework for multi-objective optimization. Our framework provides an automatic and dynamic way to aggregate different objectives in MOO problems. Furthermore, unlike the traditional loss weighting approach, our framework can capture potential nonlinear relationships among the objectives. We demonstrate that our approach achieves performance comparable to existing methods that use hand-crafted objective aggregation functions across several MOO problems, including motion tracking for simulated character and non-motion-imitation tasks. Our framework successfully enables a simulated humanoid and a simulated robot to replicate a variety of highly agile and acrobatic skills, achieving quality on par with the state-of-the-art motion tracking methods, while alleviating the need for manual reward engineering.

\section{Related Work}
Physics-based character animation enables procedural methods that can automatically synthesize realistic and responsive behaviors for virtual characters. A key challenge has been developing controllers that can reproduce the vast array of motor skills exhibited by humans and animals, while also producing life-like movements. Early efforts leveraged human insight to design skill-specific control strategies \citep{AthleticsHodgins1995,Wooten1998,Yin07,2010-TOG-gbwc}. These manually-designed controllers have been effective in replicating a large variety of complex motor skills \citep{DaSilva2008,CIO2012,Geyer2173,AlBorno2013}. However, designing skill-specific controllers often entails a lengthy development process, which can be difficult to apply to skills where domain expertise is scarce. Optimization-based methods can mitigate the reliance on manual engineering \citep{Panne94virtualwindup,delasa2010,CIO2012,AlBorno2013,Wampler2014,ClimbingNaderi2017}, but may nonetheless require carefully-designed control structures that expose a compact set of optimization parameters. Furthermore, designing suitable objective functions that lead to naturalistic behaviors for a particular skill can present a daunting challenge \citep{BipedWang2009,2013-TOG-MuscleBasedBipeds}. A recent line of work explores leveraging large language models (LLMs) to automate the design of objective functions \citep{ma2023eureka, cui2025grove}.

\paragraph{Data-driven methods:}
Meanwhile, data-driven techniques alleviate some challenges of controller engineering by imitating reference motion data, such as motion clips acquired through motion capture or artist-animated keyframes \citep{Zordan2002,Sharon2005Walking}. One of the most common approaches for imitating motion data is via motion tracking, where a controller imitates a desired behavior by explicitly tracking target poses prescribed by a reference motion clip \citep{MotionStyleLiu2005,BipedDataSok2007,Liu2010Samcon}. These tracking-based methods can reproduce a wide range of behaviors without designing skill-specific objectives for every behavior \citep{DataDrivenLee2010,2016-TOG-controlGraphs,BasketballLiu2018}. Motion tracking combined with deep reinforcement learning has led to general frameworks that can imitate diverse motor skills \citep{deep_mimic}, with systems that can reproduce hundreds of distinct behaviors using the same learning algorithm \citep{ScalableWon2020,wang2020unicon,Yuan2021SimPoESC}. The vital component to the success of motion tracking methods lies in designing sufficiently general tracking objectives that can be applied to a large variety of skills, while also producing high-quality results for each specific skill \citep{wang2020unicon,SpacetimeBoundsMa2021}. Constructing general-purpose tracking objectives can therefore entail a tedious design and tuning process.

\paragraph{Adversarial imitation learning:} Adversarial imitation learning and related inverse reinforcement learning methods provide an alternative to the manual design of objective functions by jointly learning an objective function and a corresponding policy from data through an adversarial mini-max game \citep{MaxEntIRL2008,ApprenticeAbbeel2005}. Adversarial imitation learning can be instantiated through a GAN-like framework \citep{GANGoodfellow2014}, where a discriminator is trained to differentiate between behaviors from a demonstrator and behaviors produced by an agent. The agent then aims to learn a policy that produces behaviors that maximize the prediction error of the discriminator \citep{GAIL2016}. It has been shown that this adversarial framework imitates demonstrations by optimizing a variational approximation of the divergence between the demonstrations and the agent's behavior distribution \citep{FGAN2016,ke2021imitation}. \citet{2021-TOG-AMP} leveraged adversarial imitation learning to train controllers that are able to imitate task-relevant behaviors from unstructured motion datasets containing diverse motion clips. These adversarial techniques have been able to produce high-quality motions that are comparable to state-of-the-art motion tracking methods \citep{peng2018variational,GanXu2021,2022-TOG-ASE}. However, prior adversarial imitation learning methods for motion imitation leverage a general distribution matching formulation, where an agent is only required to match the overall motion distribution of the dataset \citep{2021-TOG-AMP,GanXu2021}. This provides the agent with the flexibility to sequence motion transitions and segments in different orders, potentially dropping modes in the target motion distribution. This flexibility can be an advantage in certain applications, but it can also be detrimental in animation applications that require precise replication of a motion clip, such as in-betweening and post-processing of keyframe animations.

\paragraph{Multi-objective optimization:}
{MOO problems involve optimizing multiple objectives simultaneously. A common strategy is to combine the objectives via a weighted sum. However, the effectiveness of such methods is highly sensitive to the chosen weights, which entail meticulous manual tuning. Multi-objective evolutionary algorithms can approximate the Pareto front by evolving a population of solutions
\citep{deb2002NSGAII, prediction-guided-morl-xu}. For instance, \citet{agrawal2013pareto} adapted ($1 + \lambda$) CMA-ES to synthesize jumping controllers with optimal trade-offs between effort and jump height \citep{igel2007cmaes}. However, evolutionary algorithms are computationally expensive as they need to evaluate and maintain a large population of solutions.
\citet{meta-learning-morl} proposed a more efficient method that trains a single meta-policy that can be quickly adapted to different preference vectors. Yet, it requires manual specification of preferences that can be difficult due to varying objective scales. Prior work, such as \citet{distributional-view-moo} and \citet{composite-motion-learning}, addresses this by introducing scale-invariant multi-critic RL frameworks, in which each objective is assigned its own critic and a single policy is jointly optimized over this set of critics. In this work, we propose an adversarial MOO approach that automatically balances the different objectives without manual preference specification or multiple critics. Moreover, our method enables nonlinear combinations of objectives that can better capture complex relationships among disparate objectives and the potentially non-convex Pareto front.}

\section{Background}
In this work, we propose a multi-objective optimization method using an adversarial differential discriminator (ADD). To evaluate its effectiveness, we apply ADD to solve a range of control problems modeled as MDPs, where an agent aims to optimize multiple objectives concurrently. An MDP is defined as $M = (\mathcal{S}, \mathcal{A}, \gamma, \rho_0, \rho, r)$, with a state space $\mathcal{S}$, action space $\mathcal{A}$, discount factor $\gamma \in [0, 1]$, initial state distribution $\rho_0(\rvs)$, dynamics function $\rho(\rvs' | \rvs, \rva)$, and reward function $r$. An agent interacts with the environment by sampling an action from a policy $\rva_t \sim \pi(\rva_t|\rvs_t)$ at each timestep $t$ conditioned on state $\rvs_t$.  The agent then performs the action, resulting in a new state $\rvs_{t+1}$, sampled according to the environment dynamics $\rvs_{t+1} \sim \rho(\rvs_{t+1}|\rvs_t, \rva_t)$.  A trajectory $\tau = (\rvs_0, \rva_0, \rvs_1, \rva_1, ..., \rvs_{T-1}, \rva_{T-1}, \rvs_T)$ consists of a sequence of state $\rvs_t \in \mathcal{S}$ and action $\rva_t \in \mathcal{A}$ pairs. The goal of the agent is to learn a policy that maximizes the expected discounted return $J(\pi)$,
  \begin{align}
  \label{eq:policy_gradient}
    & J(\pi)  = \mathbb{E}_{p(\tau | \pi)}\left[\sum_{t=0}^{T-1}\gamma^t r_t\right]
  \end{align}
\noindent
where $p(\tau|\pi) = \rho_0(\rvs_0) \prod_{t=0}^{T-1} \rho(\rvs_{t+1}|\rvs_t,\rva_t) \pi(\rva_t | \rvs_t)$ represents the likelihood of a trajectory $\tau$ under a policy $\pi$.

\section{Adversarial Differential Discriminator}

In MOO problems, multiple objectives are typically expressed as loss functions, where the goal is to jointly minimize these losses. Let $l^i(\cdot)$ denote the $i$-th loss function for $1\leq i \leq n$. In the following discussion, we assume that each loss function is non-negative $l^i(\cdot) \geq 0$.
A common approach for tackling MOO problems is to aggregate the individual loss functions using a weighted sum:
\begin{equation}
    \min_{\theta} \sum_{i} w^i l^i(\theta),
\end{equation}
where $w^i$ denotes the corresponding weight assigned to the $i$-th loss function.
The goal of the optimization problem then is to find a set of model parameters $\theta$ that minimizes this linear combination of loss functions \citep{Kendall2017MultitaskLU, liu2021endtoend}.

This formulation, however, restricts the aggregation to linear combinations of the individual objectives. In this work, we propose constructing a nonlinear aggregation using an \textit{adversarial differential discriminator} $D(\Delta)$, which enables our method to automatically learn more flexible combinations of the objectives. In our framework, the loss functions $l^i(\theta)$ are assembled into a \textit{differential} vector $\Delta = 
    \left( l^1(\theta), ...,
    l^n(\theta)
    \right) $.
The differential vector can be interpreted as the error, or the difference between the ideal and actual performance, for each objective. The discriminator $D(\Delta)$ acts as a nonlinear aggregation function that combines the individual losses in $\Delta$ together into a single aggregate loss. The multiple objectives are then jointly optimized through an adversarial framework \citep{GANGoodfellow2014}, which formulates the MOO problem as a mini-max game:
\begin{align}
\label{eq:training_loss}
    \min_{\theta} \max_{D} \quad  & \log(D(\mathbf{0})) + 
    \log(1 - D(\Delta)).
\end{align}
A key distinction between ADD and prior adversarial learning frameworks is that the adversarial differential discriminator $D$ only receives a \emph{single} positive sample, a zero vector $\Delta = \mathbf{0}$ corresponding to the differential vector of an ideal solution. One of the key findings of this work is that an adversarial discriminator trained on a single positive sample can still be effective at solving a wide range of tasks.

However, simply optimizing Eq. \ref{eq:training_loss} can lead to degenerate behaviors, where the discriminator $D$ may converge to a delta function that assigns a score of 1 to the zero differential vector and a score of 0 to any non-zero differential vector. This delta function can lead to uninformative gradients, which can impair the optimization process. To mitigate this degeneracy, we follow \citet{2021-TOG-AMP} and introduce a gradient penalty (GP) regularizer:
\begin{align}
\label{eq:training_loss_w_grad_penalty}
    \min_{\theta} \max_{D} \quad  & \log(D(\mathbf{0})) + 
    \log(1 - D(\Delta)) - \lambda^{GP} \mathcal{L}^{GP}(D),
\end{align}
where the gradient penalty $\mathcal{L}^{GP}$ is given by:
\begin{equation}
    \mathcal{L}^{GP}(D) =\, \left|\left|\nabla_{\phi} D(\phi) \middle|_{\phi = \Delta}\right|\right|_2^2.
\end{equation}
When training with the objective outlined in Eq. \ref{eq:training_loss_w_grad_penalty}, the model adjusts its parameters $\theta$ to drive $\Delta$ closer to zero to fool the discriminator. Meanwhile, the discriminator dynamically attends to different objectives and hones in on the more difficult combinations of objectives to continually challenge the model.

\section{Motion Tracking with ADD}
In this section, we show how ADD can be applied in a reinforcement learning framework to train control policies that enable physically simulated characters to imitate challenging reference motions. RL-based motion tracking methods typically use a tracking reward,
\begin{equation}
    \label{Eq:tracking_reward}
    r_t = \sum_{i} w^i r_t^i,
\end{equation}
composed of a weighted sum of various reward terms $r_t^i$. Each reward term quantifies the error between the agent's motion and the reference motion for a specific motion feature \citep{MocapChentanez2018, deep_mimic, wang2020unicon}, such as joint rotations, root positions, etc. A commonly used formulation for these reward terms is as an exponentiated error:
\begin{equation}
    \label{eq:exponential_reward_subterm}
    r_t^i = \exp(-\alpha^i \| \hat{\rvq}_t^i \ominus \rvq_t^i \|_2^2),
\end{equation}
where $\rvq_t^i$ denotes a vector of features, such as position or velocities, extracted from the agent's state $\rvs_t$; $\hat{\rvq}_t^i$ are the corresponding target features specified by the reference motion; and $\alpha^i$ are manually specified scale parameters. Manually designing effective reward functions for precise imitation of a wide range of motions can be challenging. Moreover, the reward parameters $w^i$ and $\alpha^i$ can be laborious to tune, and may need to be adjusted for different types of motions.

We propose an adaptive motion tracking reward function that models motion tracking as an MOO problem and replaces the linear weighted sum in Eq. \ref{Eq:tracking_reward} with a learned adversarial differential discriminator. 
During training, the discriminator $D(\Delta)$ receives the difference between the agent's state $\rvs$ and the reference motion $\hat{\rvs}$ as negative samples (i.e., $\Delta = \hat{\rvs} \ominus \rvs$). The only positive sample provided to $D$ is $\Delta = \mathbf{0}$, representing perfect tracking with zero tracking error. The discriminator $D$ is trained using the following objective:
\begin{align}
\label{Eq:motion_tracking_disc_loss}
    \max_{D} \quad  & \log(D(\mathbf{0})) + 
    \mathbb{E}_{p(\mathbf{\rvs}  | \pi)} 
    \left[ \log(1 - D(\Delta)) \right] - \lambda^{GP} \mathcal{L}^{GP}(D).
\end{align}
Here, the gradient penalty regularizer is specified according to
\begin{equation}
    \label{eq:motion_tracking_gp}
    \mathcal{L}^{GP}(D) =\, \mathbb{E}_{p(\rvs  | \pi)} \left[ \|\nabla_{\phi} D(\phi)|_{\phi = \Delta}\|_2^2 \right],
\end{equation}
where $p(\rvs  | \pi)$ represents the marginal state distribution under the policy $\pi$.
In our framework, the gradient penalty is applied to the negative samples instead of the positive samples, as ADD receives only one positive training sample. This differs from prior work, such as \citet{2021-TOG-AMP}, where the gradient penalty is applied exclusively to the positive samples.
The reward for training the tracking policy $\pi$ is then given by:
\begin{equation}
\label{Eq:add_reward}
    r_t = - \log (1 - D(\Delta_t)),
\end{equation}
where the differential vector is simply the difference between the agent's state and the target state $\Delta_t = \hat{\rvs}_t \ominus {\rvs}_t$.

Previous adversarial imitation learning techniques adopt a distribution matching approach, where the discriminator $D(s_{t-n:t})$ classifies a sequence of $n$ states as either reference or policy-generated. This encourages the policy to produce trajectories that broadly resemble the characteristics of the reference motion. In contrast, our formulation allows for precise frame-level replication, which is essential for applications requiring high accuracy, such as motion in-betweening or animation keyframe post-processing.

\subsection{Discriminator Observations}
Following \citet{2021-TOG-AMP}, an observation map $\phi(\cdot)$ extracts features from the agent's state $\rvs$ and the reference motion $\hat{\rvs}$. The differential vector then consists of the differences between the extracted features $\Delta_t = \phi(\hat{\rvs}_t) \ominus \phi(\rvs_t)$. 
The observation map $\phi(\cdot)$ extracts a set of features similarly to those from \citet{deep_mimic}:
\begin{itemize}
    \item Global position and rotation of the root
    \item Position of each joint represented in the character's local coordinate frame
    \item Global rotation of each joint
    \item Linear and angular velocity of the root represented in the character's local coordinate frame
    \item Local velocity of each joint
\end{itemize}
where the character's local coordinate frame is specified with the origin located at the root of the character (i.e., pelvis). The x-axis of the local coordinate frame aligns with the root link's facing direction, while the positive y-axis points in the global upward direction.

\begin{figure*}[t]
\centering
\includegraphics[width=1.\textwidth]{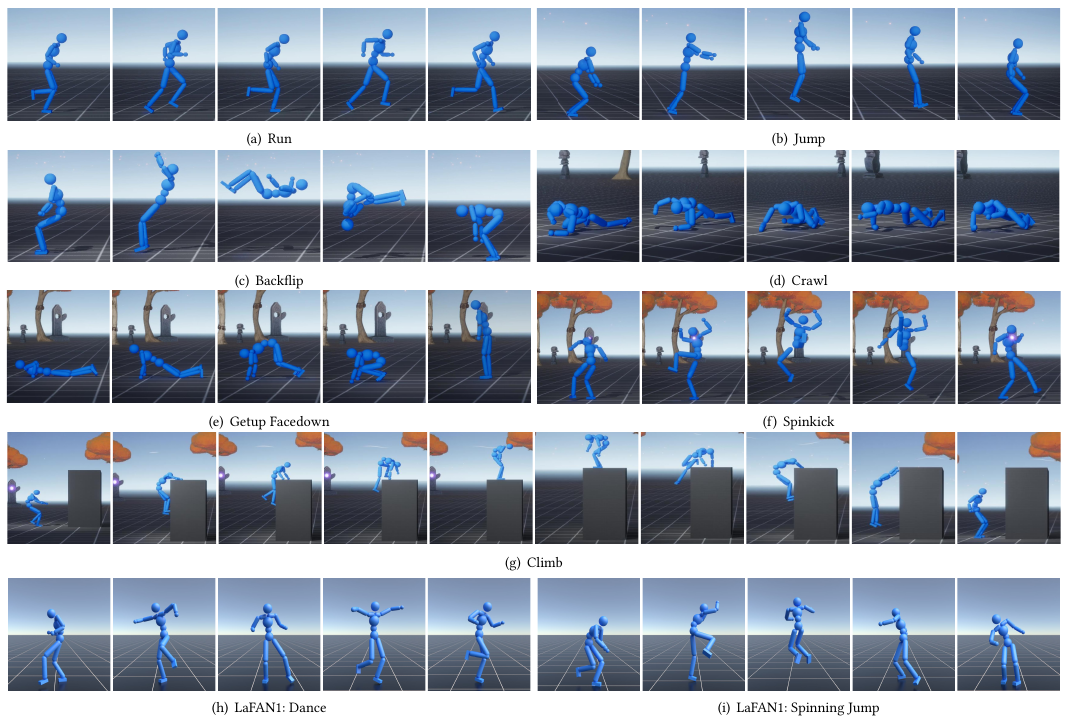}
\vspace{-0.6cm}
\caption{Snapshots of the simulated humanoid characters trained using ADD performing various skills. ADD enables characters to replicate a diverse repertoire of behaviors, achieving tracking quality comparable to state-of-the-art motion imitation methods, without manual reward engineering.}
\label{fig:motion_snapshots}
\end{figure*}

\begin{figure*}[h]
	\centering
    \subfigure[Kick]{\includegraphics[width=1.035\columnwidth]{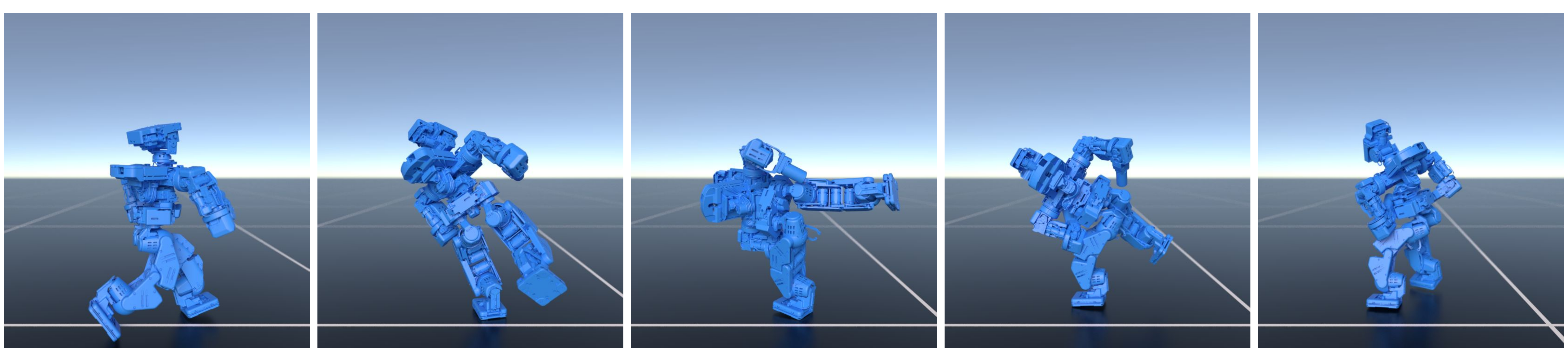}}
    \subfigure[Punch]{\includegraphics[width=1.035\columnwidth]{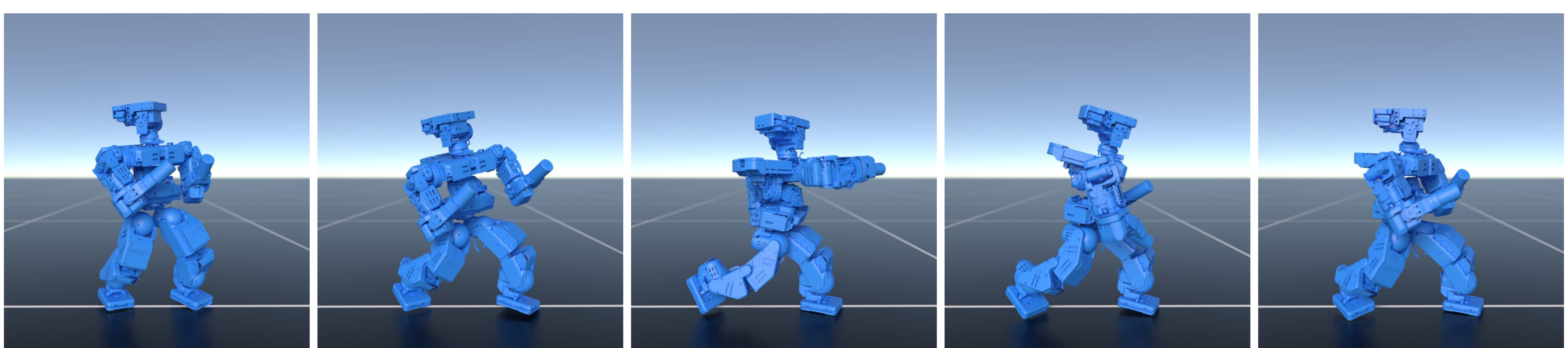}}\\
    \vspace{-0.4cm}
    \caption{Visual snapshots of the simulated EVAL robot replicating a range of target motions, using motion-tracking controllers trained with ADD. The controllers, trained using learned tracking rewards, successfully enable the robot to reproduce a set of challenging skills.}
    \label{fig:EVAL_snapshots}
\end{figure*}

\section{Motion Tracking}
To evaluate the effectiveness of ADD for motion imitation, we apply ADD to train a 28 DoF simulated humanoid and a 26 DoF simulated Sony EVAL robot \citep{Taylor2021LearningBR} to imitate a diverse suite of motion clips. We evaluate ADD on imitating individual motion clips from \citet{deep_mimic}, as well as training a single general policy on larger motion datasets, such as the DanceDB subset of AMASS and a subset of LaFAN1 \citep{AMASS_Mahmood_2019_ICCV, LAFAN_harvey2020robust}. The LaFAN1 subset is curated by excluding motions involving object and terrain interactions, which are not simulated in our environments. The LaFAN1 subset contains over an hour of various locomotion skills, including jumping, sprinting, fighting, dancing, etc.

\subsection{States and Actions}
The state $\rvs_t$ is composed of features, similar to those used by \citet{2021-TOG-AMP}, including the positions of each body link relative to the root, the rotations of the links encoded in the 6D normal-tangent representation, as well as the linear and angular velocities of each link. 
All features are recorded in the character's local coordinate frame. Target poses from the reference motion are also provided to the policy to synchronize the simulated character with the reference motion. The policy's actions $\rva_t$ specify target rotations for each joint, which are actuated using PD controllers. Spherical joint targets are represented using 3D exponential maps \citep{Grassia1998ExpMap}, while revolute joints are represented using scalar rotation angles.

\subsection{Network Architecture}
The policy $\pi$ is modeled with a neural network that maps a given state $\rvs_t$ to a Gaussian distribution over the actions, $\pi(\rva_t | \rvs_t) = \mathcal{N}\left(\rvmu(\rvs_t), \Sigma \right)$. The covariance matrix $\Sigma$ is fixed over the course of training, and it is represented by a diagonal matrix $\Sigma=\mathrm{diag}(\sigma_1, \sigma_2, ...)$ with manually specified values. The input-dependent mean $\rvmu(\rvs_t)$ is modeled by a fully connected network with two hidden layers, consisting of 1024 and 512 ReLU-activated units \citep{ReLU2010}, followed by a linear output layer. Similar architectures are adopted for the value function $V(\rvs_t)$ and discriminator $D(\Delta)$, except that their output layers consist of a single linear unit.

\begin{algorithm}[t]
\caption{ADD Training Procedure for Motion Imitation}
\label{alg:ADD}
\begin{algorithmic}[1]
\STATE{{\bf input} $\mathcal{M}$: a reference motion clip or dataset}

\STATE{$D \leftarrow$ initialize discriminator}
\STATE{$\pi \leftarrow$ initialize policy}
\STATE{$V \leftarrow$ initialize value function}
\STATE{$\mathcal{B} \leftarrow \emptyset$ \ initialize experience buffer}
\item[]
\WHILE{not done}
    \FOR{trajectory $i = 1,...,m$}
    	\STATE{$\tau^i \leftarrow \{(\rvs_t, \rva_t)_{t=0}^{T-1}, \rvs_T\}$ collect trajectory with $\pi$}
            \STATE{$\hat{\tau}^i \leftarrow \{(\hat{\rvs}_t)_{t=0}^{T}\}$ sample reference trajectory from $\mathcal{M}$}
        \FOR{time step $t = 0,...,T-1$}
            \STATE{$\Delta_{t} \leftarrow \phi(\hat{\rvs}_{t}) \ominus \phi(\rvs_{t})$}
            \STATE{$d_t \leftarrow D(\Delta_t)$}
            \STATE{$r_t \leftarrow $ calculate reward according to Equation~\ref{Eq:add_reward} using $d_t$}
            \STATE{record $r_t$ in $\tau^i$}
        \ENDFOR
        \STATE{store $\tau^i$ in $\mathcal{B}$}
    \ENDFOR
    \item[]
    \FOR{update step $= 1,...,n$}
        \STATE{$b^\pi \leftarrow$ sample batch of $K$ differentials $\{\Delta_j\}_{j=1}^K$ from $\mathcal{B}$}
        \STATE{update $D$ (Equation~\ref{Eq:motion_tracking_disc_loss}), $V$, and $\pi$ using $b^\pi$}
    \ENDFOR
\ENDWHILE
\end{algorithmic}
\end{algorithm}

\subsection{Training}
\label{sec:motion-training}
An overview of the ADD training procedure is outlined in Algorithm~\ref{alg:ADD}. Following \citet{deep_mimic}, the character is initialized to starting states randomly sampled from the reference motion. At each timestep $t$, features are extracted from the agent's state $\phi(\rvs_t)$ and the target state from the reference motion $\phi(\hat{\rvs}_t)$. The difference $\Delta_t = \phi(\hat{\rvs}_t) \ominus \phi(\rvs_t)$ is then provided as input to the discriminator. The discriminator $D(\Delta_t)$ then outputs a score, which is used to calculate the reward for the agent according to Eq.~\ref{Eq:add_reward}. 

The trajectories collected by the agent are recorded in an experience buffer $\mathcal{B}$. After a batch of data is collected, mini-batches are sampled from the buffer to update the discriminator, policy, and value function. The policy is updated using PPO \citep{ppo}, with advantages computed with GAE($\lambda$) \citep{gae}. The value function is updated using target values computed with TD($\lambda$) \citep{reinforcement_learning}, and the discriminator is updated according to Eq. \ref{Eq:motion_tracking_disc_loss}.

\subsection{Motion Imitation Results}
To benchmark ADD's motion tracking performance, we compare ADD against two well-established methods: DeepMimic and AMP \citep{deep_mimic, 2021-TOG-AMP}. DeepMimic, designed for precise motion tracking, relies on a manually-designed imitation reward function composed of multiple sub-terms,
\begin{equation}
    r^{\text{DM}}_t = w^p r_t^p + w^{jv} r_t^{jv} + w^{rv} r_t^{rv} + w^e r_t^e + w^c r_t^c.
\end{equation}
Each term is an exponentiated error (Eq. \ref{eq:exponential_reward_subterm}), with scale $\alpha^i$ and weight $w^i$ hyperparameters (see Supplementary B.2 for the full formulation). Additionally, computing the joint pose $r_t^{p}$ and velocity rewards $r_t^{jv}$ involves joint-specific weights, requiring additional tuning effort.
Similar to ADD, DeepMimic policies receive phase information in the form of target frames, synchronizing the policy with a given reference motion. In contrast, AMP adopts an adversarial imitation learning framework to imitate the general style of a motion dataset, rather than exact motion tracking.
While it is not designed for precise motion tracking, AMP is included in the comparisons because ADD builds upon a similar adversarial framework. However, by adapting AMP's imitation objective, our approach can change the learning objective from general style imitation to accurate motion tracking. Many follow-up works extend DeepMimic and AMP with architectural innovations orthogonal to the imitation objective \citep{Luo2023PerpetualHC, tessler2024maskedmimic, 2022-TOG-ASE}. However, our experiments focus on the core imitation objective. Therefore, we compare our method directly with DeepMimic and AMP. ADD can also be combined with these additional enhancements in a similar way that DeepMimic and AMP have been extended.

For a fair comparison across methods, we disable pose termination used in \citet{deep_mimic}, which terminates an episode if the character’s pose deviates significantly from the reference. Pose termination is not applicable to distribution matching techniques such as AMP, where the policy is not synchronized with the reference motion. Early termination is triggered only when the character makes undesired contact with the ground. The baselines are implemented based on publicly available code provided and tuned by \citet{2021-TOG-AMP, deep_mimic} to ensure reliable comparisons. Motion tracking performance is evaluated using the position tracking error $e^\mathrm{pos}_t$, and DoF velocity tracking error, which is an indicator of motion smoothness. $e^\mathrm{pos}_t$ measures the difference in the root position and relative joint positions between the simulated character and the reference motion:
\begin{align}
\label{Eq:pos_tracking_err}
e^\mathrm{pos}_t = \frac{1}{N^\mathrm{joint} + 1} \bigg( & \sum\limits_{j \in \mathrm{joints}} \left|\left|(\hat{\rvx}^j_t - \hat{\rvx}^\mathrm{root}_t) - (\rvx^j_t - \rvx^\mathrm{root}_t)\right|\right|_2 \nonumber \\
& + \left|\left|\hat{\rvx}^\mathrm{root}_t - \rvx^\mathrm{root}_t\right|\right|_2 \bigg).
\end{align}
Here, $\rvx^j_t$ and $\hat{\rvx}^j_t$ represent the 3D Cartesian position of joint $j$ from the simulated character and the reference motion, respectively. $N^\mathrm{joint}$ denotes the number of joints in the character. Detailed hyperparameter settings are available in Supplementary B.4.

\begin{table*}[t]
\renewcommand{\arraystretch}{1.15} 
\centering
\caption{Motion tracking performance of simulated humanoid characters trained using AMP \citep{2021-TOG-AMP}, DeepMimic \citep{deep_mimic}, and our method ADD. Position (Eq. \ref{Eq:pos_tracking_err}) and DoF Velocity tracking errors are averaged $\pm$ 1 std across 5 models initialized with random seeds. Due to computational constraints, 1 model is trained for the LaFAN1 subset. For each model, errors are averaged across 4096 test episodes. ADD achieves tracking performance comparable to DeepMimic when imitating individual motion clips and larger motion datasets, while alleviating the need for manual reward engineering.}
\label{tab:humanoid_motion_comp}
\begin{tabular}{|>{\centering\arraybackslash}m{2.5cm}|>{\centering\arraybackslash}m{1.5cm}|>{\centering\arraybackslash}m{1.7cm}|>{\centering\arraybackslash}m{1.7cm}|>{\centering\arraybackslash}m{1.7cm}|>{\centering\arraybackslash}m{1.7cm}|>{\centering\arraybackslash}m{1.7cm}|>{\centering\arraybackslash}m{1.7cm}|}
\hline
\textbf{Skill} & \textbf{Length (s)} & \multicolumn{3}{c|}{\textbf{Position Tracking Error [m]}} & \multicolumn{3}{c|}{\textbf{DoF Velocity Tracking Error [rad/s]}} \\ \hline
               &                     & \textbf{AMP} & \textbf{DeepMimic} & \textbf{ADD (ours)} & \textbf{AMP} & \textbf{DeepMimic} & \textbf{ADD (ours)} \\ \hline
Run            & 0.80 & $0.163^{\pm 0.008}$ & \cellcolor{blue!10}\(\mathbf{0.013^{\pm 0.002}}\)  & $0.165 ^{\pm 0.017}$ & $2.811 ^{\pm 0.048}$ & $0.584 ^{\pm 0.054}$ & \cellcolor{blue!10}\(\mathbf{0.478 ^{\pm 0.007}}\) \\ \hline
Jog            & 0.83 & $0.120^{\pm 0.007}$ & \cellcolor{blue!10}\(\mathbf{0.021^{\pm 0.000}}\) & $0.024^{\pm 0.004}$ & $2.017 ^{\pm 0.052}$ & $0.575 ^{\pm 0.007}$ & \cellcolor{blue!10}\(\mathbf{0.507 ^{\pm 0.010}}\) \\ \hline
Sideflip       & 2.44 & $0.387^{\pm 0.011}$ & \cellcolor{blue!10}$\mathbf{0.138^{\pm 0.004}}$ & $0.145^{\pm 0.006}$ & $2.276 ^{\pm 0.014}$ & \cellcolor{blue!10}$\mathbf{1.118 ^{\pm 0.034}}$ & $1.350 ^{\pm 0.049}$ \\ \hline
Crawl          & 2.93 & $0.050^{\pm 0.006}$ & \cellcolor{blue!10}$\mathbf{0.027^{\pm 0.000}}$ & $0.028^{\pm 0.002}$ & $0.646 ^{\pm 0.089}$ & $0.430 ^{\pm 0.006}$ & \cellcolor{blue!10}$\mathbf{0.283 ^{\pm 0.002}}$ \\ \hline
Roll           & 2.00 & $0.141^{\pm 0.031}$ & \cellcolor{blue!10}$\mathbf{0.115^{\pm 0.132}}$ & $0.152^{\pm 0.005}$ & $1.576 ^{\pm 0.318}$ & \cellcolor{blue!10}$\mathbf{0.994 ^{\pm 0.051}}$ & $1.330 ^{\pm 0.101}$ \\ \hline
Double Kong    & 5.17 & $0.214^{\pm 0.014}$ & $0.223^{\pm 0.006}$ & \cellcolor{blue!10}$\mathbf{0.030^{\pm 0.001}}$ & $1.309 ^{\pm 0.043}$ & $0.784 ^{\pm 0.013}$ & \cellcolor{blue!10}$\mathbf{0.473 ^{\pm 0.004}}$ \\ \hline
Getup Facedown & 3.03 & $0.096^{\pm 0.018}$ & $0.023^{\pm 0.001}$ & \cellcolor{blue!10}$\mathbf{0.022^{\pm 0.001}}$ & $0.838 ^{\pm 0.029}$ & $0.433 ^{\pm 0.008}$ & \cellcolor{blue!10}$\mathbf{0.325 ^{\pm 0.005}}$ \\ \hline
Spinkick       & 1.28 & $0.064^{\pm 0.010}$ & $0.078^{\pm 0.062}$ & \cellcolor{blue!10}$\mathbf{0.025^{\pm 0.000}}$ & $1.453 ^{\pm 0.327}$ & $1.222 ^{\pm 0.233}$ & \cellcolor{blue!10}$\mathbf{0.774 ^{\pm 0.007}}$ \\ \hline
Cartwheel      & 2.71 & $0.076^{\pm 0.006}$ & $0.144^{\pm 0.153}$ & \cellcolor{blue!10}$\mathbf{0.017^{\pm 0.000}}$ & $0.722 ^{\pm 0.020}$ & $0.659 ^{\pm 0.160}$ & \cellcolor{blue!10}$\mathbf{0.317 ^{\pm 0.002}}$ \\ \hline
Backflip       & 1.75 & $0.267^{\pm 0.015}$ & $0.111^{\pm 0.054}$ & \cellcolor{blue!10}$\mathbf{0.062^{\pm 0.001}}$ & $2.243 ^{\pm 0.113}$ & $1.103 ^{\pm 0.024}$ & \cellcolor{blue!10}$\mathbf{0.878 ^{\pm 0.013}}$ \\ \hline
Climb          & 9.97 & $0.185^{\pm 0.021}$ & $0.066^{\pm 0.004}$ & \cellcolor{blue!10}$\mathbf{0.046^{\pm 0.031}}$ & $1.190 ^{\pm 0.012}$ & $0.567 ^{\pm 0.015}$ & \cellcolor{blue!10}$\mathbf{0.374 ^{\pm 0.071}}$ \\ \hline
Dance A        & 1.62 & $0.065^{\pm 0.009}$ & $0.065^{\pm 0.029}$ & \cellcolor{blue!10}$\mathbf{0.028^{\pm 0.007}}$ & $0.895 ^{\pm 0.108}$ & $0.830 ^{\pm 0.090}$ & \cellcolor{blue!10}$\mathbf{0.428 ^{\pm 0.014}}$ \\ \hline
Walk           & 0.96 & $0.132^{\pm 0.021}$ & $0.009^{\pm 0.001}$ & \cellcolor{blue!10}$\mathbf{0.009^{\pm 0.001}}$ & $1.394 ^{\pm 0.123}$ & $0.286 ^{\pm 0.005}$ & \cellcolor{blue!10}$\mathbf{0.213 ^{\pm 0.003}}$ \\ \hline
AMASS DanceDB & 2,804& $0.299^{\pm 0.001}$ & $0.045 ^{\pm 0.002}$ & \cellcolor{blue!10}$\mathbf{0.044 ^{\pm 0.001}}$ & $1.156 ^{\pm 0.009}$ & $0.504 ^{\pm 0.019}$ & \cellcolor{blue!10}$\mathbf{0.387 ^{\pm 0.010}}$ \\ \hline
LaFAN1 subset & 5,470 & $0.438^{\pm 0.000}$ & $0.029 ^{\pm 0.000}$ & \cellcolor{blue!10}$\mathbf{0.028 ^{\pm 0.000}}$ & $1.302 ^{\pm 0.000}$ & $0.511 ^{\pm 0.000}$ & \cellcolor{blue!10}$\mathbf{0.393 ^{\pm 0.000}}$ \\ \hline
\end{tabular}
\end{table*}

\begin{table}[t]
\caption{Position tracking errors (Eq.~\ref{Eq:pos_tracking_err}) of simulated EVAL robots trained using ADD, AMP \citep{2021-TOG-AMP}, and DeepMimic \citep{deep_mimic}. Only the position tracking errors are reported for brevity, showing the mean $\pm$ 1 standard deviation across three random seeds, with 4096 test episodes per seed. The results show that ADD maintains strong tracking performance comparable to DeepMimic on a different character morphology.}
\label{tab:EVALMotionTracking}
\begin{tabular}{|>{\centering\arraybackslash}m{1.3cm}|>{\centering\arraybackslash}m{1.9cm}|>{\centering\arraybackslash}m{1.9cm}|>{\centering\arraybackslash}m{1.9cm}|}
\hline
\textbf{Skill} & \multicolumn{3}{c|}{\textbf{Position Tracking Error [m]}} \\ \hline
              & \textbf{AMP} & \textbf{DeepMimic} & \textbf{ADD (ours)} \\ \hline
Walk          & $0.020^{\pm 0.001}$ & \cellcolor{blue!10}$\mathbf{0.013^{\pm 0.002}}$ & $0.036^{\pm 0.002}$ \\ \hline
Kick          & $0.030^{ \pm 0.00}$ & $0.012^{ \pm 0.000}$ & \cellcolor{blue!10}$\mathbf{0.008^{ \pm 0.000}}$  \\ \hline
Punch         & $0.020^{\pm 0.003}$ & $0.007 ^{\pm 0.000}$ & \cellcolor{blue!10}$\mathbf{0.005 ^{\pm 0.000}}$ \\ \hline
\end{tabular}
\end{table}

Figures \ref{fig:motion_snapshots} and \ref{fig:EVAL_snapshots} showcase behaviors learned by the humanoid and EVAL robot trained via ADD. The behaviors are best viewed in the supplementary video. ADD can closely imitate individual motion clips and larger motion datasets with different embodiments, successfully reproducing a diverse set of agile and acrobatic skills. This includes challenging parkour skills, such as Climb and Double Kong, which require particularly high motion tracking accuracy to replicate intricate contacts with the environment. A qualitative comparison of ADD against other methods is available in the supplementary video.

Table~\ref{tab:humanoid_motion_comp} and Table~\ref{tab:EVALMotionTracking} summarize quantitative comparisons of the different methods. AMP exhibits poor tracking performance, since the policies are trained using a general distribution-matching objective. AMP's susceptibility to mode collapse also makes it less effective at tracking larger motion datasets. Both ADD and DeepMimic are able to accurately track a wide variety of reference motions. However, there are important distinctions in the robustness of the reward function and ease of deployment. The reliance on manually-designed reward functions, to a degree, limits DeepMimic’s ability to effectively imitate a broad spectrum of motions, as it can be challenging to craft a general and effective reward function that can imitate a diverse variety of behaviors. For instance, DeepMimic policies failed to reproduce some of the challenging parkour motions. For the Double Kong motion, DeepMimic policies fail to jump over the boxes and learn to simply mimic running in place after tripping. In contrast, ADD successfully enables characters to clear the obstacles, replicating the jumps and intricate contacts. Policies trained using DeepMimic also exhibit notable jitteriness when tracking the DanceDB dataset, whereas ADD consistently produces smooth behaviors across diverse motions, as indicated by its lower DoF velocity tracking errors. Additional reward tuning, especially for DanceDB, may further improve DeepMimic's performance. But by automatically learning to balance different objectives instead of relying on fixed manually-specified parameters, ADD offers a more general and adaptive approach for imitating more diverse motions.

\begin{figure}[t]
	\centering
        \captionsetup{skip=0pt}
    \subfigure{\includegraphics[height=0.34\columnwidth]{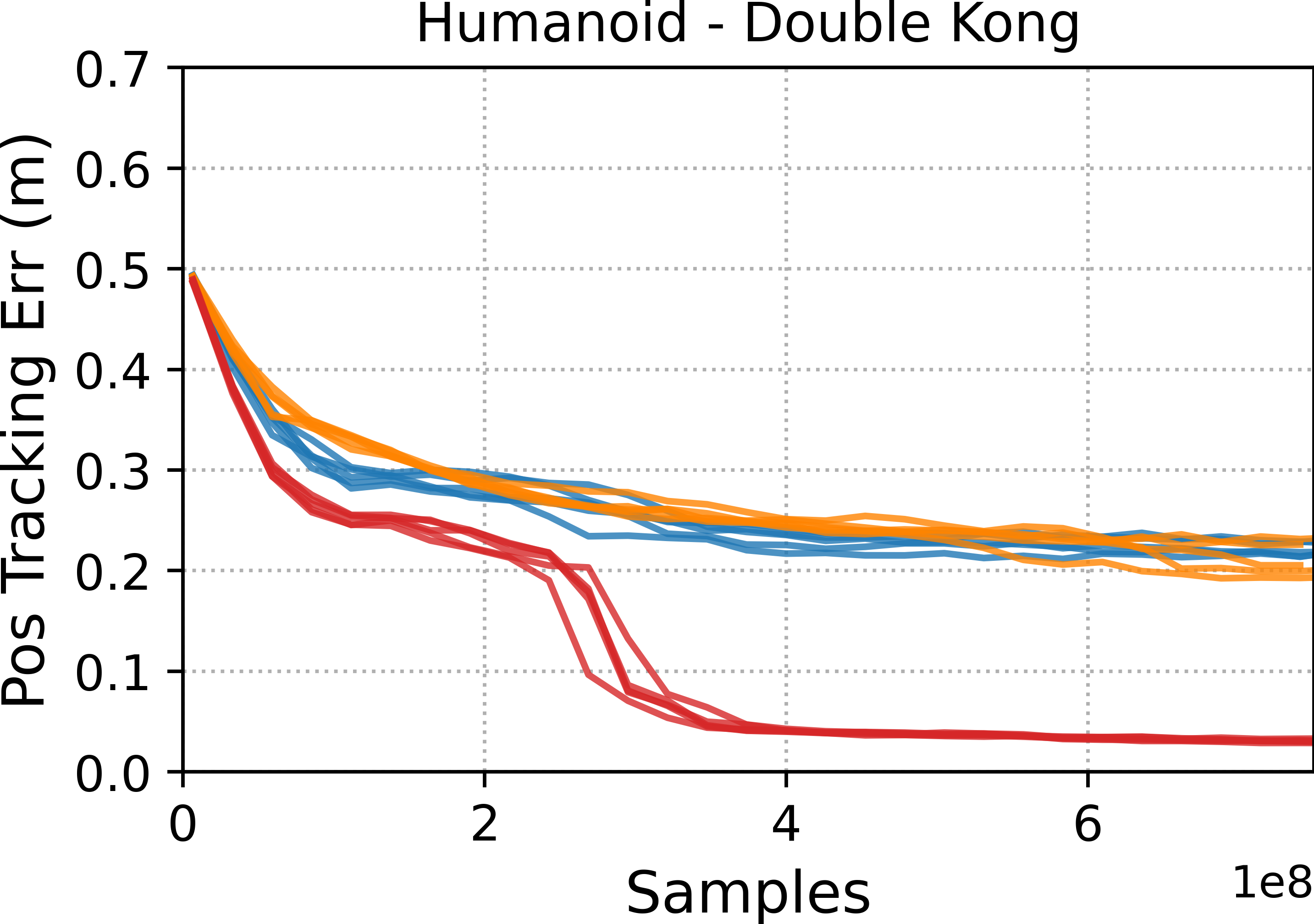}}
    \subfigure{\includegraphics[height=0.34\columnwidth]{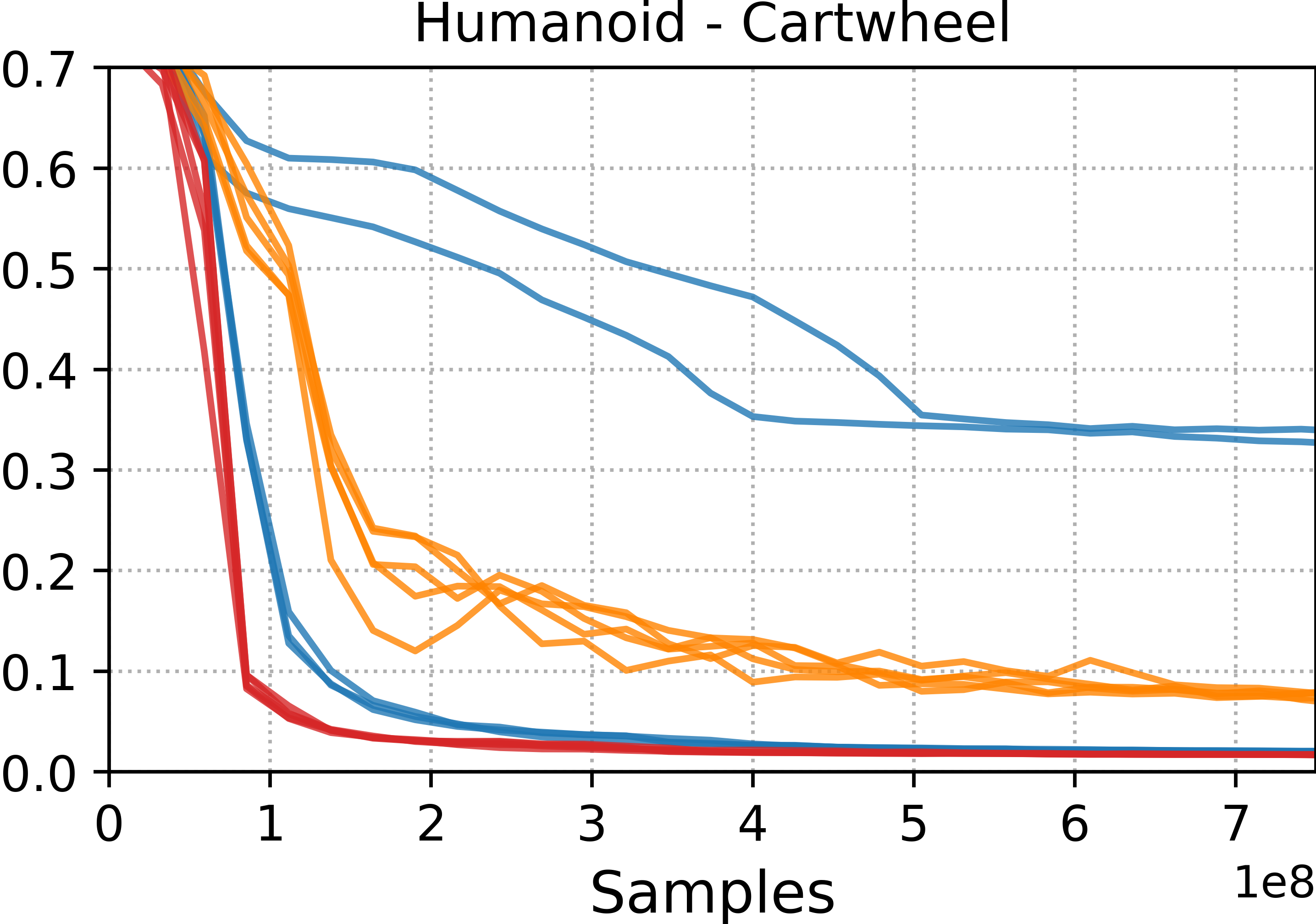}}\\ 
    \subfigure{\includegraphics[height=0.34\columnwidth]{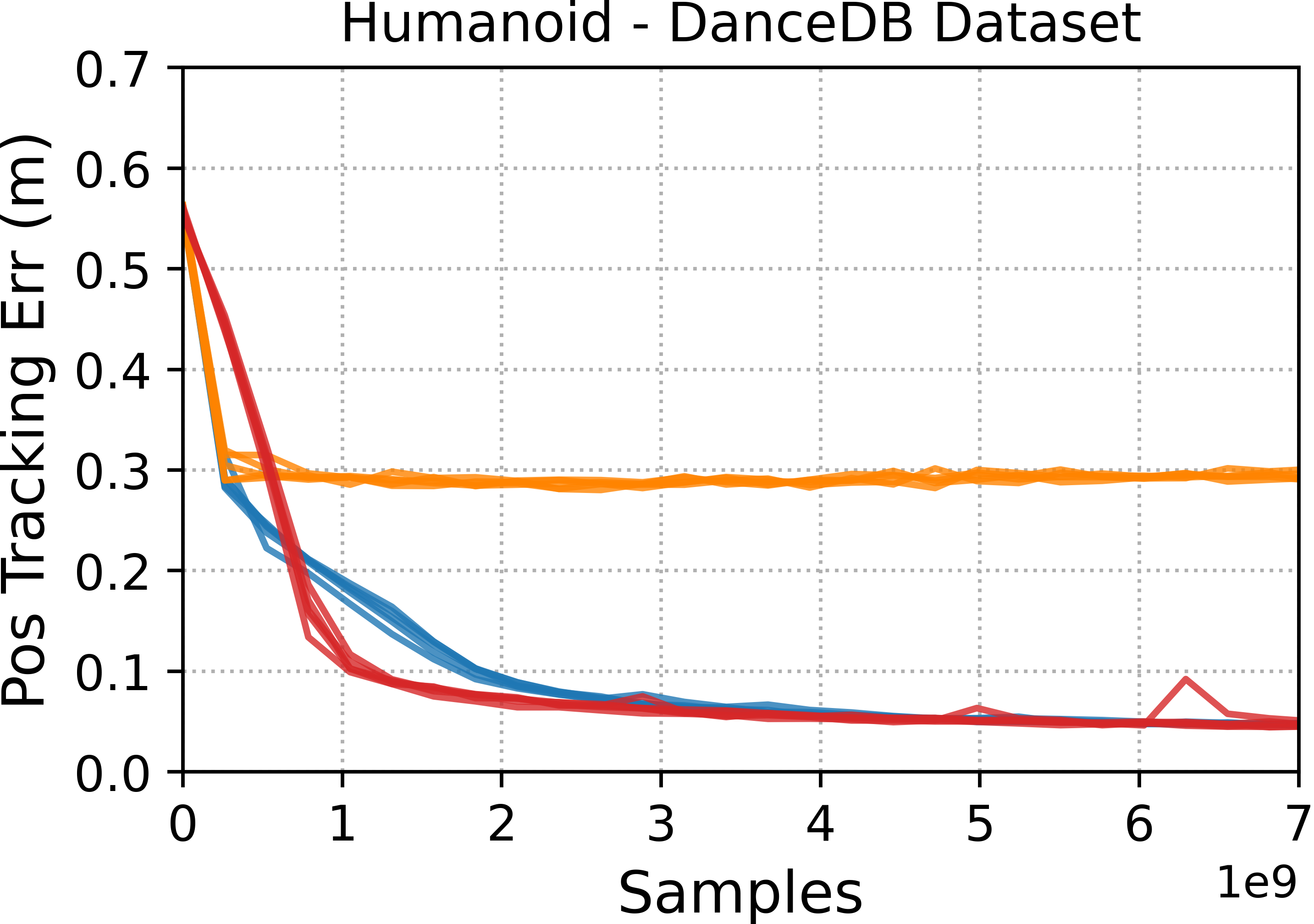}}
    \subfigure{\includegraphics[height=0.34\columnwidth]{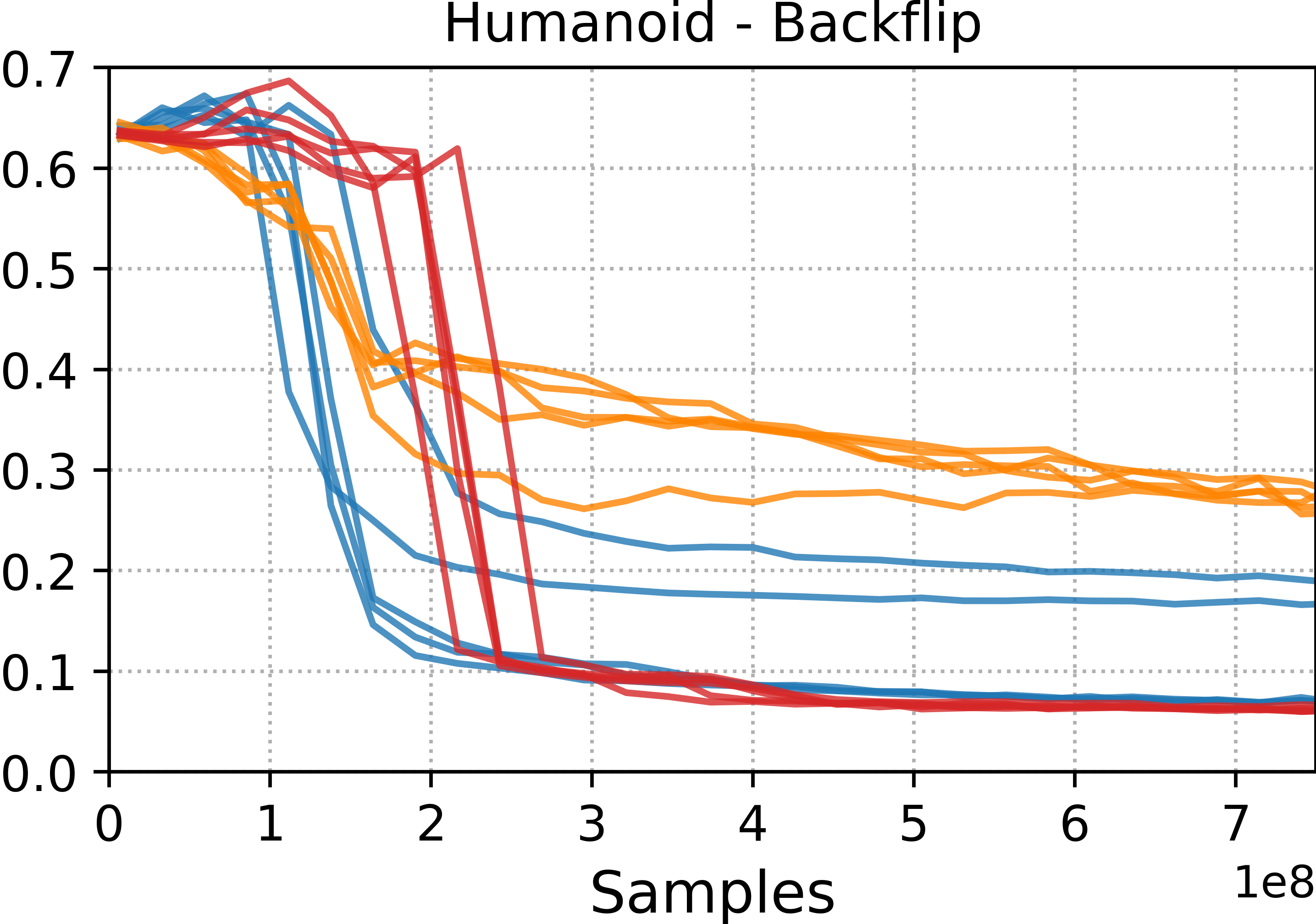}}\\
    \vspace{-0.2cm}
    \subfigure{\includegraphics[width=0.6\linewidth]{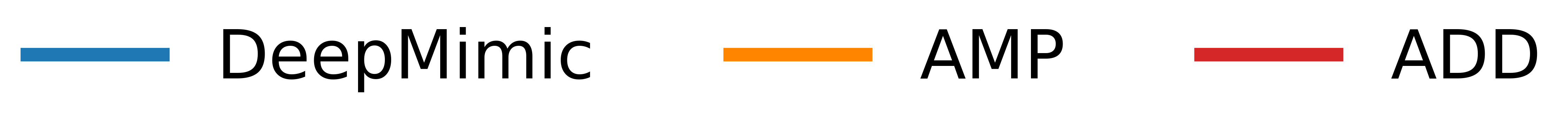}}\\
    \vspace{-0.1cm}
\caption{Learning curves of AMP, DeepMimic, and ADD, each trained with 5 different random seeds. ADD demonstrates better consistency than DeepMimic across different seeds. DeepMimic policies converge to suboptimal behaviors in half of the seeds when tracking the Backflip and Cartwheel motions.}
\label{fig:motion_imitation_learning_curve_selected}
\end{figure}

Figure~\ref{fig:motion_imitation_learning_curve_selected} compares the learning curves of humanoid characters trained using ADD, AMP, and DeepMimic. ADD demonstrates better consistency than DeepMimic across different seeds. DeepMimic policies converge to local optima in half of the seeds when imitating the Backflip and Cartwheel motions. Learning curves for all skills and characters can be found in Supplementary B.5.

Achieving high-quality motion tracking with DeepMimic often requires careful tuning of numerous reward parameters. This tuning process can require significant domain knowledge and time. Some parameter settings might improve a subset of the objectives, but degrade the performance of other objectives. Poorly selected parameters can result in low motion quality and inferior sample efficiency, which presents a challenge for applying DeepMimic to new reference motions or characters. In our experiments, DeepMimic's hyperparameters required careful re-tuning when switching from the humanoid character to the EVAL robot to maintain good tracking performance (Supplementary B.4). In comparison, ADD can automatically weigh the different objectives, reducing the burden of manual reward engineering. The sensitivity analysis of DeepMimic in Supplementary B.3 provides further experiments that highlight the challenges of DeepMimic’s reward design and the robustness of ADD.

However, ADD can struggle to reproduce certain highly dynamic motions (e.g., sideflip and roll), often converging to suboptimal behaviors. In the case of the sideflip, the policy learns to stand on the ground without executing the full flip. Prior methods address this by incorporating additional early termination heuristics \citep{deep_mimic, Luo2023PerpetualHC}, which we deliberately omitted to ensure a fair comparison and focus our evaluations on the effects of different imitation objectives. Additionally, ADD is less precise than DeepMimic on some forward locomotion skills, such as humanoid running and robot walking. The performance gap stems from ADD's difficulty in accurately tracking the root position, which is crucial for imitating forward locomotion behaviors. We hypothesize that this shortcoming arises due to the root position only accounting for three elements within the high-dimensional differential vector $\Delta$, and the gradient penalty restricts the discriminator from placing significantly greater emphasis on the root position errors.

\subsection{Tasks}
\label{sec:motion_tracking_task}

\begin{table}[t]
\renewcommand{\arraystretch}{1.15} 
\caption{Performance of various methods when applied to a composite task that combines motion imitation with a target steering objective. Motion imitation performance is measured via the position tracking error, while task performance is measured by the velocity error $\left|\left|\rvv_t - v^* \rvd^*_t\right|\right|$, where $\rvv_t$ is the 2D root velocity of the character and $v^* \rvd^*_t$ the 2D target velocity. ADD, via automatically balancing the different objectives, achieved optimal performance on both objectives.}
\label{tab:steering_task_table}
\centering
\begin{tabular}{|>{\centering\arraybackslash}m{1.4cm}|>{\centering\arraybackslash}m{1.35cm}|>{\centering\arraybackslash}m{1.35cm}|>{\centering\arraybackslash}m{1.35cm}|>{\centering\arraybackslash}m{1.35cm}|}
\hline
 & \multicolumn{2}{c|}{\textbf{Pos Tracking Err [m]}} & \multicolumn{2}{c|}{\textbf{Target Vel Err [m/s]}} \\ \hline
\diagbox[width=17.5mm]{Method}{Skill}  & {Run} & {Walk} & {Run} & {Walk} \\ \hline
AMP             & $0.188^{\pm 0.001}$ & $0.117^{\pm 0.001}$ & $0.810^{\pm 0.015}$ & $0.374^{\pm 0.042}$ \\ \hline
DeepMimic             & $0.048^{ \pm 0.000}$ & $0.034^{ \pm 0.003}$ & $0.882^{ \pm 0.038}$ & $0.411^{\pm 0.049}$ \\ \hline
ADD (ours)             & \cellcolor{blue!10}$\mathbf{0.029^{ \pm 0.001}}$ & \cellcolor{blue!10}$\mathbf{0.024^{ \pm 0.001}}$ & \cellcolor{blue!10}$\mathbf{0.803^{ \pm 0.009}}$ & \cellcolor{blue!10}$\mathbf{0.274^{ \pm 0.003}}$ \\ \hline
\end{tabular}
\end{table}

{In this section, we evaluate the effectiveness of ADD on training control policies that can both imitate motion clips and accomplish additional task objectives. Our experiments focus on a target steering task, where the objective is to move at a target speed $v^*$ along a target direction $\rvd^{*}_t$, specified by a 2D unit vector on the horizontal plane. For ADD, these task objectives are incorporated directly into the differential vector $\Delta$, whereas for AMP and DeepMimic, a separate task reward is introduced alongside the imitation reward (Supplementary B.6). The target direction and speed are provided as observations to the policy and are randomized during training. Table~\ref{tab:steering_task_table} summarizes the performance of various methods. ADD is able to automatically balance the various objectives, enabling policies to follow the steering commands accurately while closely imitating the desired reference motion.}

\begin{figure}[t]
    \centering
    \captionsetup{skip=0pt}
    \subfigure[ADD (Ours)]{\includegraphics[width=\linewidth]{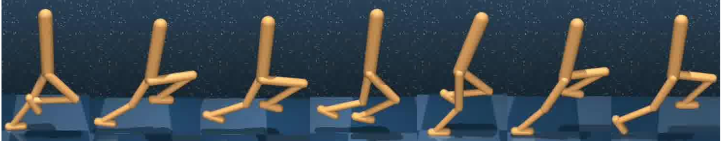}} \\
    \vspace{-0.2cm}
     \subfigure[Manual Reward]{\includegraphics[width=\linewidth]{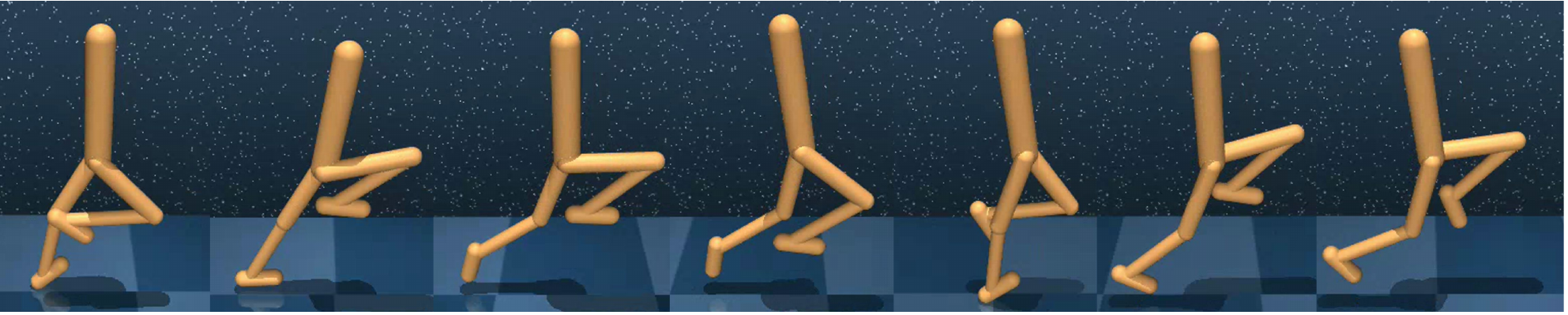}} \\
    \caption{Qualitative results of ADD on the Walker task, a standard RL benchmark. The walker trained using ADD exhibits behaviors comparable in quality to those learned with manually designed reward functions from \citet{2018DMControl}.}
\label{fig:walker_snapshot}
\end{figure}

\begin{figure}[t]
    \centering
    \captionsetup{skip=0pt}
    \subfigure[ADD (Ours)]{\includegraphics[width=\linewidth]{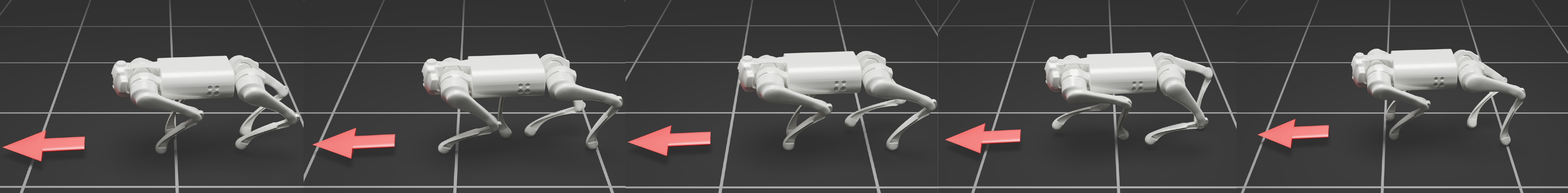}} \\
    \vspace{-0.2cm}
     \subfigure[Manual Reward]{\includegraphics[width=\linewidth]{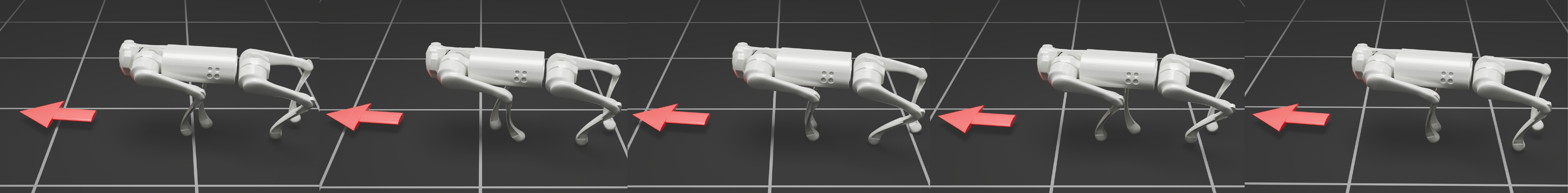}} \\
    \caption{Qualitative results of ADD on training a UniTree Go1 quadruped to walk. The arrow denotes the steering command. Compared to controllers trained with manually tuned rewards \citep{legged-gym-pmlr-v164-rudin22a}, the ADD Go1 policy displays more natural gaits, with greater foot lift and longer strides.}
\label{fig:go1_snapshot}
\end{figure}

\section{Non-Motion-Imitation Tasks}
To demonstrate that ADD is effective beyond motion imitation tasks, we evaluate ADD on the Walker task \citep{2018DMControl}, a standard RL benchmark task, and a UniTree Go1 walking task from a widely-used framework for robotic locomotion \citep{legged-gym-pmlr-v164-rudin22a}. The Walker task consists of 3 different objectives, whereas the Go1 task presents a much more complex reward function with 12 objectives, including 3 dynamic steering commands. These experiments demonstrate ADD's ability to balance a number of competing objectives.

\begin{figure}[t]
	\centering
        \captionsetup{skip=0pt}
    \subfigure{\includegraphics[height=0.34\columnwidth]{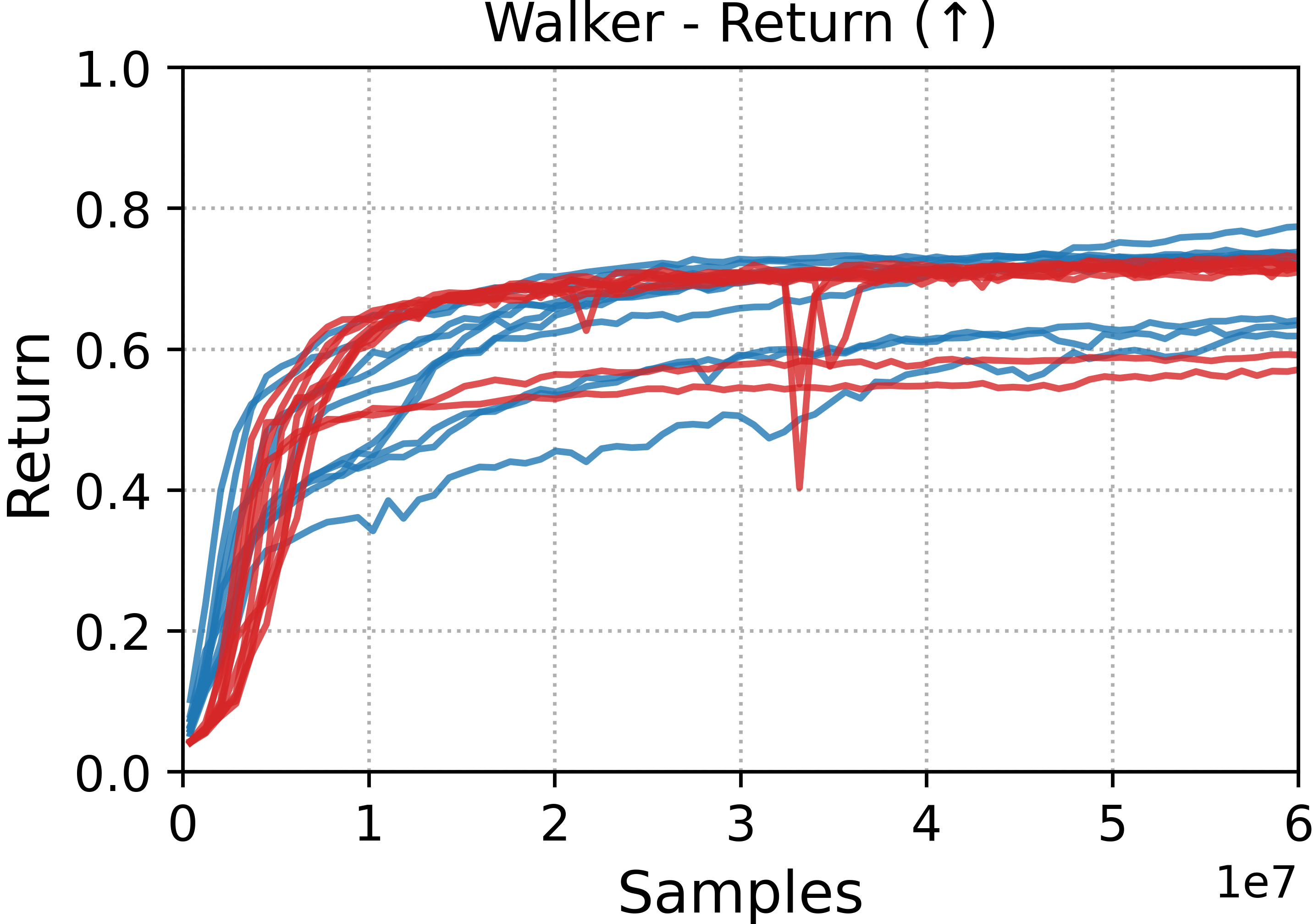}}
    \subfigure{\includegraphics[height=0.34\columnwidth]{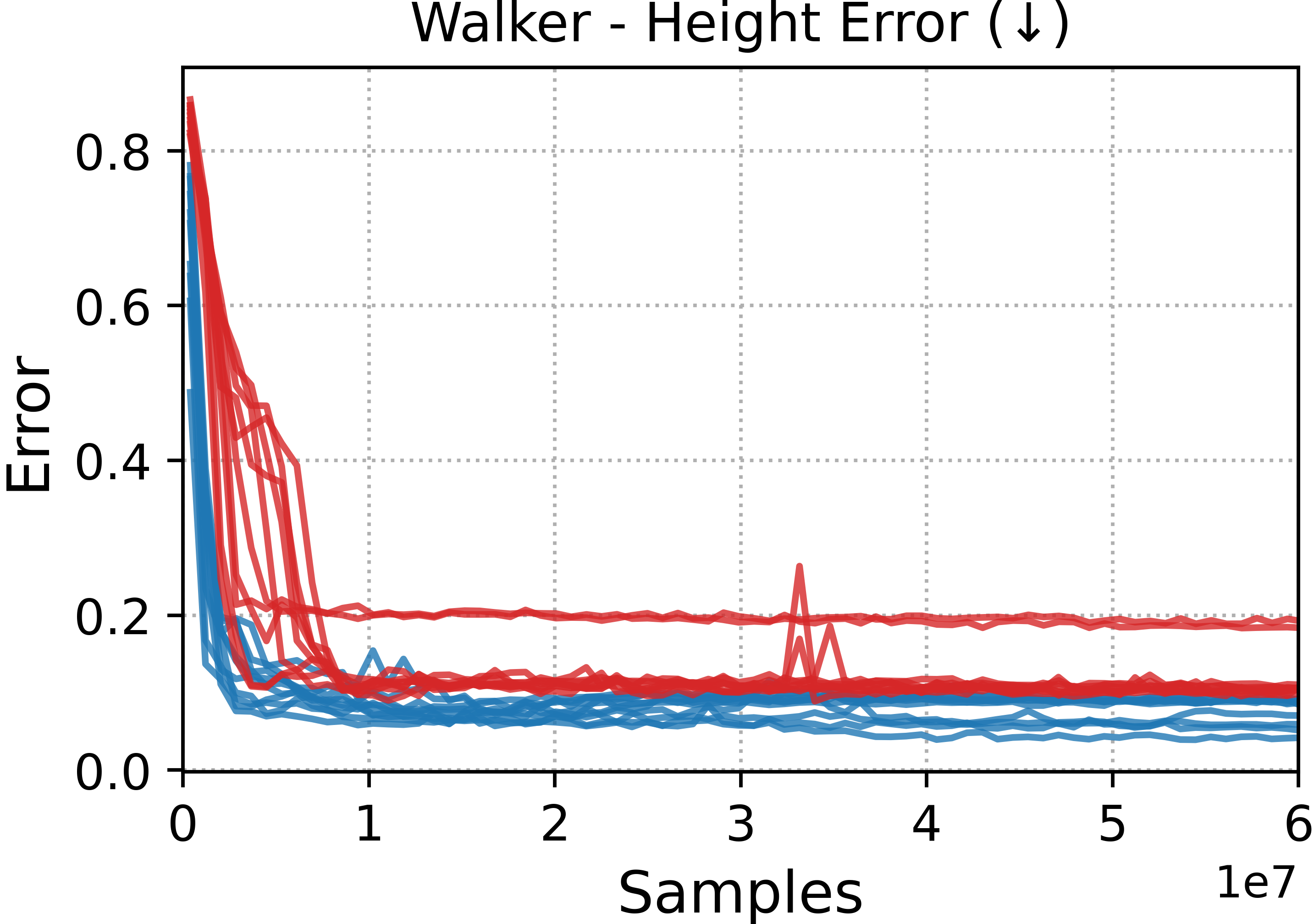}}\\
    \vspace{-0.23cm}
    \subfigure{\includegraphics[height=0.34\columnwidth]{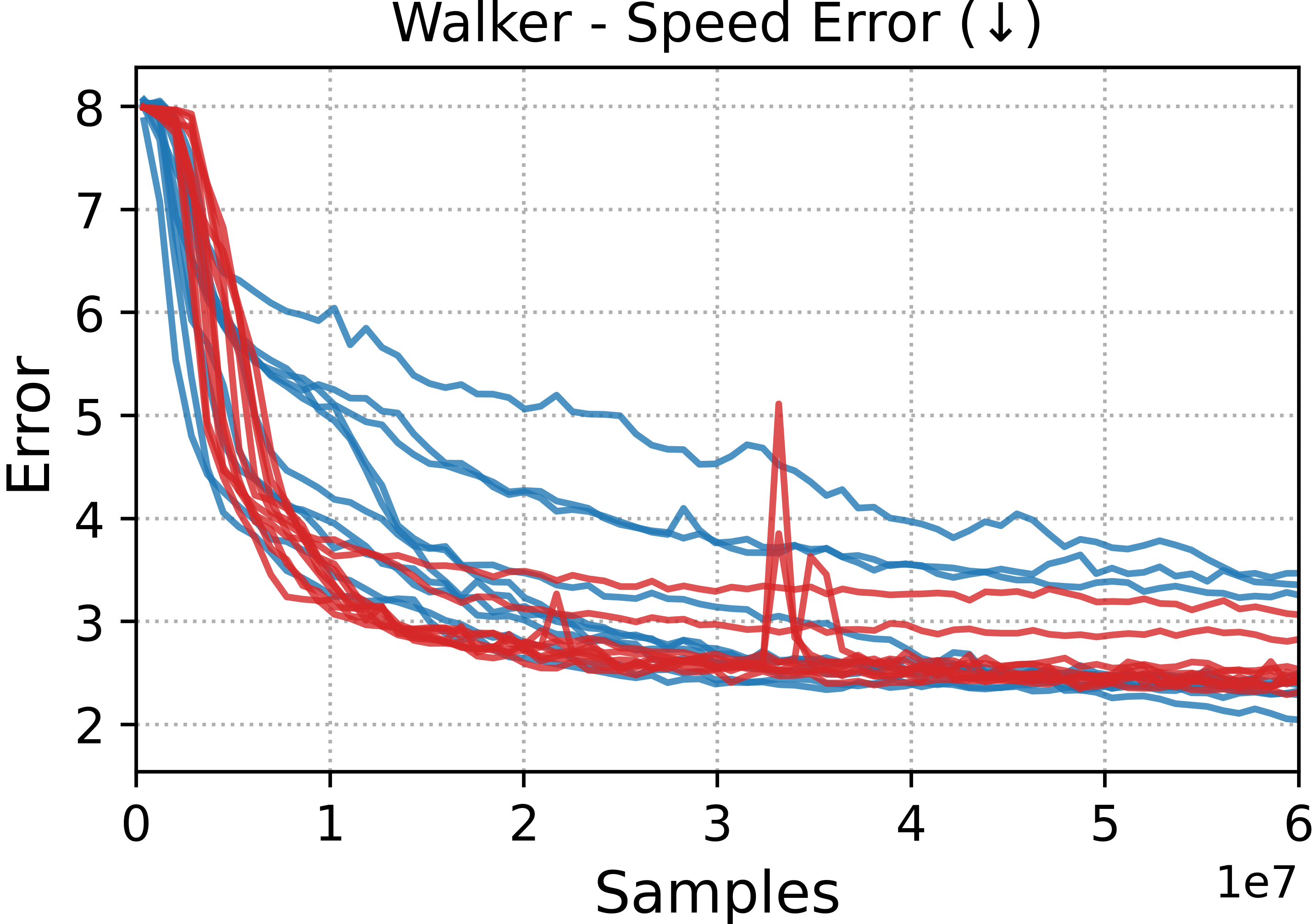}} 
    \subfigure{\includegraphics[height=0.34\columnwidth]{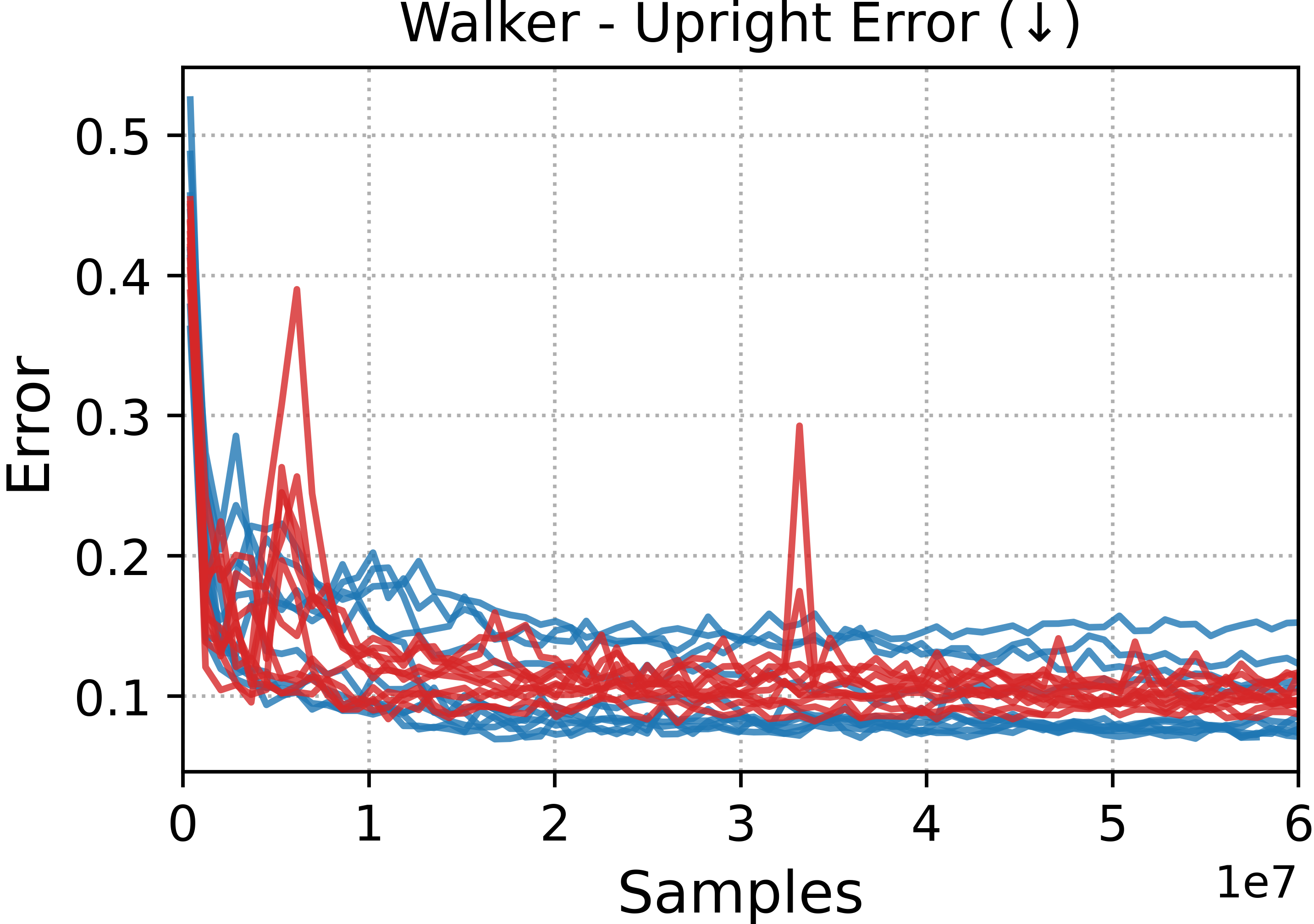}}\\
    \vspace{-0.2cm}
    \subfigure{\includegraphics[width=0.5\linewidth]{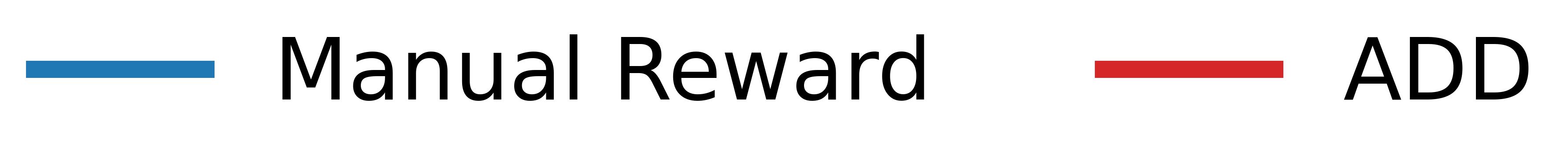}}\\
\caption{Learning curves comparing ADD to the manually designed reward function from \citet{2018DMControl} on the 2D Walker task. Returns are calculated according to the reward function from \citet{2018DMControl} and normalized between the minimum and maximum possible returns per episode. Statistics are computed over 10 random seeds. ADD demonstrates comparable performance and sample efficiency to the hand-crafted reward function, despite requiring no manual reward engineering or tuning.}
\label{fig:walker_learning_curves}
\end{figure}

\begin{figure}[t]
	\centering
        \captionsetup{skip=0pt}
    \subfigure{\includegraphics[height=0.34\columnwidth]{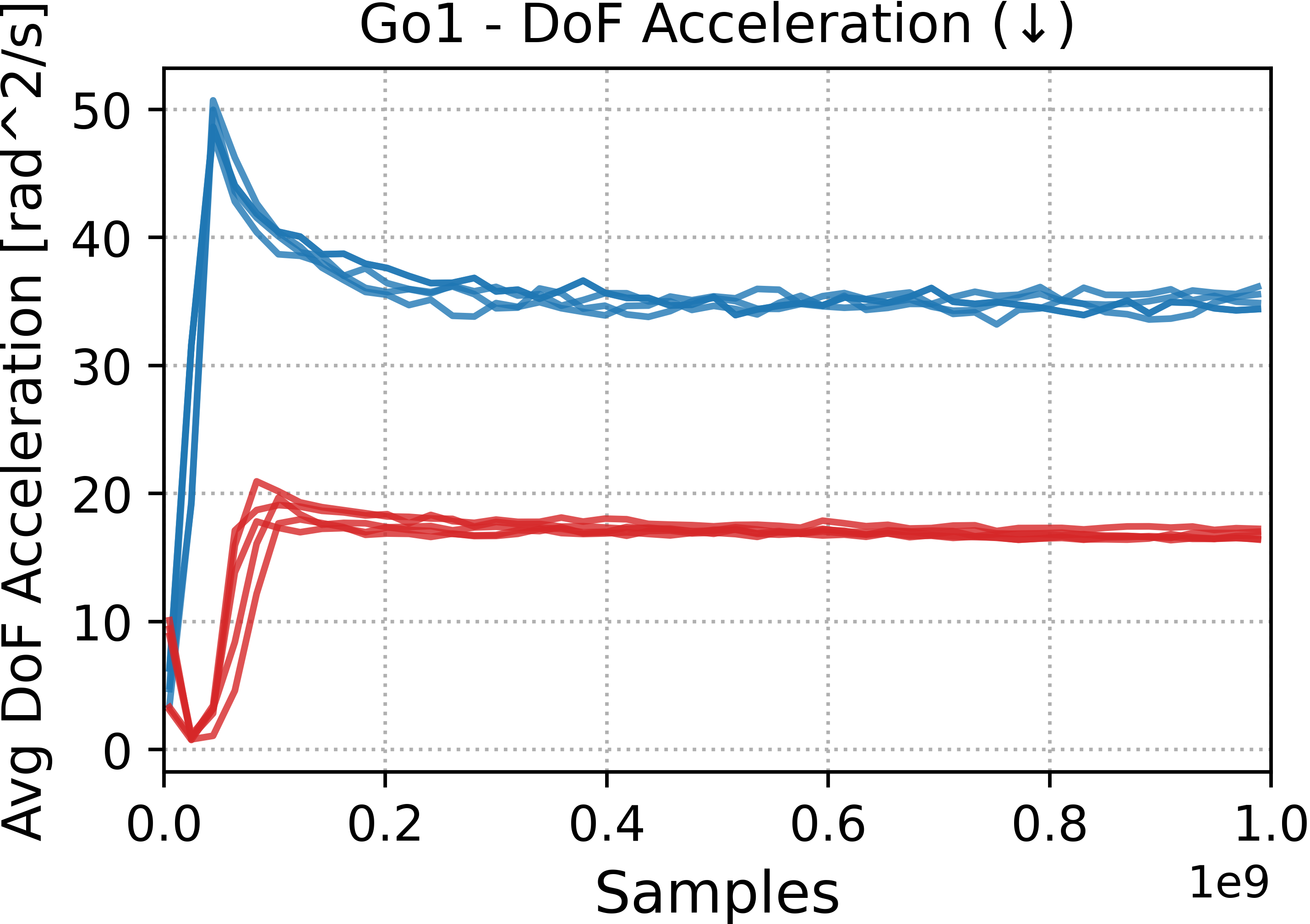}}
    \subfigure{\includegraphics[height=0.34\columnwidth]{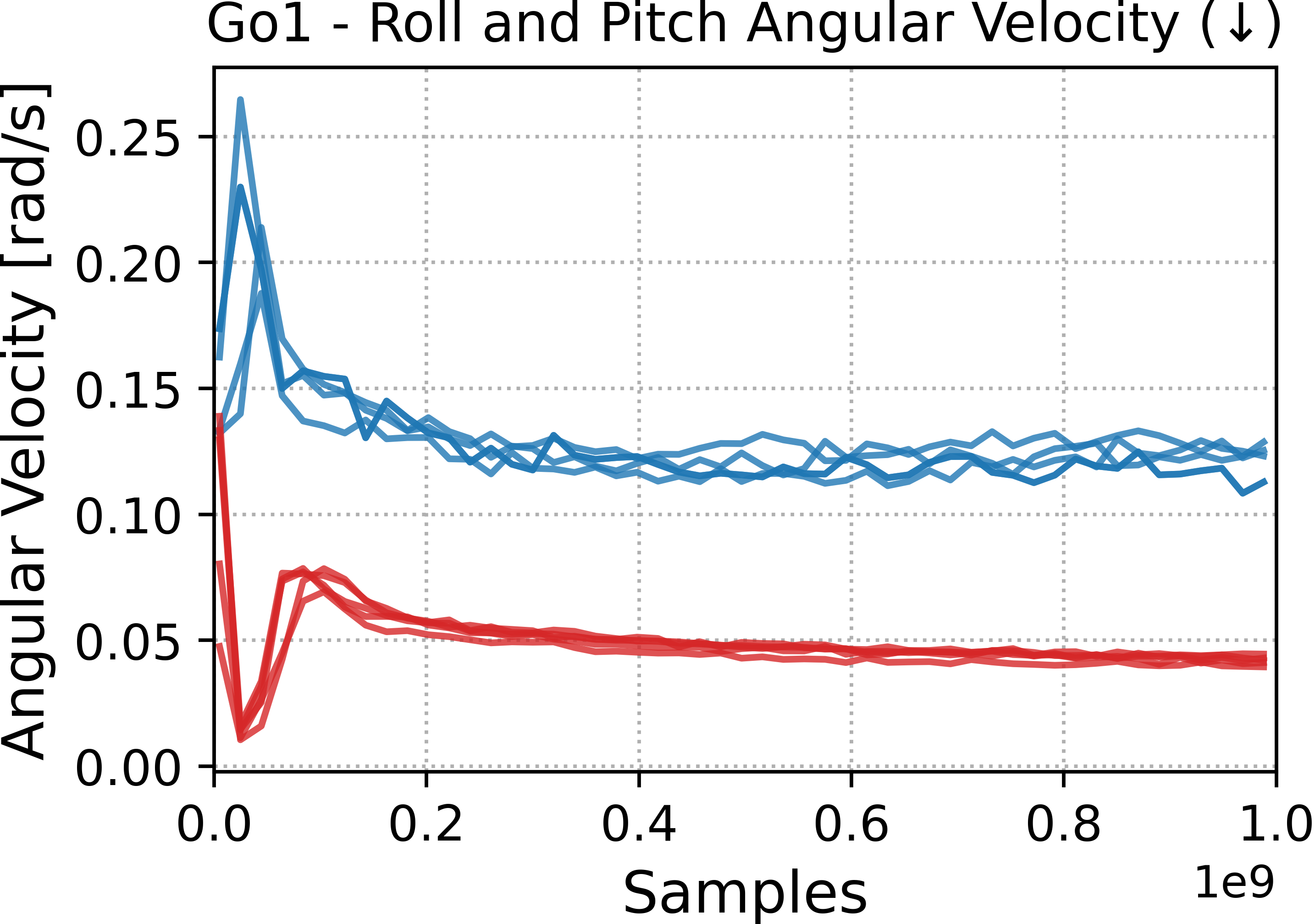}}\\
    \vspace{-0.23cm}
    \subfigure{\includegraphics[height=0.34\columnwidth]{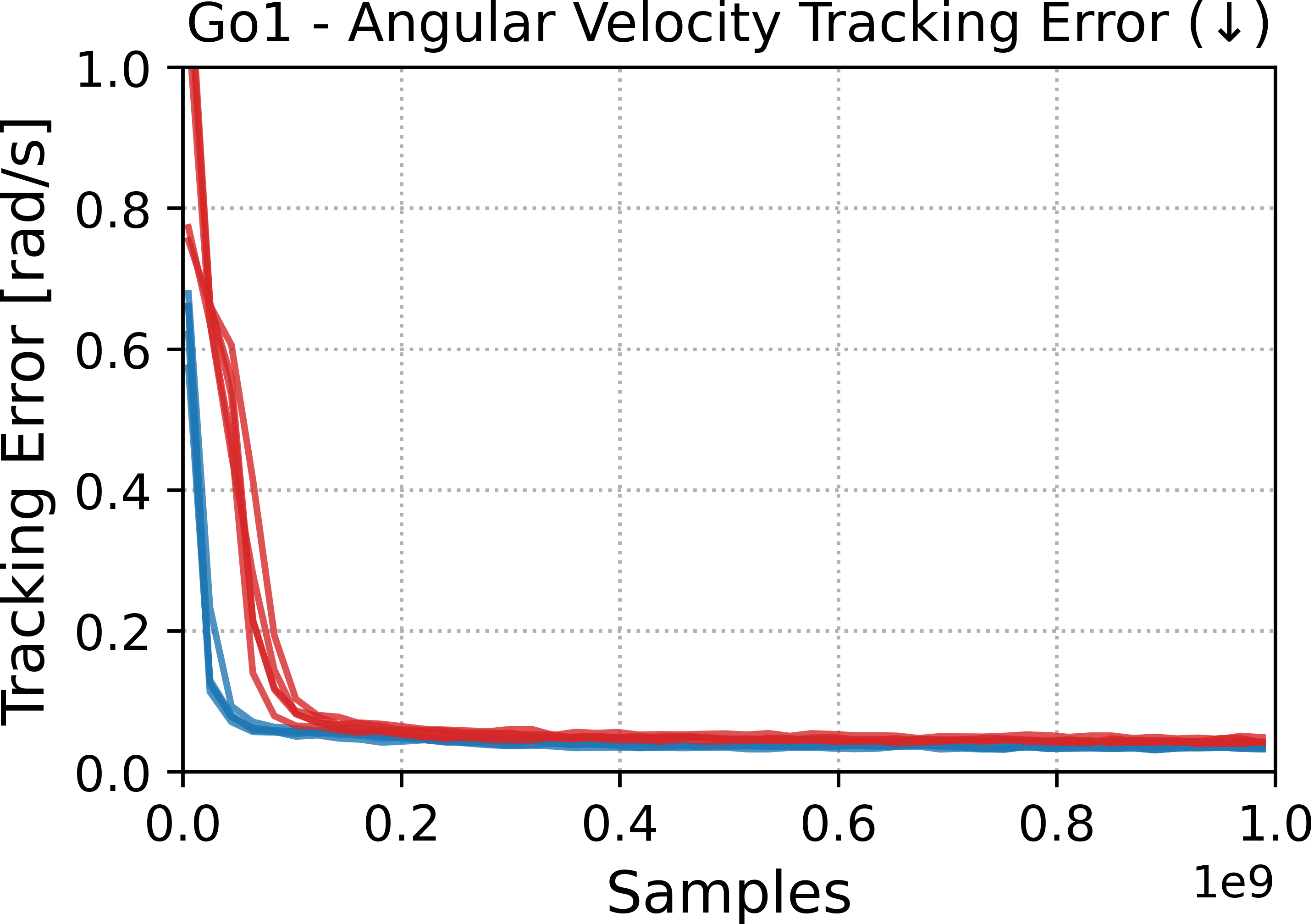}} 
    \subfigure{\includegraphics[height=0.34\columnwidth]{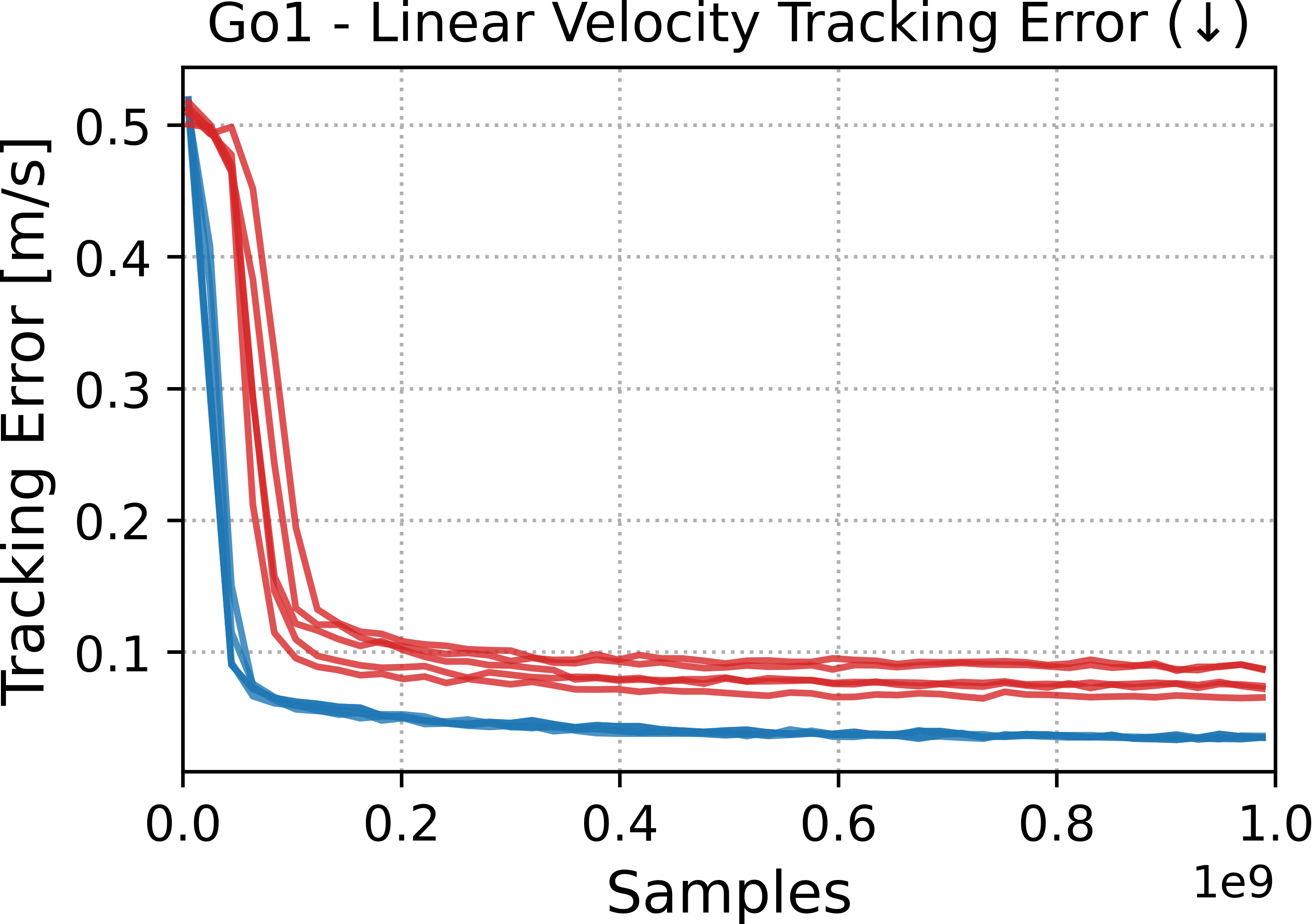}}\\
    \vspace{-0.2cm}
    \subfigure{\includegraphics[width=0.5\linewidth]{figures/walker_learning_curve_legend.png}}\\
\caption{Learning curves comparing ADD to the manually tuned reward function from \citet{legged-gym-pmlr-v164-rudin22a} on training a Go1 quadruped to move. Results are shown across 5 random seeds per method. While ADD performs slightly worse in following linear velocity commands, it achieves lower roll and pitch angular velocities--indicating a more stable robot base--and lower DoF accelerations, which means smoother control over time.}
\label{fig:go1_learning_curves}
\end{figure}

\subsection{Training}
Policies are trained using PPO with either ADD or manually-designed reward functions from \citet{2018DMControl} and \citet{legged-gym-pmlr-v164-rudin22a}. The ADD training procedure largely follows Algorithm~\ref{alg:ADD}, with the key difference being the absence of reference motions. Instead, the target state $\hat{\rvs}_t$ is composed of target values specified by the task, such as velocity commands or target orientations. Implementation details for each task are available in Supplementary C and D.

\subsection{Results}
Figures~\ref{fig:walker_snapshot} and \ref{fig:go1_snapshot} present qualitative comparisons between ADD and manually-designed reward functions on the two tasks: training a 2D walker to run and a quadrupedal Go1 robot to walk while following steering commands. Examples of the learned behaviors are available in the supplementary video. 

As shown in Figure~\ref{fig:walker_snapshot}, the ADD policy produces upright and fast running behaviors of comparable performance to those produced by the manually-designed reward function. This observation is supported by the learning curves in Figure~\ref{fig:walker_learning_curves}, which compares the two methods on the returns and the three different training objectives. ADD achieves similar final performance and sample efficiency as the manually-designed rewards. ADD produces a final return of $0.691 \pm 0.053$, whereas the manual rewards result in a final return of $0.705 \pm 0.051$. Furthermore, Figure~\ref{fig:walker_learning_curves} shows that ADD demonstrates better consistency across training runs.

On the Go1 task, ADD enables the Go1 robot to develop more natural gaits, characterized by greater foot lift and longer strides. In comparison, the manually-designed reward produces less natural behaviors that exhibit more jittering and small shuffling steps. Figure~\ref{fig:go1_learning_curves}, which compares ADD and \citet{legged-gym-pmlr-v164-rudin22a} on key objectives, shows that while ADD performs slightly worse on following linear velocity commands, it achieves lower roll and pitch angular velocities, indicating a more stable robot base. Additionally, ADD produces smoother movements with lower DoF accelerations, a desirable property for real-world robotic deployment. A comprehensive set of learning curves for all training objectives is available in Supplementary D.3.

Across different tasks and embodiments, ADD consistently matches the performance of carefully designed reward functions with respect to performance, motion quality, consistency, and sample efficiency, all while alleviating the dependence on manual reward engineering. While manually tuned methods can be effective when well-calibrated, they often require significant domain knowledge and effort to balance different objectives. These results collectively highlight ADD's generality as an effective solution for a diverse range of multi-objective RL tasks. 

\begin{figure}[t]
	\centering
        \captionsetup{skip=0pt}
    \subfigure{\includegraphics[height=0.34\columnwidth]{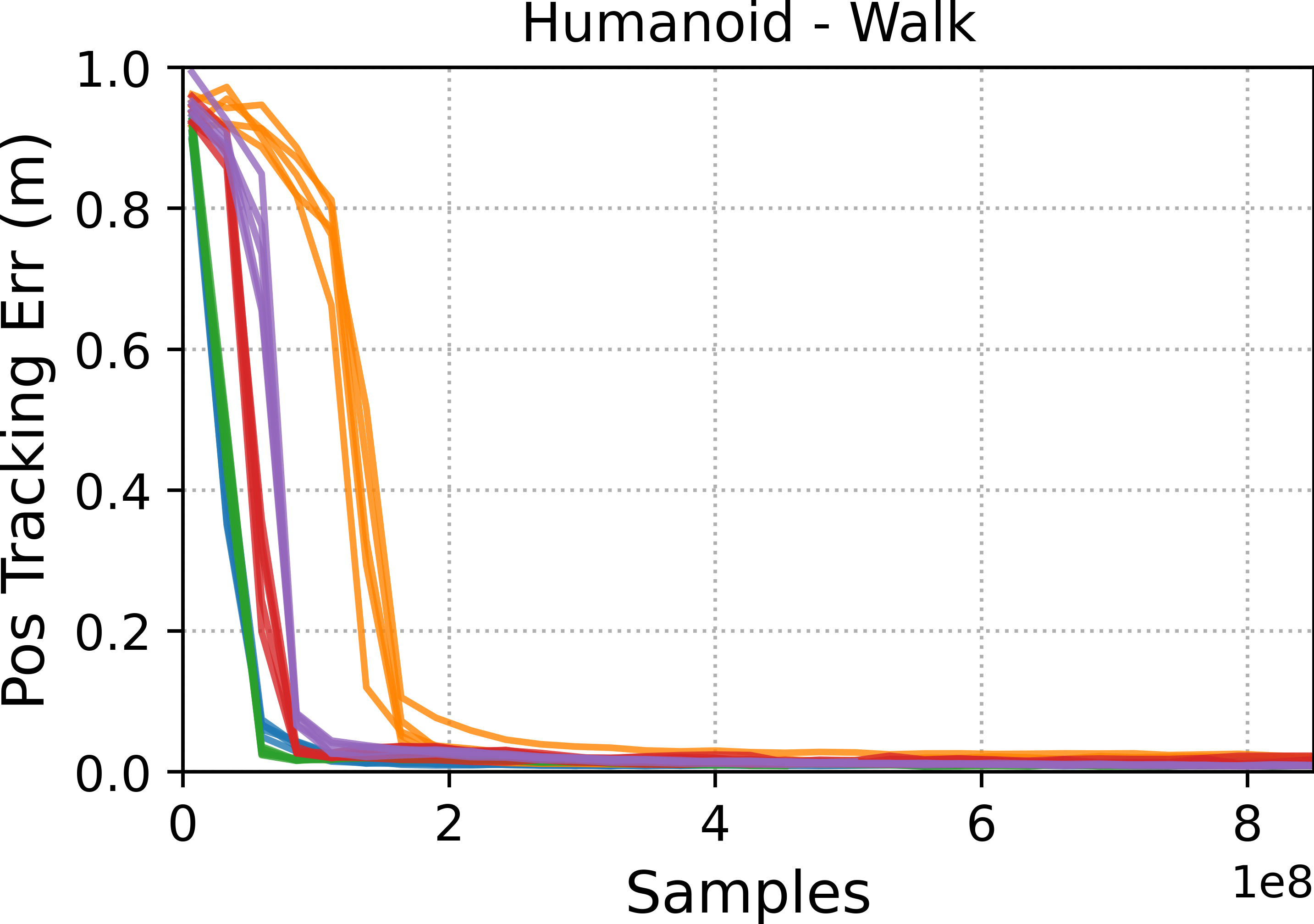}}
    \subfigure{\includegraphics[height=0.339\columnwidth]{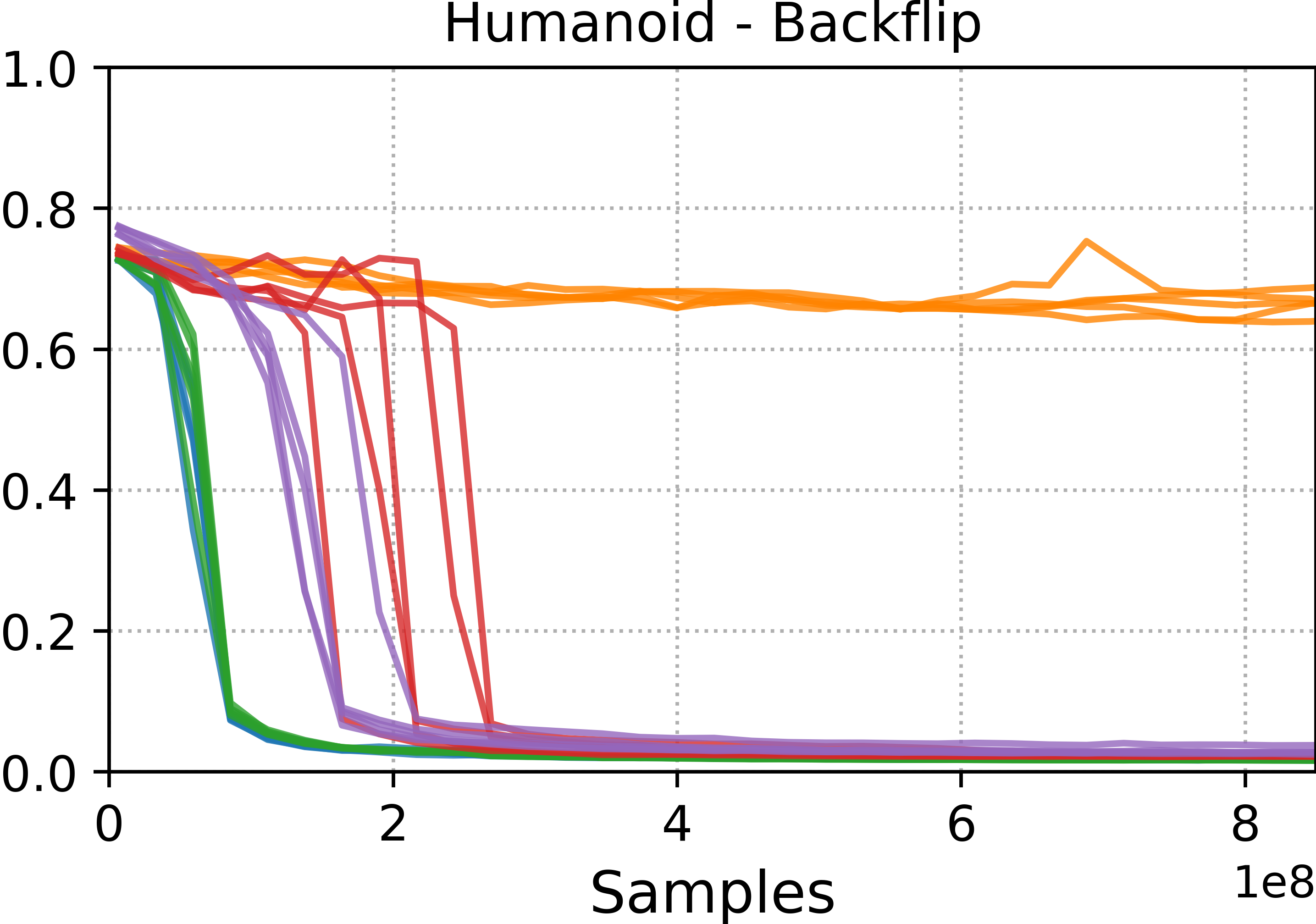}}\\
    \vspace{-0.2cm}
    \subfigure{\includegraphics[height=0.34\columnwidth]{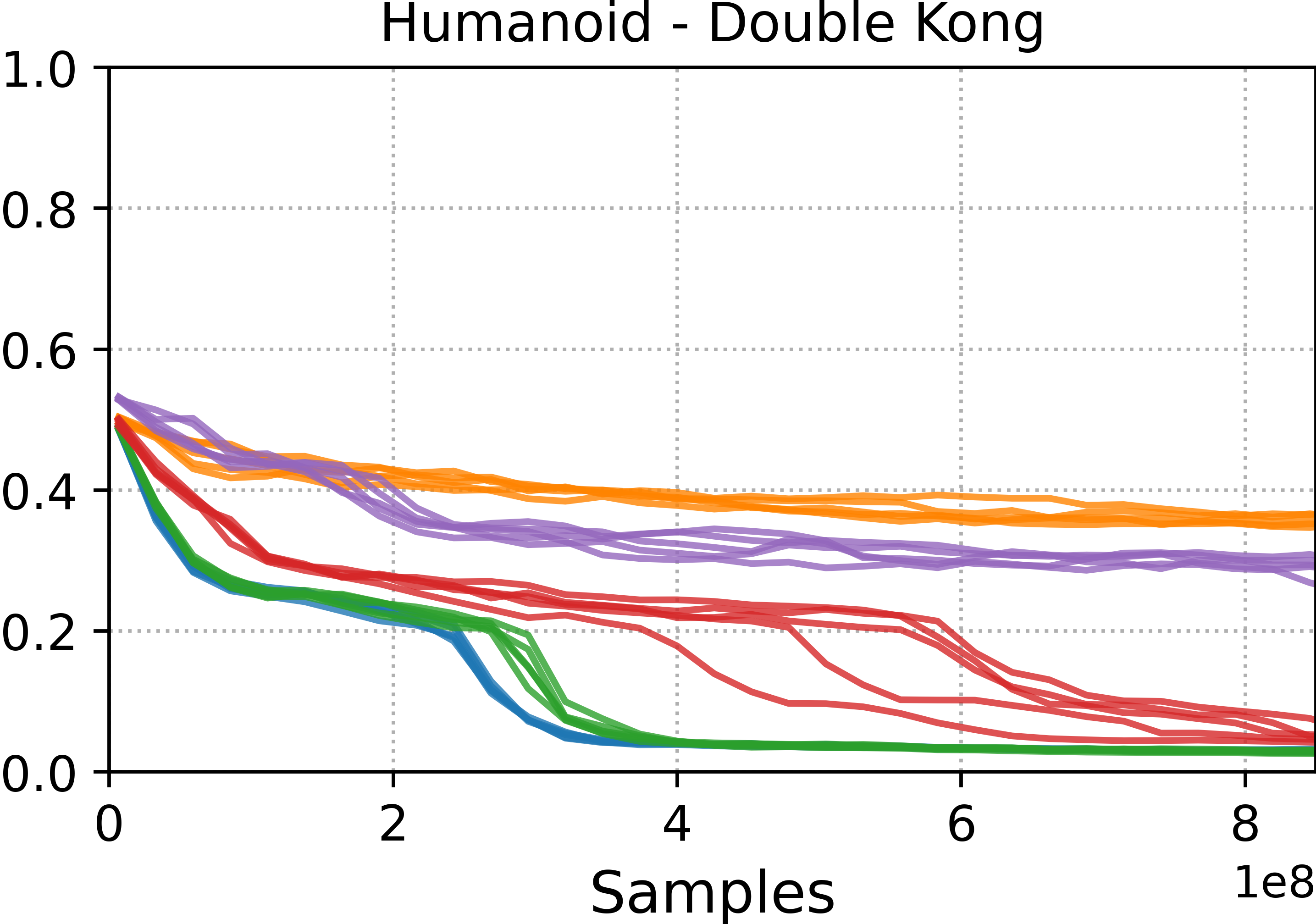}} 
    \subfigure{\includegraphics[height=0.339\columnwidth]{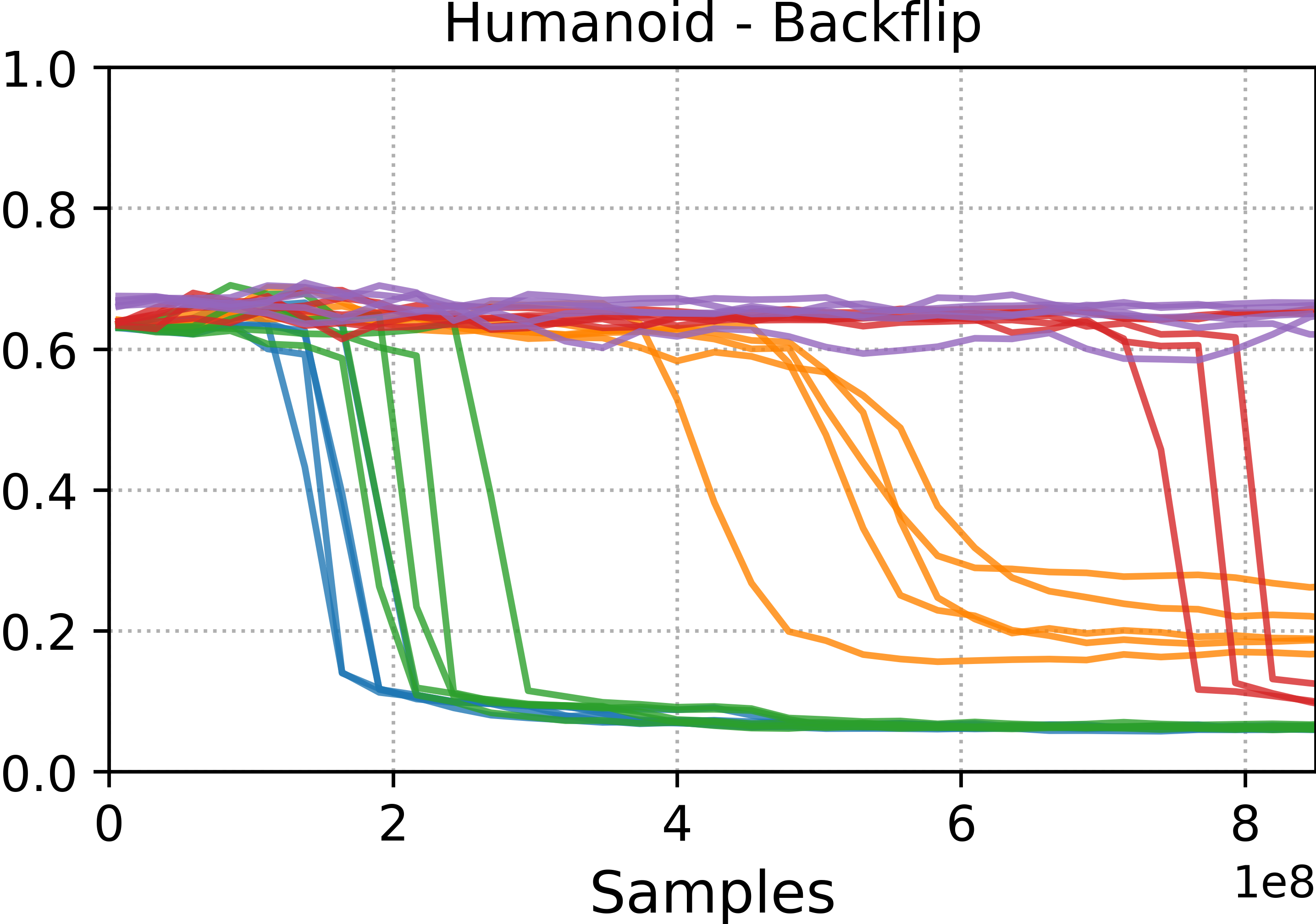}}\\
    \vspace{-0.2cm}
    \subfigure{\includegraphics[width=0.9\linewidth]{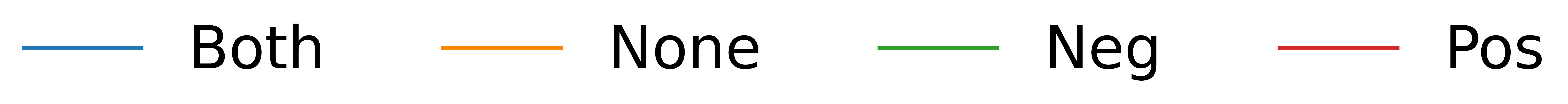}}\\
\caption{Learning curves of ADD trained, under varying gradient penalty configurations, to track different reference motions. We conduct ablations over five gradient penalty configurations: none (None), penalty applied only to negative samples (Neg), only to positive samples (Pos), to both sample types (Both), and to interpolations between positive and negative samples as proposed in \citet{wgan-gp} (WGAN-GP). Results indicate that Neg and Both significantly outperform the other settings in tracking performance, with Neg and Both achieving comparable results. These results highlight the importance of applying the gradient penalty to negative samples for ADD.}
\label{fig:gp_ablation}
\end{figure}

\section{Gradient Penalty Ablation}
{Gradient penalty is an effective regularizer in adversarial imitation learning methods \citep{pmlr-v80-mescheder18a,2021-TOG-AMP}. In this section, we investigate the impact of different gradient penalty configurations on the effectiveness of ADD. We examine five configurations --- no gradient penalty at all (None), gradient penalty applied exclusively to negative samples (Neg), exclusively to positive samples (Pos), to both positive and negative samples (Both), and to random interpolations between positive and negative samples as proposed in \citet{wgan-gp} (WGAN-GP). In the WGAN-GP setting, following \citet{wgan-gp}, the gradient penalty is modified to enforce 1-Lipschitz continuity,
\begin{equation}
    \mathcal{L}^{GP}(D) =\, \mathbb{E}_{p(\rvs  | \pi)} \left[ (\|\nabla_{\phi} D(\phi)|_{\phi = \Delta} \|_2 - 1)^2 \right],
\end{equation}
as opposed to Eq.~\ref{eq:motion_tracking_gp}.
Figure~\ref{fig:gp_ablation} shows that models trained without regularization (None) or with WGAN-GP exhibit significantly higher tracking errors. Although the WGAN-GP has proven effective for prior adversarial methods, it does not appear to be suitable for ADD. Models trained under the Pos configuration, while notably more accurate than the None and WGAN-GP settings, result in higher tracking errors and worse sample efficiency compared to Neg and Both settings. Models trained under the Neg and Both settings produce comparable results in terms of tracking error, with the Both configuration showing a slight advantage with regard to sample efficiency. Nonetheless, these findings validate the importance of applying the gradient penalty to negative samples for the adversarial differential discriminator, instead of only positive samples as done in prior work \citep{2021-TOG-AMP}. Moreover, these results underscore the critical role of the gradient penalty in stabilizing training and enabling effective learning, as models trained without it often fail to learn effective policies.}

\section{Discussion and Future Work}
In this work, we present an adversarial multi-objective optimization technique that is broadly applicable to a variety of tasks, including motion imitation for physics-based character animation. Our technique enables both a simulated humanoid and a simulated EVAL robot to accurately replicate a diverse suite of challenging skills, achieving comparable performance to state-of-the-art motion tracking methods, without requiring manual reward engineering. Despite its effectiveness, our method has certain limitations. One major limitation is that ADD is susceptible to converging to locally optimal behaviors. For example, when attempting to reproduce the highly dynamic rolling motion, characters trained with ADD learn to simply lie down and get up, rather than completing a full roll. Prior motion tracking methods typically mitigate the chances of converging to such behaviors by incorporating additional heuristics, such as early termination based on tracking error \citep{deep_mimic, Luo2023PerpetualHC}. Furthermore, recent studies show that training controllers on massive motion datasets is a scalable approach to enable physically simulated characters to acquire diverse behaviors \citep{Luo2023PerpetualHC, tessler2024maskedmimic, yao2024MoConVQ}. We would like to explore, in future work, applying ADD to train a general controller that does not rely on hand-crafted reward functions, a common dependency in existing frameworks.
Finally, another interesting avenue for future work is to apply ADD to multi-objective optimization or multi-task problems in other domains, such as computer vision and natural language processing.

\begin{acks}
We would like to express our gratitude to Michael Xu for the help with data processing, Beyond Capture Studios for providing the high-quality parkour motion data, and Yvette Chen for the 3D artwork that elevated the overall visual presentation of our work. This work was supported by NSERC via a Discovery Grant (RGPIN-2015-04843).
\end{acks}

\bibliographystyle{ACM-Reference-Format}
\bibliography{main}

\clearpage
\appendix

\section*{{Supplementary Material}}
\section*{Table of Contents}
\startcontents
\printcontents{}{1}{}

\section{Didactic Example}

\begin{figure}[t]
    \centering
    \captionsetup{skip=0pt}
    \includegraphics[width=\linewidth]{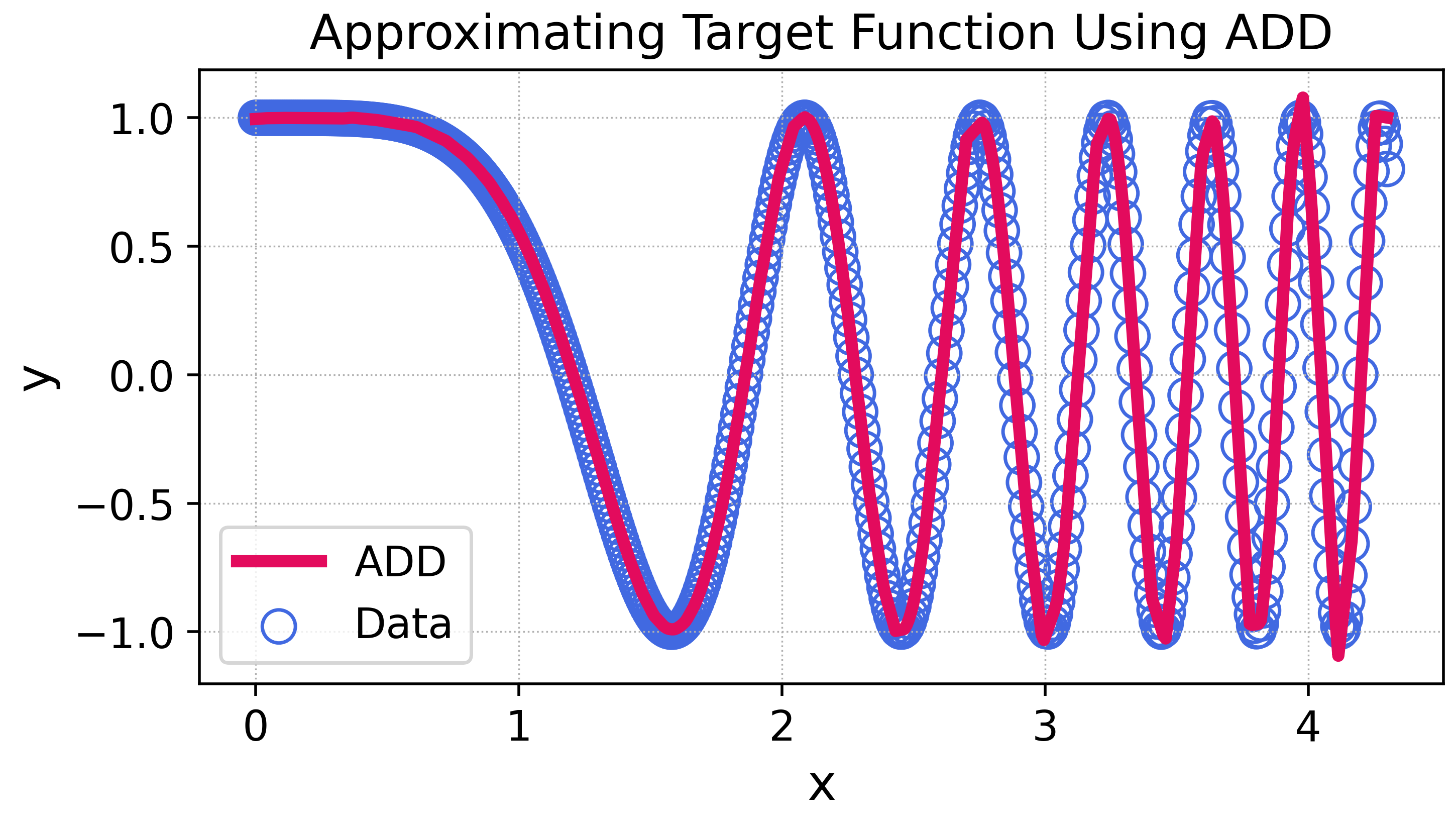}
    \caption{Visualization of predictions made by ADD (red line) using data points (blue circles) sampled from the target function $f(x) = \mathrm{cos}(x^{2.5})$. ADD approximates the ground truth data adequately.}
    \label{fig:add_pred}
\end{figure}

We include a didactic example to analyze the behavior of ADD. In this experiment, we apply ADD to a simple regression task, where the goal is to approximate a 1D target function $f(x) = \mathrm{cos}(x^{2.5})$, visualized in Figure \ref{fig:add_pred}. The target function is designed to present a function approximator with varying levels of difficulty in different regions of the input space. Near the origin, where the oscillation frequency is low, $f(x) = \mathrm{cos}(x^{2.5})$ is easy to model. However, as the frequency increases farther from the origin, the function becomes progressively difficult to model. This variation in the smoothness of the target function presents a testbed for examining the adaptive behavior of ADD to attend to different sub-objectives based on their relative difficulty.

\subsection{Experimental Setup}
We uniformly sample $N = 512$ points from $f(x)$ within the interval $[0,4.3]$ to build the training dataset. A network $G(x)$ is trained to predict an output $y_i$ that approximates $\hat{y} = f(x)$ for every $x_i$ in the dataset where $1 \leq i \leq N$. The prediction errors for the entire dataset are then aggregated into a vector,
\[
\Delta = \begin{bmatrix}
\hat{y}_1 - y_1 \\
\vdots \\
\hat{y}_N - y_N
\end{bmatrix} ,
\]
and provided as input to the discriminator $D(\Delta)$.
Both $G(x)$ and $D(\Delta)$ consist of 2 hidden layers with 1024 and 512 units, respectively. Table \ref{tab:suppADDRegressionParams} provides a detailed documentation of the hyperparameters used in this experiment.

\begin{table}[t]
{ \centering  
\caption{ADD regression experiment hyperparameters.}
\label{tab:suppADDRegressionParams}
\begin{tabular}{|l|c|}
\hline
{\bf Parameter} & {\bf Value}  \\ \hline
    $\lambda^\mathrm{GP}$ Gradient Penalty &  $0.1$  \\ \hline
    Generator Batch Size &  $512$  \\ \hline
    $D$ Discriminator Stepsize &  $10^{-5}$  \\ \hline
    $G$ Generator Stepsize &  $10^{-4}$  \\ \hline
\end{tabular}
}
\end{table}

\subsection{Results}
\begin{figure}[t]
	\centering
        \captionsetup{skip=0pt}
    \subfigure{\includegraphics[height=0.365\columnwidth]{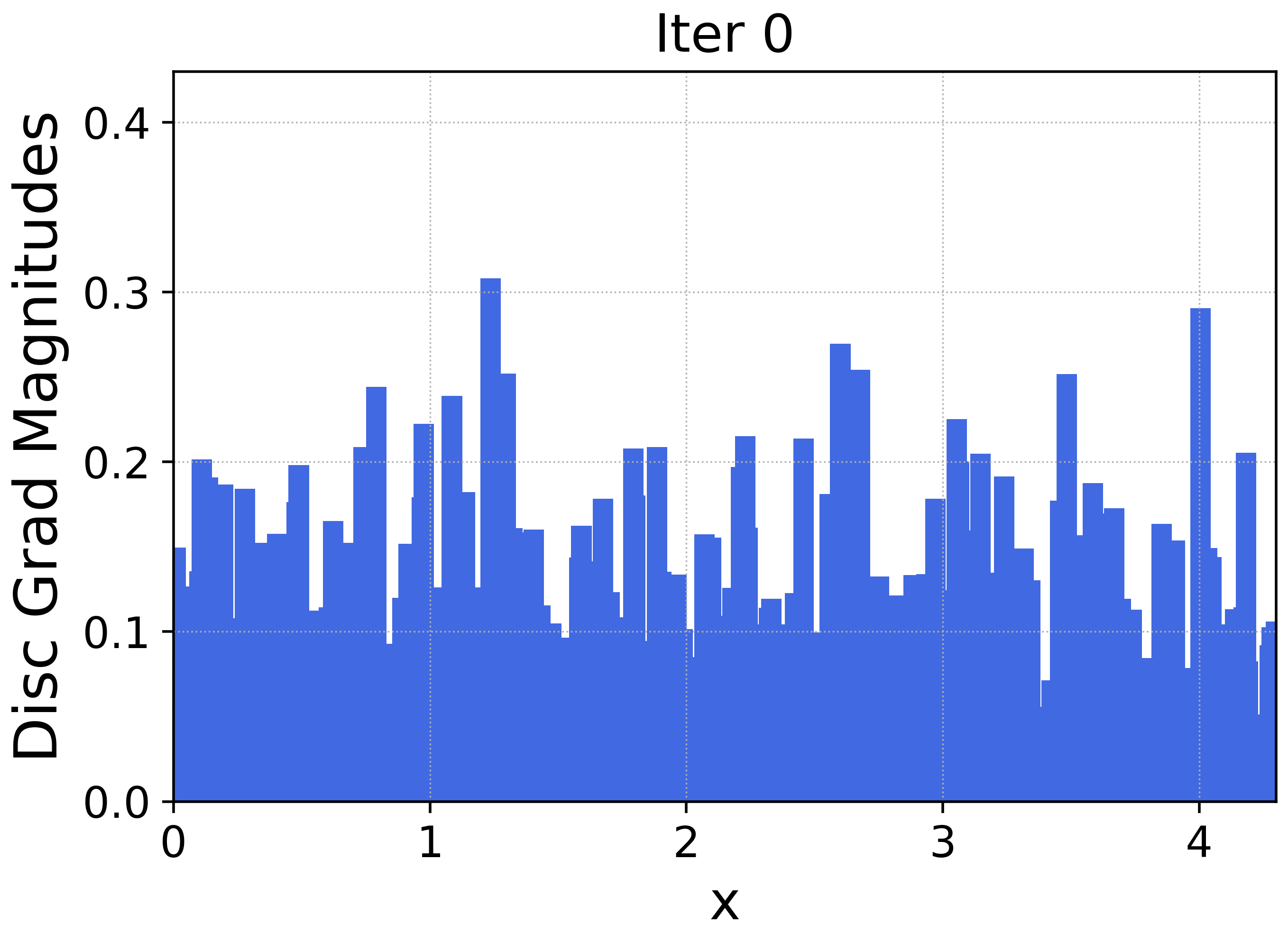}}
    \subfigure{\includegraphics[height=0.365\columnwidth]{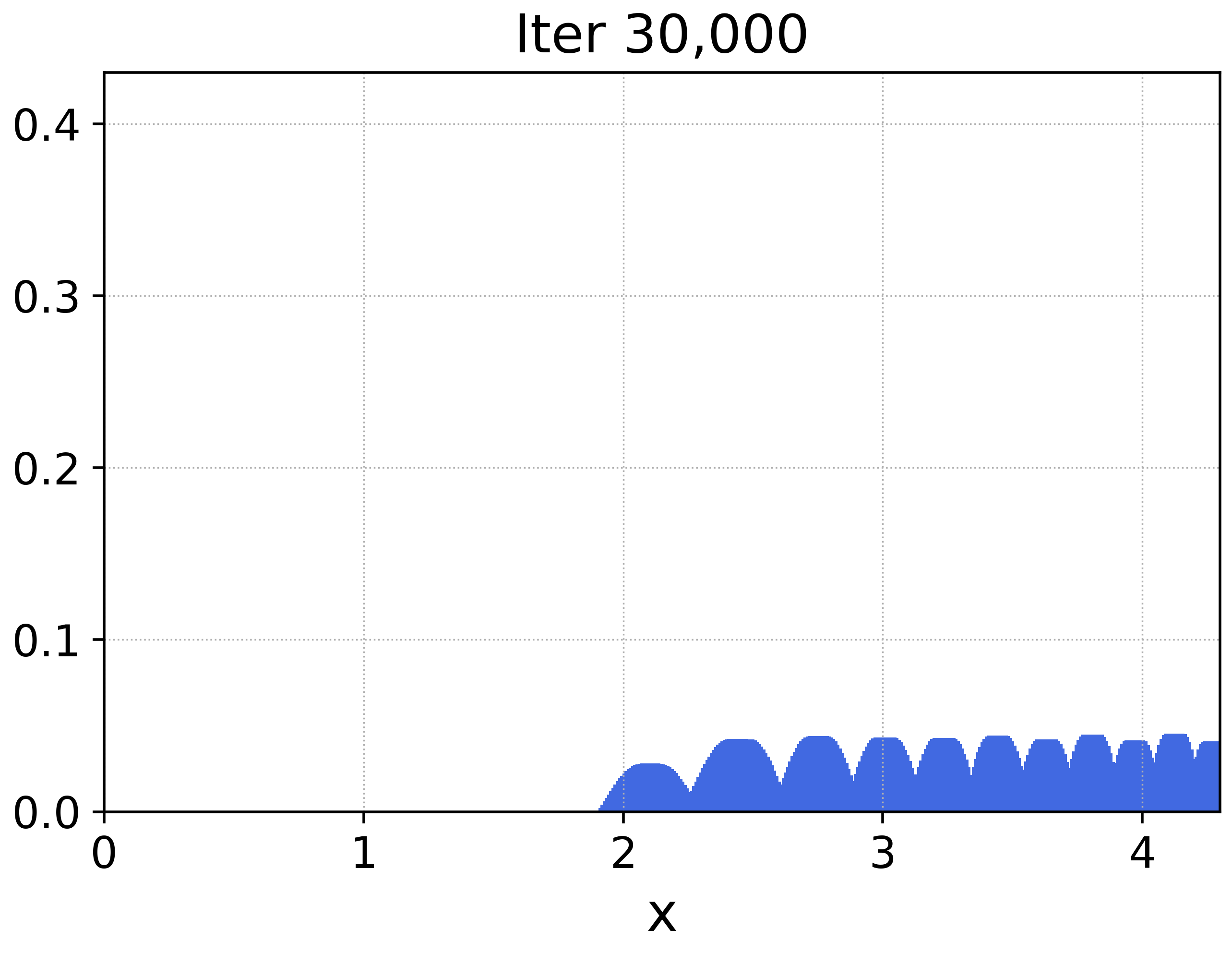}}\\
    \vspace{-0.57cm}
    \subfigure{\includegraphics[height=0.365\columnwidth]{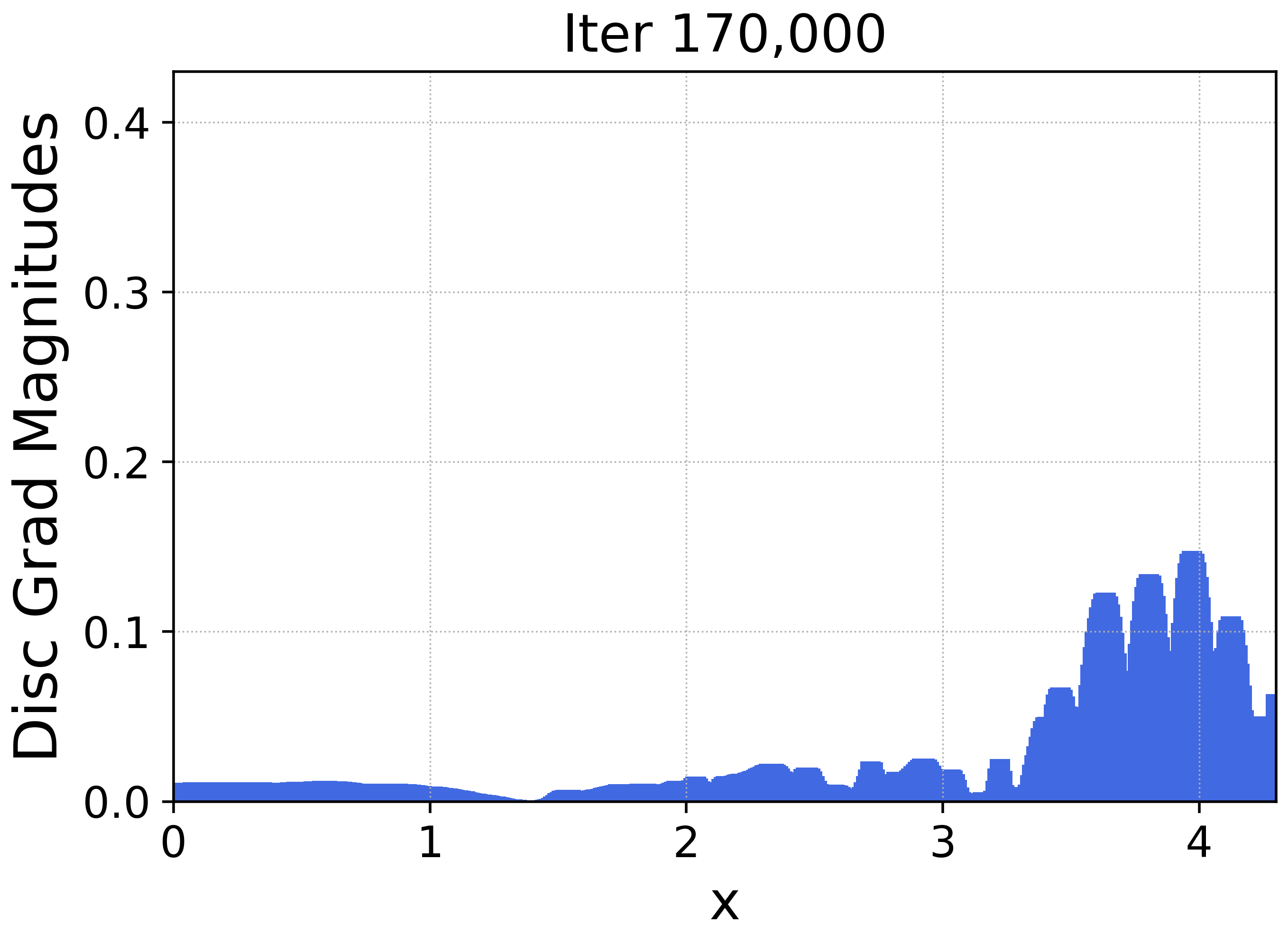}} 
    \subfigure{\includegraphics[height=0.365\columnwidth]{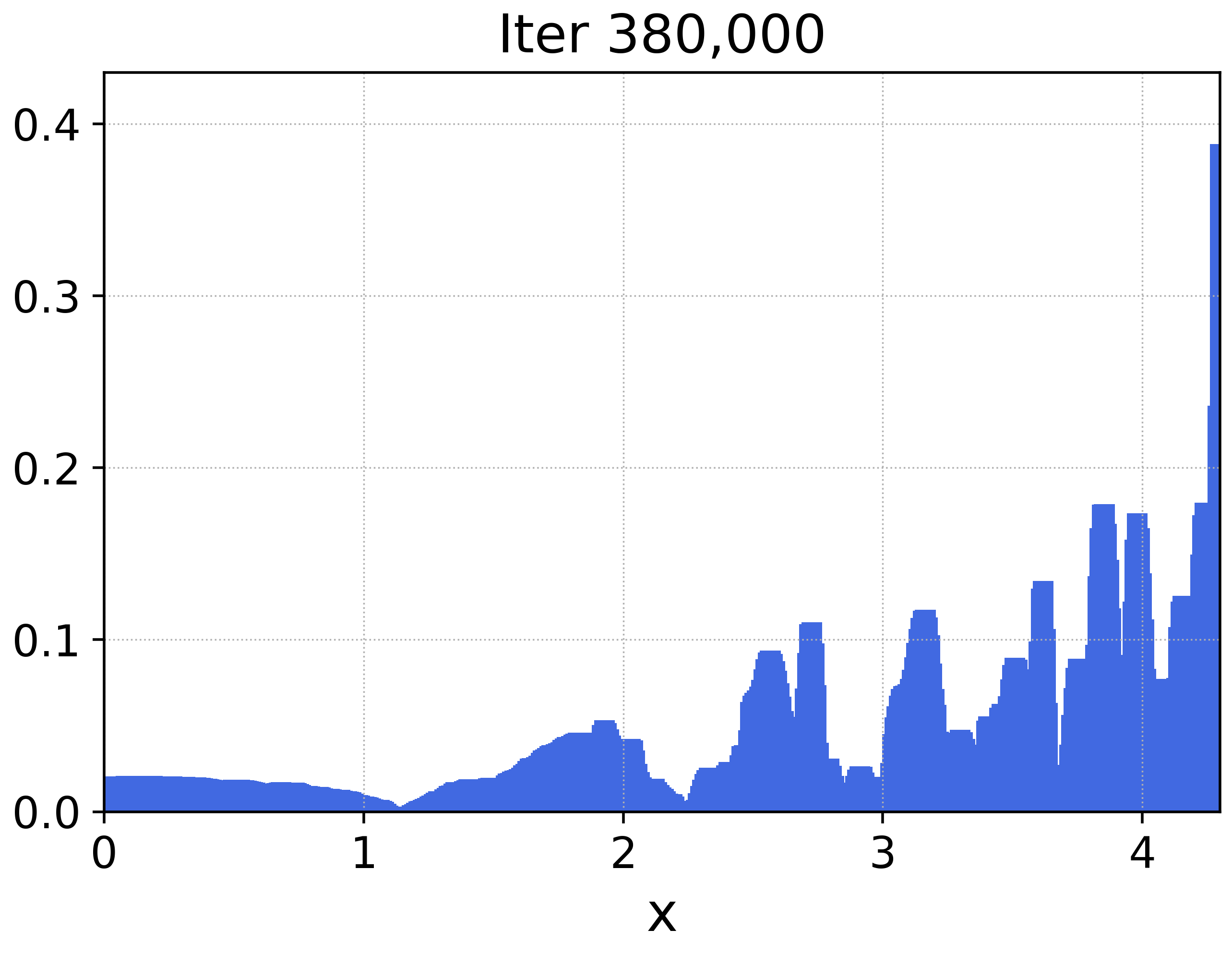}}\\
\caption{Magnitudes of the gradients of the discriminator's output with respect to the prediction error for each sample at different training iterations. As training progresses and $G(x)$ gradually learns to approximate samples near the origin, the discriminator assigns higher and higher weights to samples farther from the origin, essentially honing in on the more difficult objectives.}
\label{fig:regression_grad}
\end{figure}

Figure~\ref{fig:add_pred} demonstrates that the regression model trained through ADD provides a reasonable approximation of the ground truth target function. To analyze the discriminator’s behavior in dynamically weighting objectives over the course of training, Figure~\ref{fig:regression_grad} depicts the gradients of the discriminator's output with respect to the prediction error for each sample $\nabla_{\Delta} D(\Delta)$ at different training iterations. The $N$ prediction errors are the objectives of this MOO problem, and the gradients $\nabla_{\Delta} D(\Delta)$ can be interpreted as the weights assigned by $D(\Delta)$ to each objective. At the beginning of training, the discriminator assigns random weights to the objectives. As training progresses, and $G(x)$ learns how to fit $f(x)$ accurately near the origin, the discriminator automatically begins to assign higher weights to regions farther from the origin, which exhibit larger prediction errors. ADD's adaptive weighting of different objectives ensures that more difficult objectives receive more focus as training progresses, preventing the easier objectives from dominating the overall loss.

\section{Motion Tracking Additional Details}
\label{supp:motion_tracking_additional_details}

\subsection{Experimental Setup and Computational Cost}
All experiments are conducted on Nvidia A100 GPUs, with physics simulations implemented using Isaac Gym~\citep{issac-gym}. Humanoid character simulations run at \SI{120}{\hertz}, and the policy is queried at \SI{30}{\hertz}. For most motion imitation tasks on the humanoid character (excluding DanceDB and LaFAN1), policies are trained with approximately 750–850 million samples, requiring about \SI{9}{\hour} for ADD, \SI{12}{\hour} for AMP, and \SI{6.5}{\hour} for DeepMimic. On the larger DanceDB dataset \citep{AMASS_Mahmood_2019_ICCV}, all methods are trained with roughly 9 billion samples, taking 5 days for ADD, 6 days for AMP, and 3.5 days for DeepMimic. For the LaFAN1 subset \citep{LAFAN_harvey2020robust}, training is extended to 18 billion samples, requiring 10 days for ADD, 7 days for DeepMimic, and 11 days for AMP.

Each EVAL robot policy is trained with approximately 900 million samples, requiring a wall-clock time of around \SI{16}{\hour} for ADD, \SI{18}{\hour} for AMP, and \SI{13}{\hour} for DeepMimic. The robot simulations are performed at a simulation frequency of \SI{150}{\hertz}, with the control frequency set to \SI{50}{\hertz}.

\subsection{DeepMimic's Imitation Reward Function}
\label{supp:DM_Imitation_Reward_Fn}

\citet{deep_mimic} employ a manually-designed motion imitation reward function composed of five sub-terms,
\begin{equation}
    r^{\text{DM}}_t = w^p r_t^p + w^{jv} r_t^{jv} + w^{rv} r_t^{rv} + w^e r_t^e + w^c r_t^c.
\end{equation}
The pose reward $r_t^p$ computes the difference between the joint orientation quaternions of the simulated character $\rvq_t^j$ and those of the reference motion $\hat{{\rvq}}_t^j$,
\begin{equation}
r_t^p = \exp\left[ -\alpha^p \left( \sum_j \left\| \hat{{\rvq}}_t^j \ominus {\rvq}_t^j \right\|^2 \right) \right].
\end{equation}
Here, $\alpha^p$ is a manually-tuned hyperparameter.
The joint velocity reward $r_t^{jv}$ is computed from the difference of local joint velocities, with $\dot{\rvq}_t^j$ being the angular velocity of the $j$-th joint,
\begin{equation}
r_t^{jv} = \exp\left[ -\alpha^{jv} \left( \sum_j \left\| \hat{\dot{\rvq}}_t^j - \dot{\rvq}_t^j \right\|^2 \right) \right].
\end{equation}
The root velocity reward $r_t^{rv}$ measures the difference between the root velocities of the simulated character and that of the reference motion, with $\rvv_t^r$ representing the linear and angular root velocity,
\begin{equation}
r_t^{rv} = \exp\left[ -\alpha^{rv} \left( \left\| \hat{{\rvv}}_t^r - \rvv_t^r \right\|^2 \right) \right].
\end{equation}
The end-effector reward $r_t^e$ calculates the difference in the end-effector positions of the simulated character $\rvp_t^e$ and those of the reference motion $\hat{\rvp}_t^e$,
\begin{equation}
r_t^e = \exp\left[ -\alpha^{e} \left( \sum_e \left\| \hat{\rvp}_t^e - \rvp_t^e \right\|^2 \right) \right].
\end{equation}
Finally, $r_t^c$ encourages the character's center-of-mass $\rvp_t^c$ to match the center-of-mass of the reference motion $\hat{\rvp}_t^c$,
\begin{equation}
r_t^c = \exp\left[ -\alpha^c \left( \left\| \hat{\rvp}_t^c - \rvp_t^c \right\|^2 \right) \right].
\end{equation}
In the implementation, every joint orientation or velocity error is also multiplied by a joint-specific weight in the calculation of $r_t^p$ and $r_t^{jv}$.

\subsection{DeepMimic Sensitivity Analysis}
\label{supp:dm_sensitivity_analysis}

\begin{table}[t]
\centering
\caption{DeepMimic sensitivity analysis parameter settings. Reward weights are listed in the order of $w^p, w^{jv}, w^{rv}, w^e, w^c$, and reward scales are listed in the order of $\alpha^p, \alpha^{jv}, \alpha^{rv}, \alpha^e, \alpha^c$. * denotes the final parameter setting used in the experiments.}
\label{tab:deepmimic_sensitivity_settings}
\begin{tabular}{|c|c|c|}
\hline
\textbf{Setting} & \multicolumn{2}{c|}{\textbf{Parameters}} \\
\hline
 & \textbf{Reward Weights} & \textbf{Reward Scales} \\
\hline
Setting 1 & 0.2, 0.2, 0.2, 0.2, 0.2 & 1, 1, 1, 1, 1 \\ \hline
Setting 2 & 0.5, 0.1, 0.15, 0.1, 0.15 & 4, 10, 0.2, 1, 0.1 \\ \hline
Setting 3 & 0.5, 0.1, 0.15, 0.1, 0.15 & 0.2, 0.05, 3, 1.5, 8 \\ \hline
Setting 4 & 0.5, 0.1, 0.15, 0.1, 0.15 & 10, 0.04, 100, 7.5, 75 \\ \hline
Setting 5 & 0.2, 0.1, 0.2, 0.05, 0.45 & 0.25, 0.01, 5, 1, 10 \\ \hline
Default* & 0.5, 0.1, 0.15, 0.1, 0.15 & 0.25, 0.01, 5, 1, 10 \\
\hline
\end{tabular}
\end{table}

\begin{figure}[t]
	\centering
        \captionsetup{skip=0pt}
    \subfigure{\includegraphics[height=0.315\columnwidth]{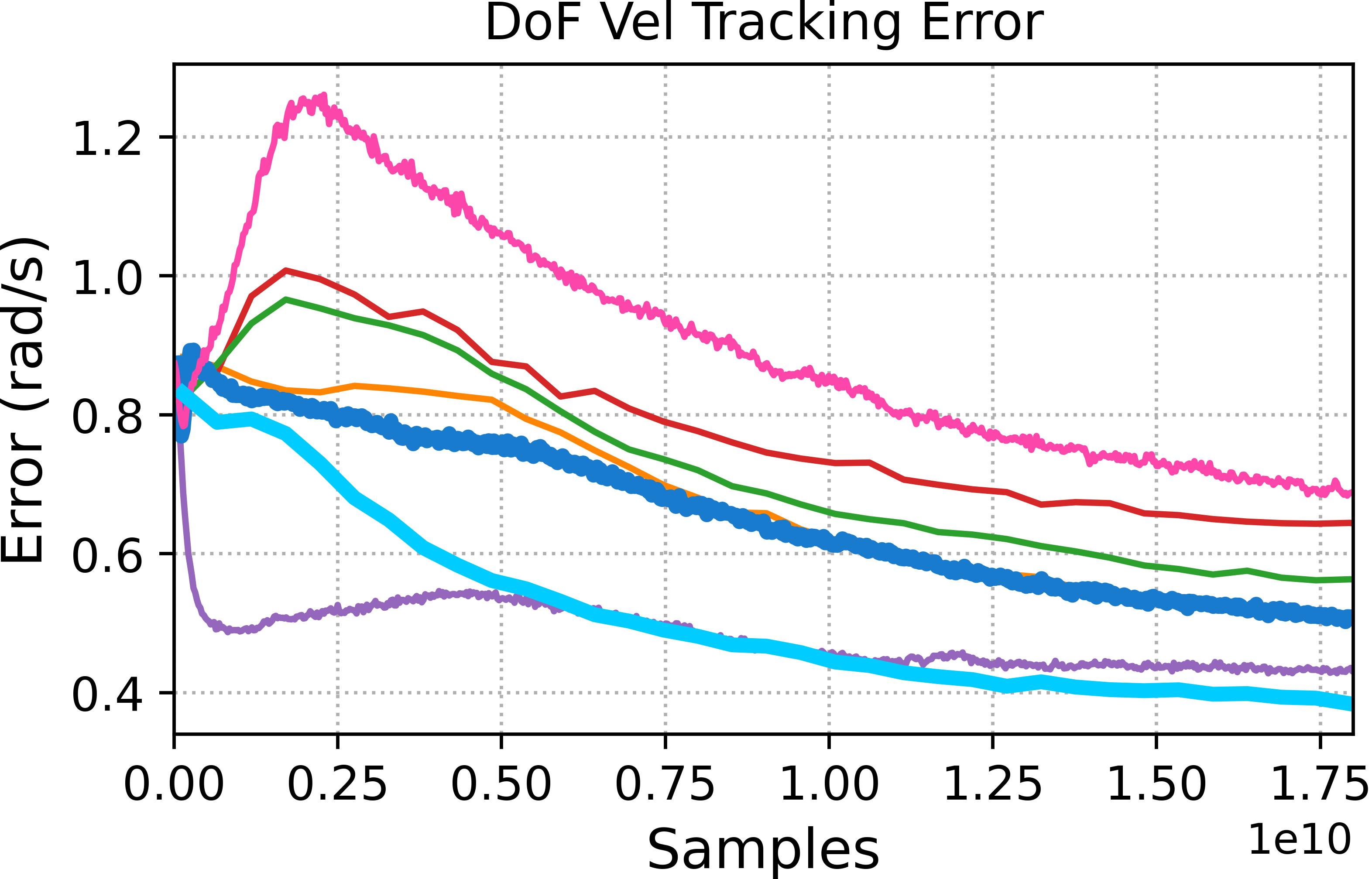}}
    \subfigure{\includegraphics[height=0.315\columnwidth]{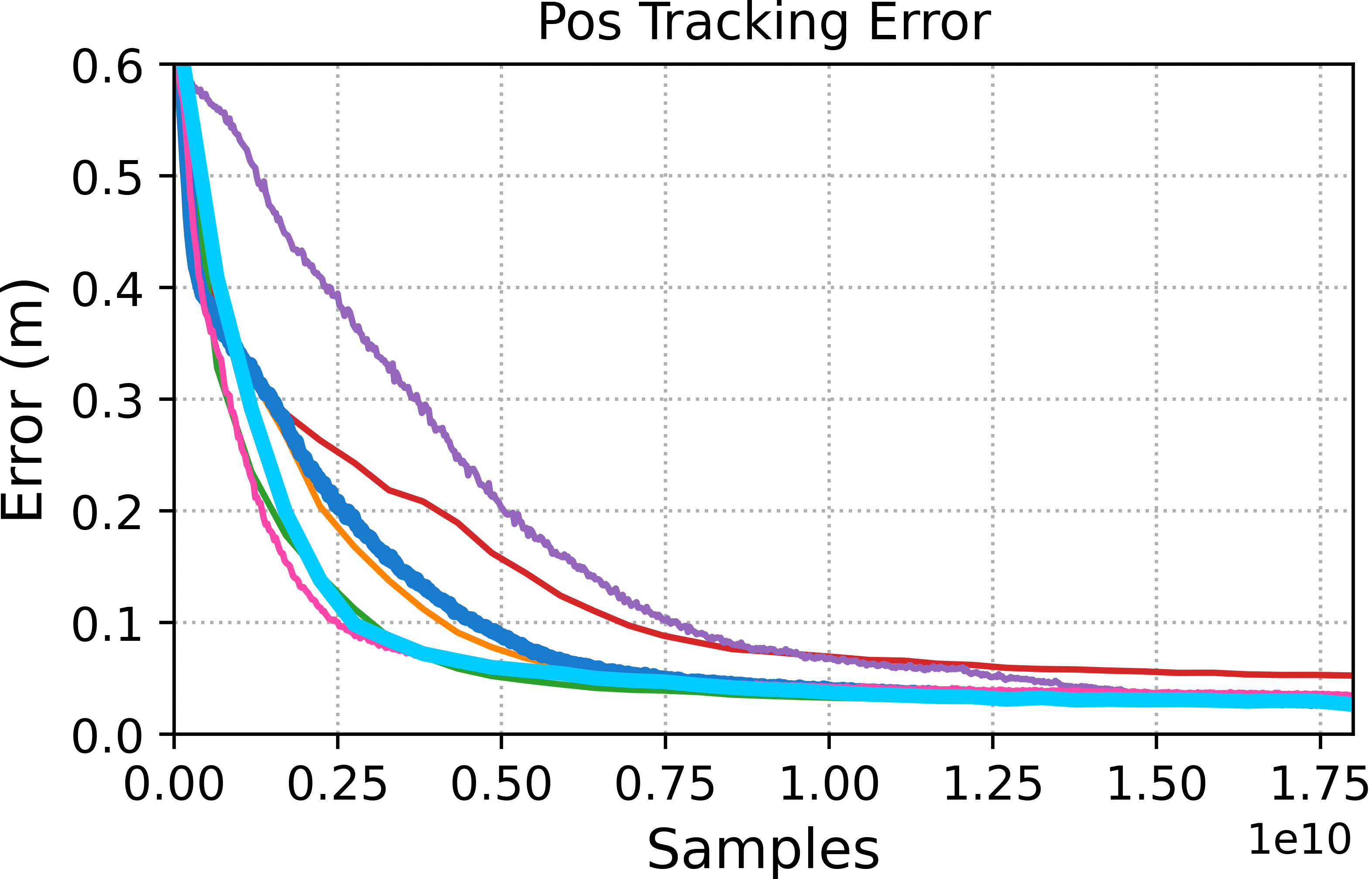}}\\
    \vspace{-0.2cm}
    \subfigure{\includegraphics[height=0.31\columnwidth]{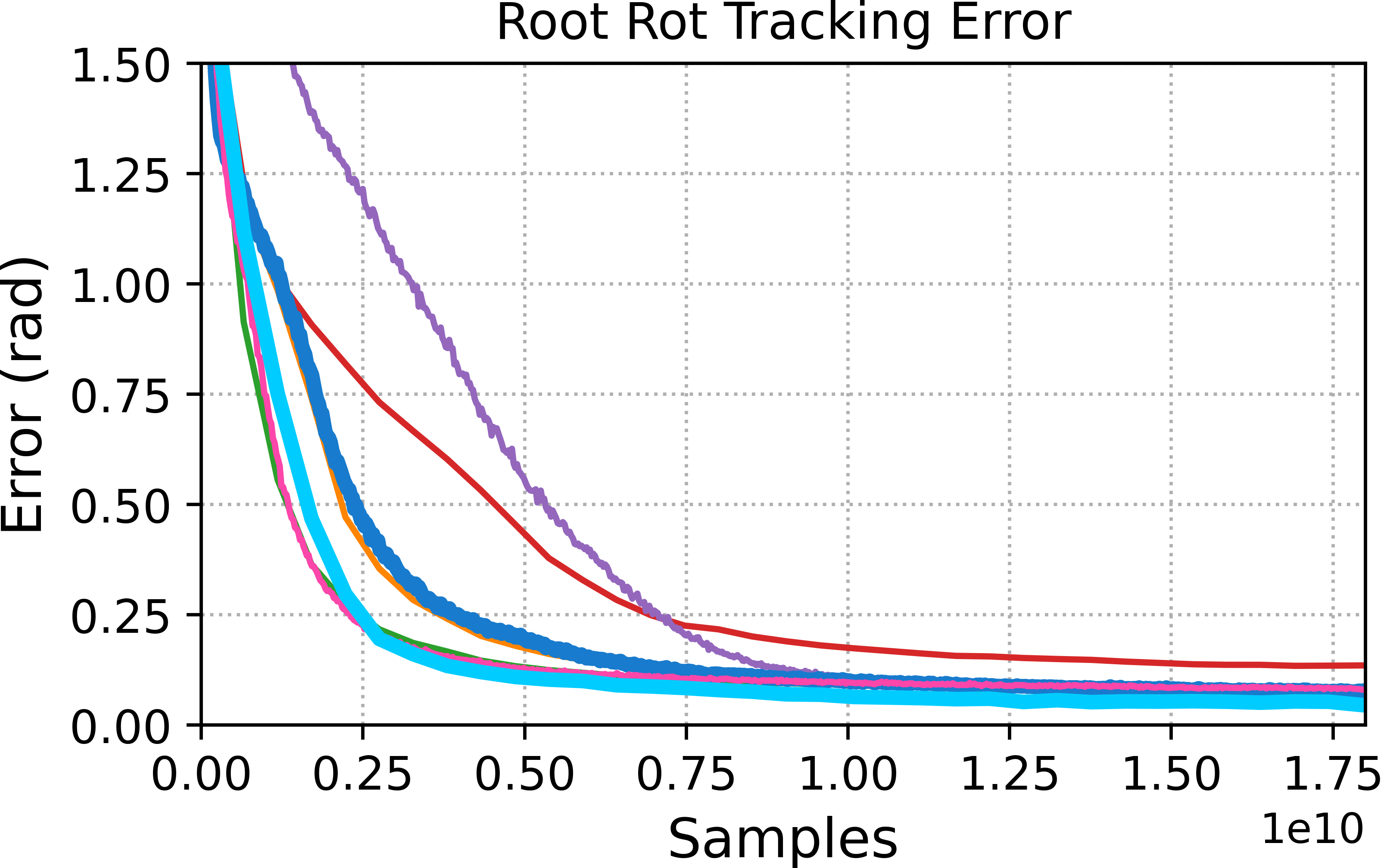}} 
    \subfigure{\includegraphics[height=0.31\columnwidth]{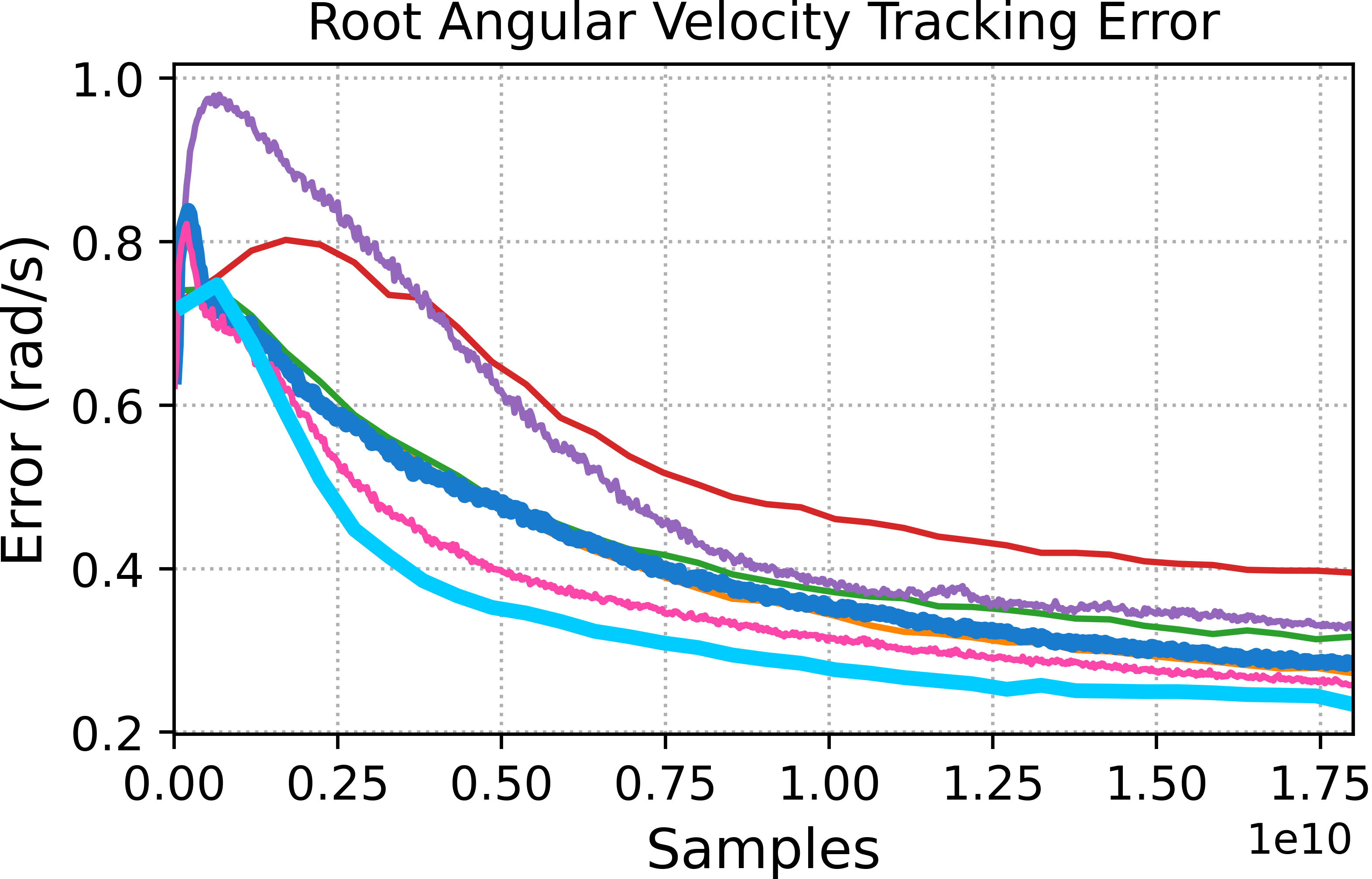}}\\
    \vspace{-0.2cm}
    \subfigure{\includegraphics[width=\linewidth]{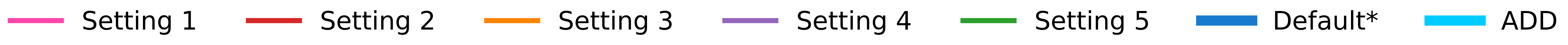}}\\
\caption{Learning curves comparing the tracking performance of DeepMimic, under different parameter settings, with ADD on a subset of the LaFAN1 dataset \citep{LAFAN_harvey2020robust}. Table \ref{tab:deepmimic_sensitivity_settings} provides more details on the parameter settings. DeepMimic's performance is sensitive to its hyperparameters. Some configurations can lead to poor tracking performance and sample efficiency. Some settings lead to improvement on certain metrics but degrade others. Selecting an optimal set of hyperparameters requires domain knowledge and significant manual tuning. In contrast, ADD automatically balances competing objectives during training, consistently achieving superior tracking accuracy and sample efficiency across metrics.}
\label{fig:dm_sensitivity_curves}
\end{figure}

To evaluate how sensitive DeepMimic is to its reward hyperparameters, we use DeepMimic to train simulated humanoid characters to imitate motions from the LaFAN1 subset using six different parameter settings. Details of the six settings are provided in Table~\ref{tab:deepmimic_sensitivity_settings}, and the resulting learning curves are shown in Figure~\ref{fig:dm_sensitivity_curves}.

The results show considerable variation in both tracking performance and sample efficiency across settings. As shown in Figure \ref{fig:dm_sensitivity_curves}, Setting 4 achieves strong tracking accuracy, but demonstrates poor sample efficiency. Setting 1 produces motions with accurate root rotation, root angular velocity, and position tracking, but suffers from high DoF velocity tracking errors, which lead to visibly jittery motions. Setting 2 performs poorly across all metrics and learns slowly, suggesting that certain parameter choices can significantly degrade both sample efficiency and tracking quality. In contrast, Settings 3, 5, and default*—sharing similar parameter configurations—consistently achieve great tracking accuracy and learning efficiency. Based on this observation, Setting default* is selected for the humanoid motion imitation experiments. Meanwhile, ADD achieves superior tracking accuracy and sample efficiency across all metrics without requiring manual tuning.

These findings highlight a key limitation of DeepMimic: its performance is highly sensitive to the choice of hyperparameters. While some configurations may yield strong results, finding them often requires extensive domain knowledge and iterative tuning. Poorly chosen parameters can result in degraded motion quality or inefficient learning, which presents a challenge for generalizing DeepMimic to new tasks or embodiments. In contrast, ADD's ability to automatically balance objectives leads to more robust and efficient learning across metrics, reducing the burden of manual design.

\subsection{Hyperparameter Settings}
\label{supp:motion_imitation_hyperparam}

\begin{table}[t]
    \centering
    \caption{ADD humanoid character motion imitation experiment hyperparameters.}
    \label{tab:suppADDMotionImitationParams}
    \begin{tabular}{|l|c|}
        \hline
        \textbf{Parameter} & {\bf Value} \\ \hline
        \multirow{2}{*}{$\lambda^{\mathrm{GP}}$ Gradient Penalty}
          & 0.1 (Climb) \\ 
        \cline{2-2}
          & 1 (Other) \\
        \hline  
        $K$ Update Minibatch Size &  $4096 \times 4$  \\ \hline
        $\pi$ Policy Stepsize&  $1 \times 10^{-4}$  \\ \hline
        $V$ Value Stepsize &  $1 \times 10^{-4}$  \\ \hline
        {$D$ Discriminator Stepsize}
          & $5 \times 10^{-4}$\\
        \hline
        $\mathcal{B}$ Experience Buffer Size &  $4096 \times 32$  \\ \hline
        $\gamma$ Discount&  $0.99$  \\ \hline
        SGD Momentum &  $0.9$  \\ \hline
        GAE($\lambda$) &  $0.95$  \\ \hline
        TD($\lambda$) &  $0.95$  \\ \hline
        PPO Clip Threshold &  $0.2$  \\ \hline
    \end{tabular}
\end{table}

\begin{table}[t]
    \centering
    \caption{DeepMimic humanoid character motion imitation experiment hyperparameters.}
    \label{tab:suppDMMotionImitationParams}
    \begin{tabular}{|l|c|}
        \hline
        \textbf{Parameter} & {\bf Value} \\ \hline
        $K$ Update Minibatch Size &  $4096\times4$  \\ \hline
        $\pi$ Policy Stepsize&  $5 \times 10^{-5}$  \\ \hline
        $V$ Value Stepsize &  $5 \times 10^{-5}$  \\ \hline
        $\mathcal{B}$ Experience Buffer Size &  $4096\times32$  \\ \hline
        $\gamma$ Discount&  $0.99$  \\ \hline
        SGD Momentum &  $0.9$  \\ \hline
        GAE($\lambda$) &  $0.95$  \\ \hline
        TD($\lambda$) &  $0.95$  \\ \hline
        PPO Clip Threshold &  $0.2$  \\ \hline
        $w^p, w^{jv}, w^{rv}, w^e, w^c$ Reward Weights &  $0.5, 0.1, 0.15, 0.1, 0.15$  \\ \hline
        $\alpha^p, \alpha^{jv}, \alpha^{rv}, \alpha^e, \alpha^c$ Reward Scales &  $0.25, 0.01, 5.0, 1.0, 10.0$  \\ \hline
        Joint Weights & \makecell[c]{1.0, 0.6, 0.6, 0.4, 0.0, \\ 0.6, 0.4, 0.0, 1.0, 0.6, \\ 0.4, 1.0, 0.6, 0.4} \\ \hline
    \end{tabular}
\end{table}

\begin{table}[t]
    \centering
    \caption{AMP humanoid character motion imitation experiment hyperparameters.}
    \label{tab:suppAMPMotionImitationParams}
    \begin{tabular}{|l|c|}
        \hline
        \textbf{Parameter} & {\bf Value} \\ \hline
        $\lambda^{\mathrm{GP}}$ Gradient Penalty & 5.0 \\ \hline 
        $K$ Update Minibatch Size &  $4096 \times 4$  \\ \hline
        $\pi$ Policy Stepsize&  $5 \times 10^{-5}$  \\ \hline
        $V$ Value Stepsize &  $5 \times 10^{-5}$  \\ \hline
        {$D$ Discriminator Stepsize}
          & $2.5 \times 10^{-4}$\\
        \hline
        $\mathcal{B}$ Experience Buffer Size &  $4096 \times 32$  \\ \hline
        $\gamma$ Discount&  $0.99$  \\ \hline
        SGD Momentum &  $0.9$  \\ \hline
        GAE($\lambda$) &  $0.95$  \\ \hline
        TD($\lambda$) &  $0.95$  \\ \hline
        PPO Clip Threshold &  $0.2$  \\ \hline
    \end{tabular}
\end{table}

Tables \ref{tab:suppADDMotionImitationParams}, \ref{tab:suppDMMotionImitationParams}, \ref{tab:suppAMPMotionImitationParams} list the hyperparameters of ADD, DeepMimic, and AMP used in the humanoid character motion imitation experiments. Similarly, Tables \ref{tab:suppADDEVALMotionImitationParams}, \ref{tab:suppDMEVALMotionImitationParams}, \ref{tab:suppAMPEVALMotionImitationParams} provide the hyperparameters used in the EVAL robot motion imitation experiments.

DeepMimic has significantly more hyperparameters to be tuned to achieve good tracking performance than ADD and AMP. Moreover, these weights need to be re-tuned when switching to characters of different morphologies, as shown in Tables \ref{tab:suppDMMotionImitationParams} and \ref{tab:suppDMEVALMotionImitationParams}. In contrast, many fewer hyperparameters need to be adjusted for ADD when switching to different characters.

\begin{table}[t]
    \centering
    \caption{ADD EVAL robot motion imitation experiment hyperparameters.}
    \label{tab:suppADDEVALMotionImitationParams}
    \begin{tabular}{|l|c|}
        \hline
        \textbf{Parameter} & {\bf Value} \\ \hline
        {$\lambda^{\mathrm{GP}}$ Gradient Penalty}  & $1$ \\ \hline  
        $\mathcal{B}$ Experience Buffer Size  &  $4096\times 32 $ \\ \hline
        $K$ Update Minibatch Size &  $4096\times 4$  \\ \hline
        $\pi$ Policy Stepsize&  $8 \times 10^{-5}$  \\ \hline
        $V$ Value Stepsize &  $8 \times 10^{-5}$  \\ \hline
        {$D$ Discriminator Stepsize} & $4 \times 10^{-4}$ \\  \hline
        $\gamma$ Discount&  $0.99$  \\ \hline
        SGD Momentum &  $0.9$  \\ \hline
        GAE($\lambda$) &  $0.95$  \\ \hline
        TD($\lambda$) &  $0.95$  \\ \hline
        PPO Clip Threshold &  $0.2$  \\ \hline
    \end{tabular}
\end{table}

\begin{table}[t]
    \centering
    \caption{DeepMimic EVAL robot motion imitation experiment hyperparameters.}
    \label{tab:suppDMEVALMotionImitationParams}
    \begin{tabular}{|l|c|}
        \hline
        \textbf{Parameter} & {\bf Value} \\ \hline
        $K$ Update Minibatch Size &  $4096\times4$  \\ \hline
        $\pi$ Policy Stepsize&  $1 \times 10^{-4}$  \\ \hline
        $V$ Value Stepsize &  $1 \times 10^{-4}$  \\ \hline
        $\mathcal{B}$ Experience Buffer Size &  $4096\times32$  \\ \hline
        $\gamma$ Discount&  $0.99$  \\ \hline
        SGD Momentum &  $0.9$  \\ \hline
        GAE($\lambda$) &  $0.95$  \\ \hline
        TD($\lambda$) &  $0.95$  \\ \hline
        PPO Clip Threshold &  $0.2$  \\ \hline
        $w^p, w^{jv}, w^{rv}, w^e, w^c$ Reward Weights &  $0.5, 0.05, 0.5, 0.15, 0.1$  \\ \hline
        $\alpha^p, \alpha^{jv}, \alpha^{rv}, \alpha^e, \alpha^c$ Reward Scales &  $1.0, 0.01, 10.0, 1.0, 10.0$  \\ \hline
        Joint Weights & \makecell[c]{1.0, 1.0, 1.0, 1.0, 1.0, \\ 1.0, 1.0, 1.0, 1.0, 1.0, \\ 1.0, 1.0, 1.0, 1.0, 1.0, \\ 1.0, 1.0, 1.0, 1.0, 1.0, \\ 1.0, 1.0, 1.0, 1.0, 1.0, \\ 1.0
} \\ \hline
    \end{tabular}
\end{table}

\begin{table}[t]
    \centering
    \caption{AMP EVAL robot motion imitation experiment hyperparameters.}
    \label{tab:suppAMPEVALMotionImitationParams}
    \begin{tabular}{|l|c|}
        \hline
        \textbf{Parameter} & {\bf Value} \\ \hline
        $\lambda^{\mathrm{GP}}$ Gradient Penalty & 0.1 \\ \hline 
        $K$ Update Minibatch Size &  $4096 \times 4$  \\ \hline
        $\pi$ Policy Stepsize&  $8 \times 10^{-5}$  \\ \hline
        $V$ Value Stepsize &  $8 \times 10^{-5}$  \\ \hline
        {$D$ Discriminator Stepsize}
          & $8 \times 10^{-5}$\\
        \hline
        $\mathcal{B}$ Experience Buffer Size &  $4096 \times 32$  \\ \hline
        $\gamma$ Discount&  $0.99$  \\ \hline
        SGD Momentum &  $0.9$  \\ \hline
        GAE($\lambda$) &  $0.95$  \\ \hline
        TD($\lambda$) &  $0.95$  \\ \hline
        PPO Clip Threshold &  $0.2$  \\ \hline
    \end{tabular}
\end{table}

\begin{figure*}[t]
	\centering
        \captionsetup{skip=0pt}
    \subfigure{\includegraphics[height=0.349\columnwidth]{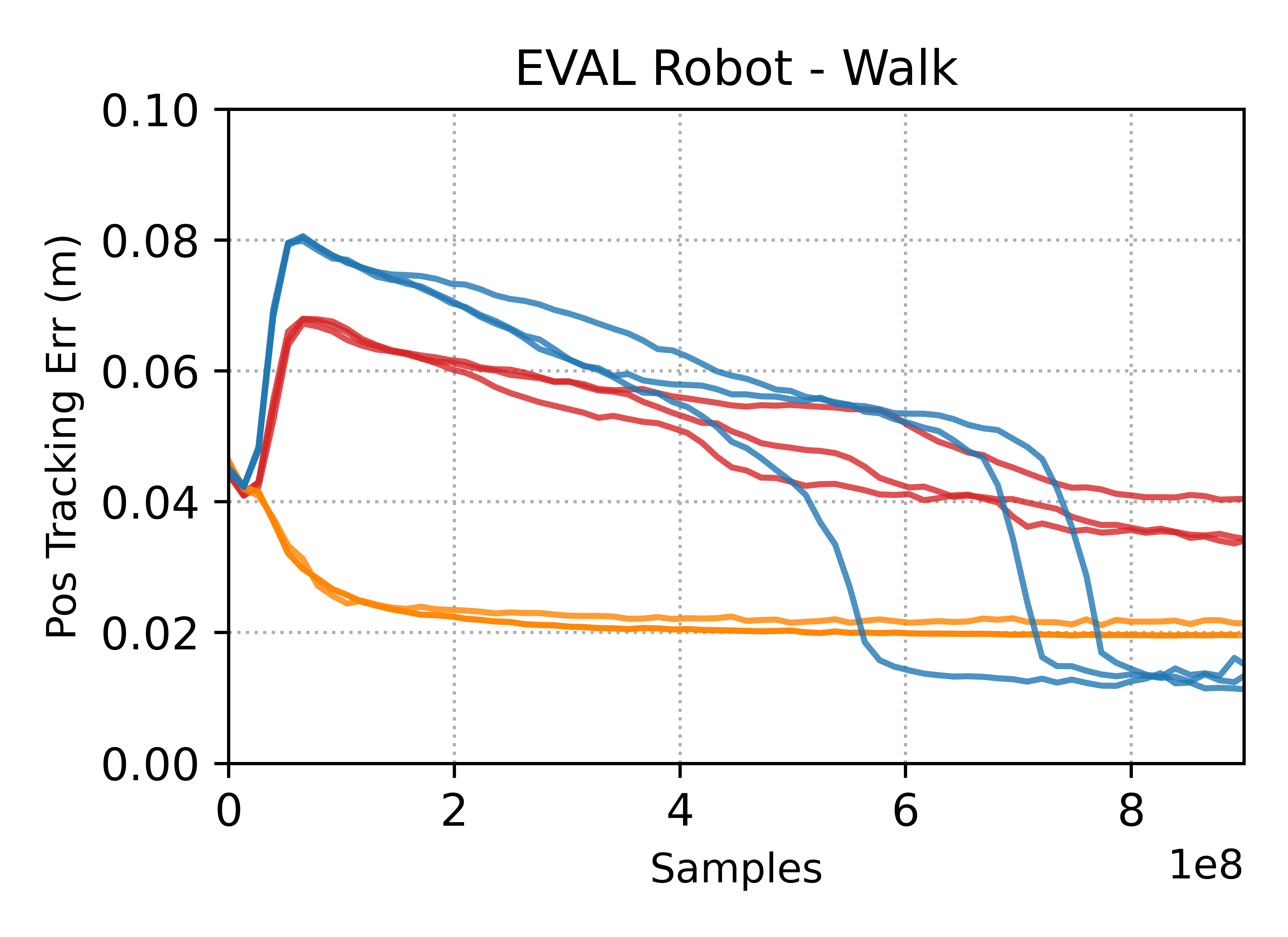}}
    \subfigure{\includegraphics[height=0.347\columnwidth]{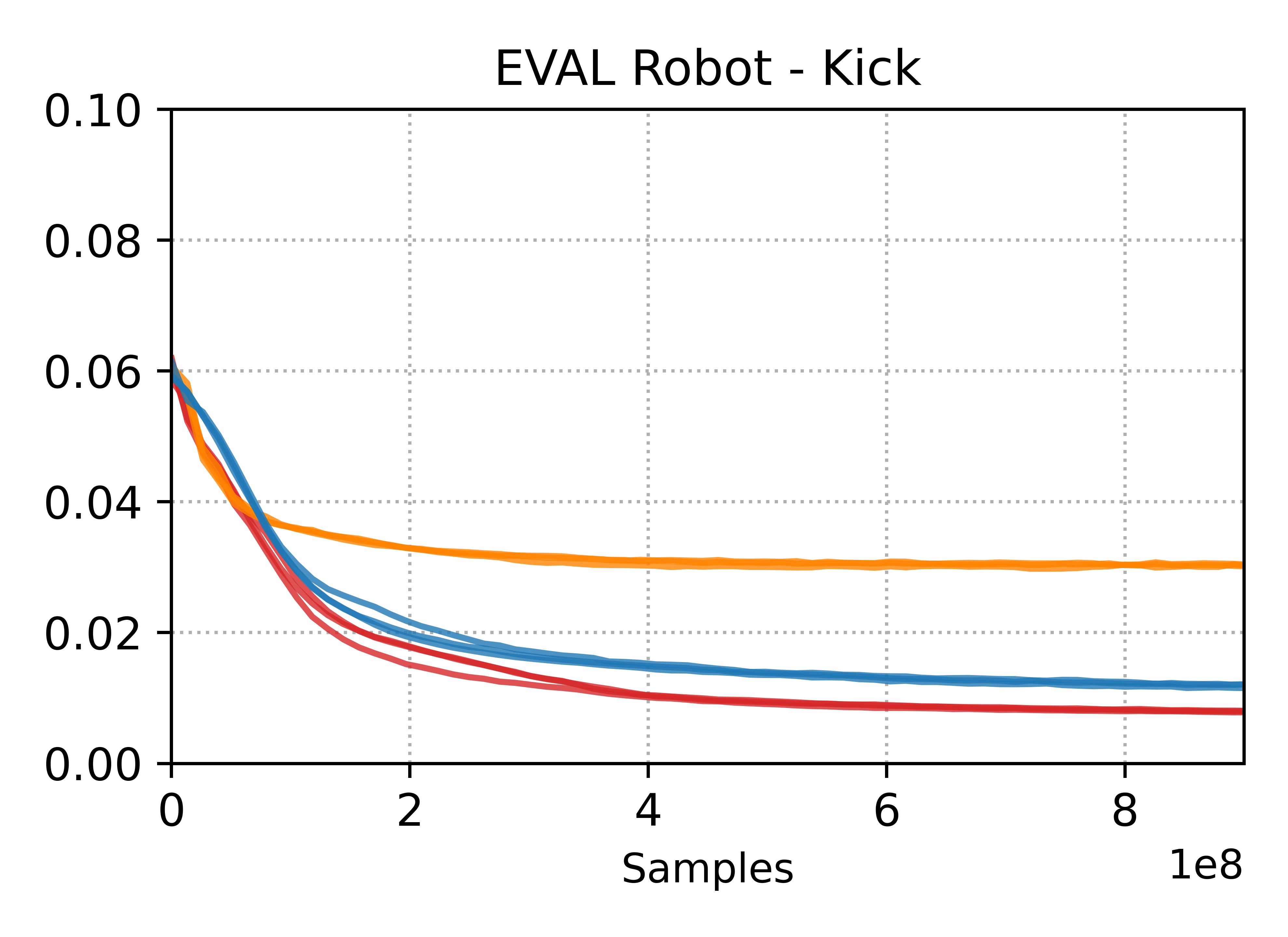}}
    \subfigure{\includegraphics[height=0.349\columnwidth]{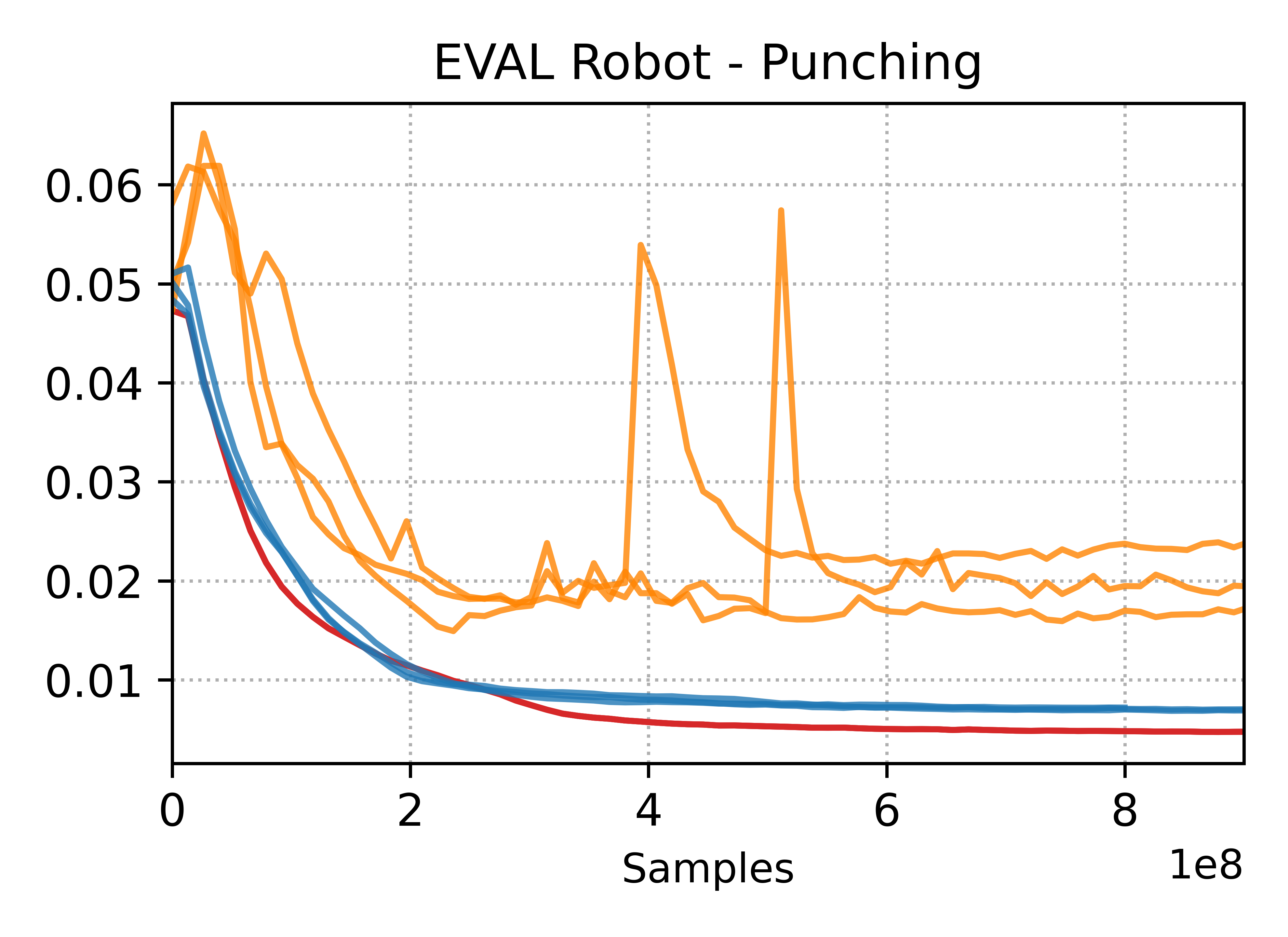}}\\
    \vspace{-0.2cm}
    \subfigure{\includegraphics[width=0.34\linewidth]{figures/motion_comparison_learning_curves/imitation_learning_curve_legend.png}}\\
\caption{Learning curves illustrating the position tracking errors of simulated robots trained using AMP \citep{2021-TOG-AMP}, DeepMimic \citep{deep_mimic}, and our method ADD. Statistics over 3 training runs, initialized with different random seeds, are depicted. Without relying on manually designed reward functions, ADD enables simulated robots to reproduce a set of challenging skills with quality on par with DeepMimic, a state-of-the-art motion-tracking method.}
\label{fig:EVAL-Learning Curve}
\end{figure*}

\begin{figure*}[t]
	\centering
        \captionsetup{skip=0pt}
    \subfigure{\includegraphics[height=0.16\linewidth]{figures/motion_comparison_learning_curves/double_kong_learning_curve.png}}
    \subfigure{\includegraphics[height=0.16\linewidth]{figures/motion_comparison_learning_curves/backflip_learning_curve.png}}
    \subfigure{\includegraphics[height=0.16\linewidth]{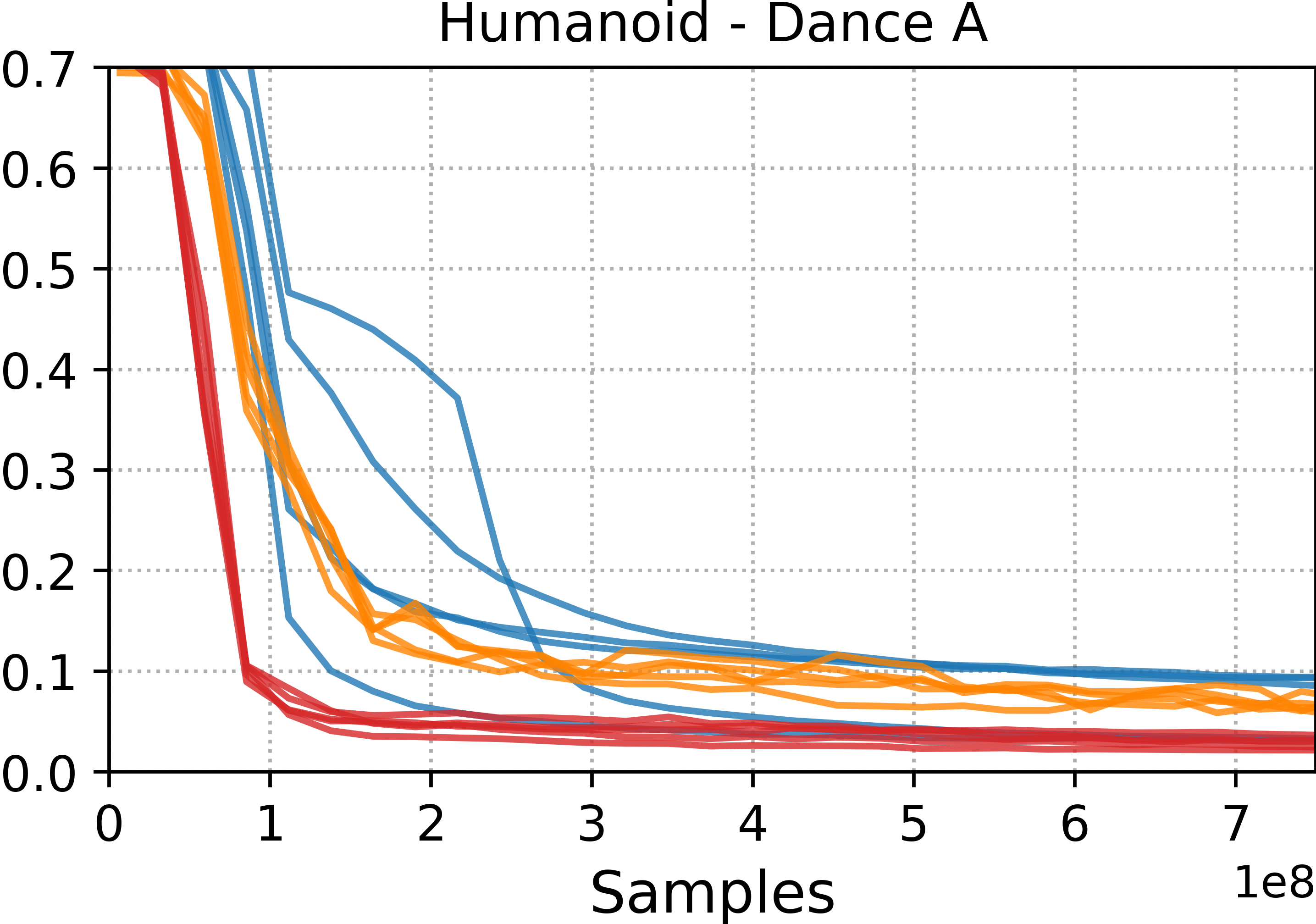}}
    \subfigure{\includegraphics[height=0.16\linewidth]{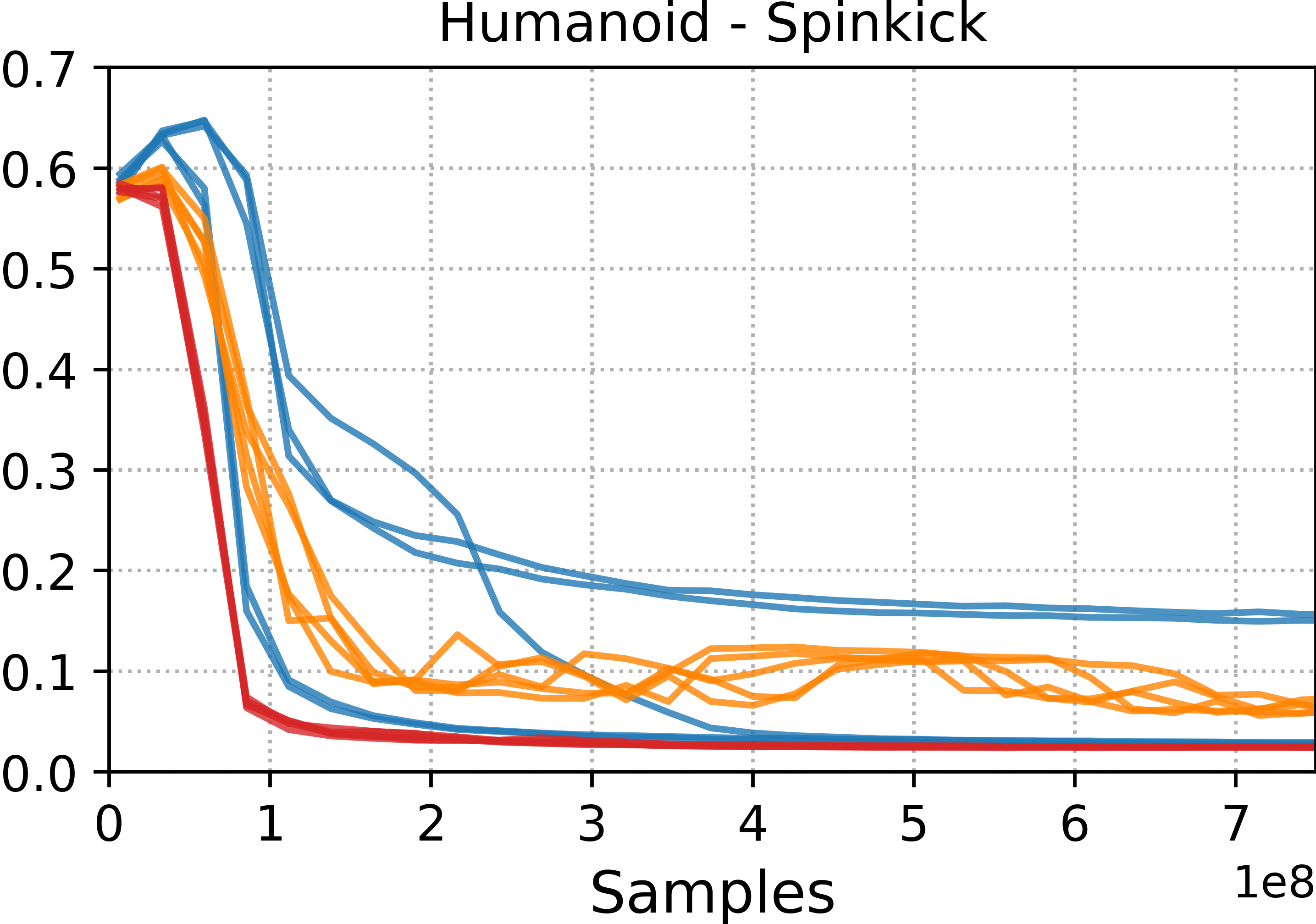}}\\
    \vspace{-0.2cm}
    \subfigure{\includegraphics[height=0.16\linewidth]{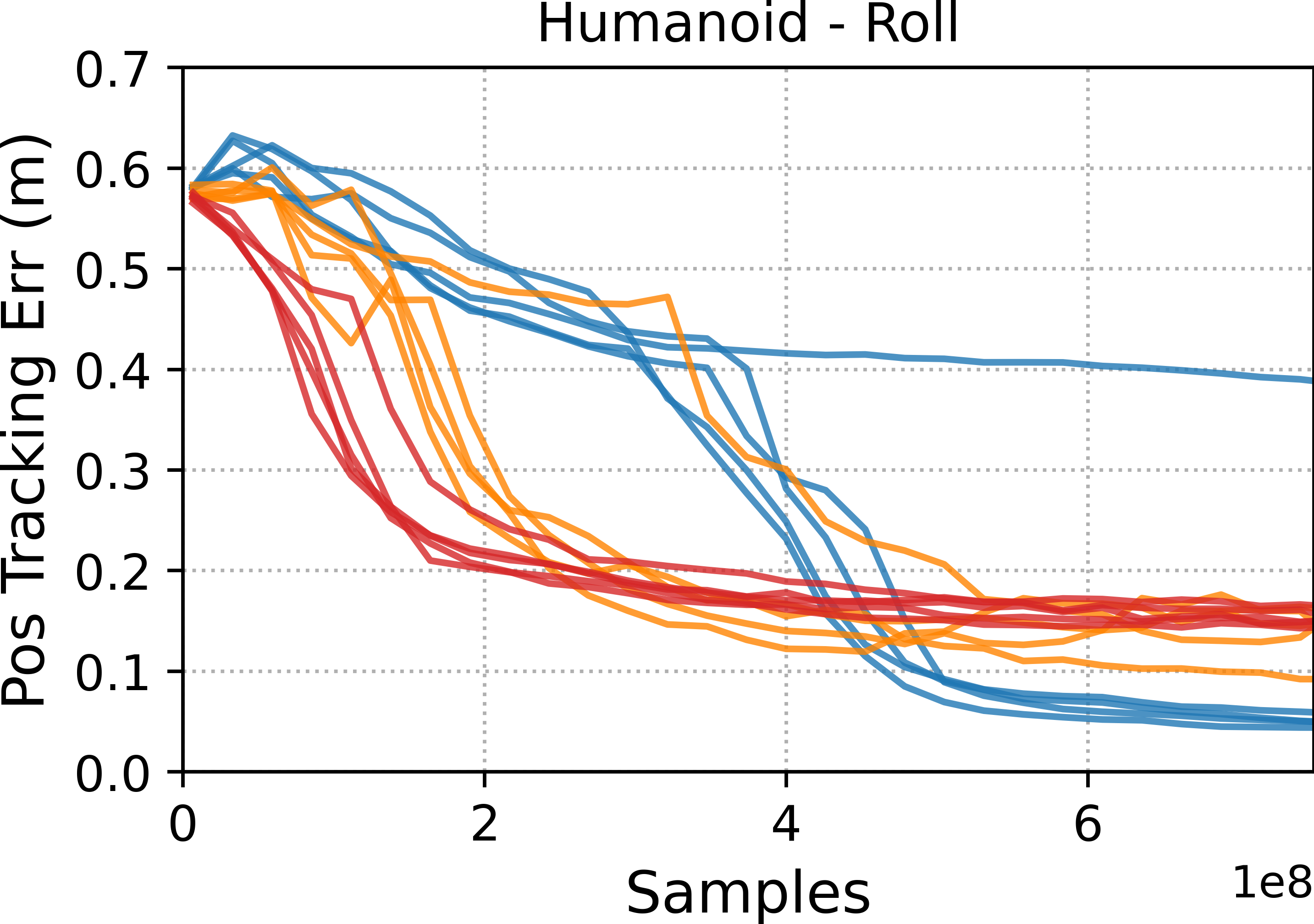}} 
    \subfigure{\includegraphics[height=0.16\linewidth]{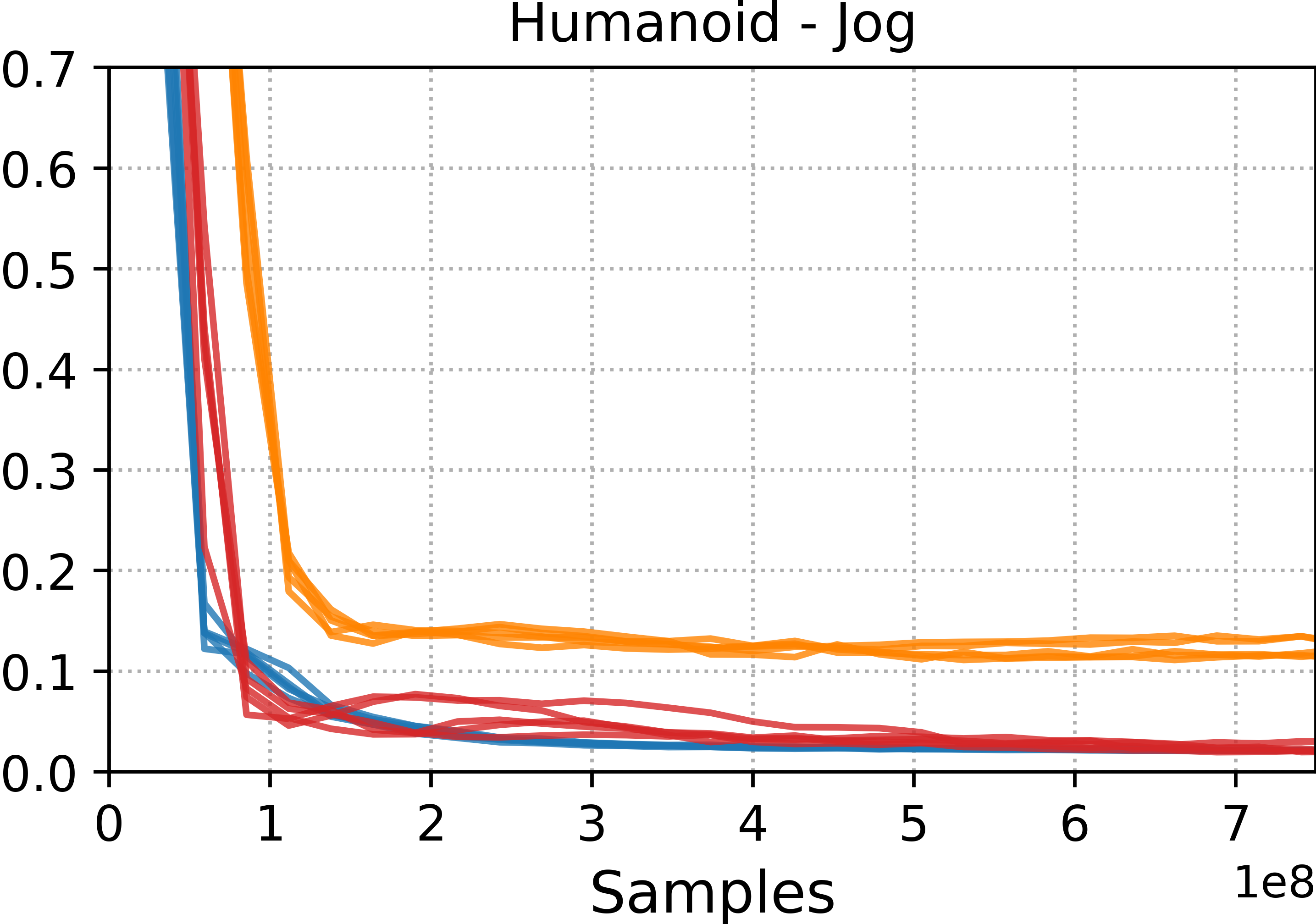}}
    \subfigure{\includegraphics[height=0.16\linewidth]{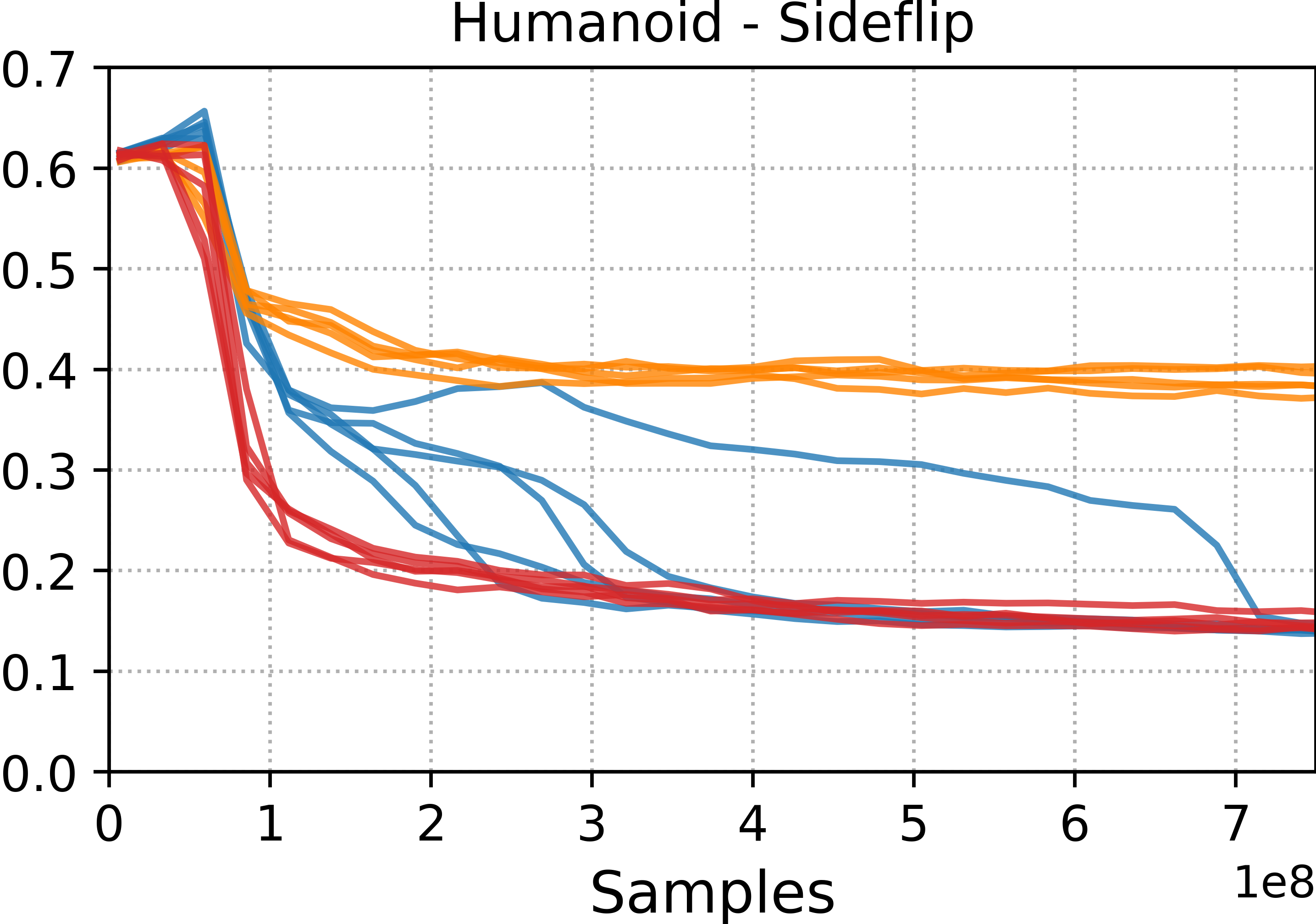}}
    \subfigure{\includegraphics[height=0.16\linewidth]{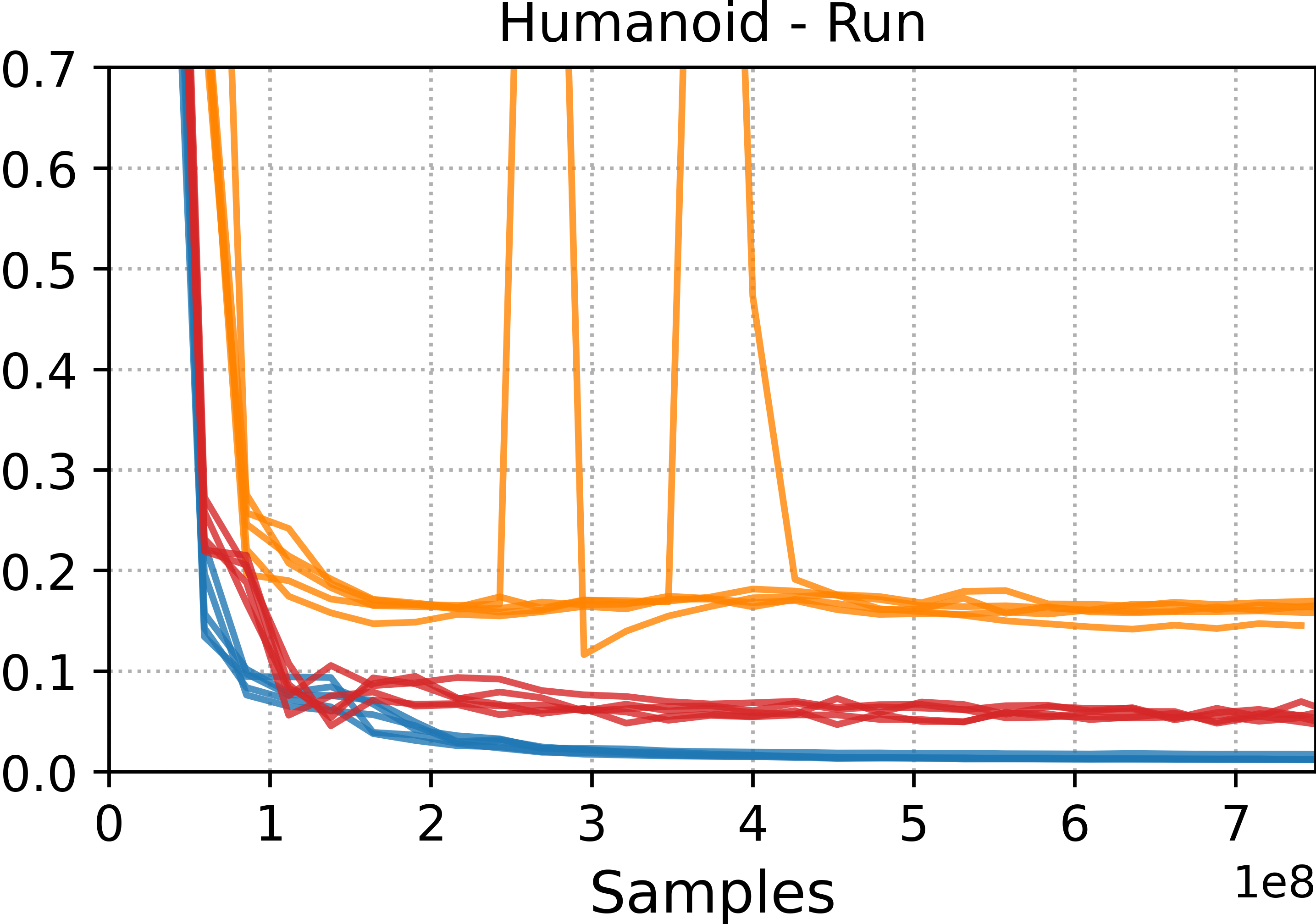}}\\
    \vspace{-0.2cm}
    \subfigure{\includegraphics[height=0.16\linewidth]{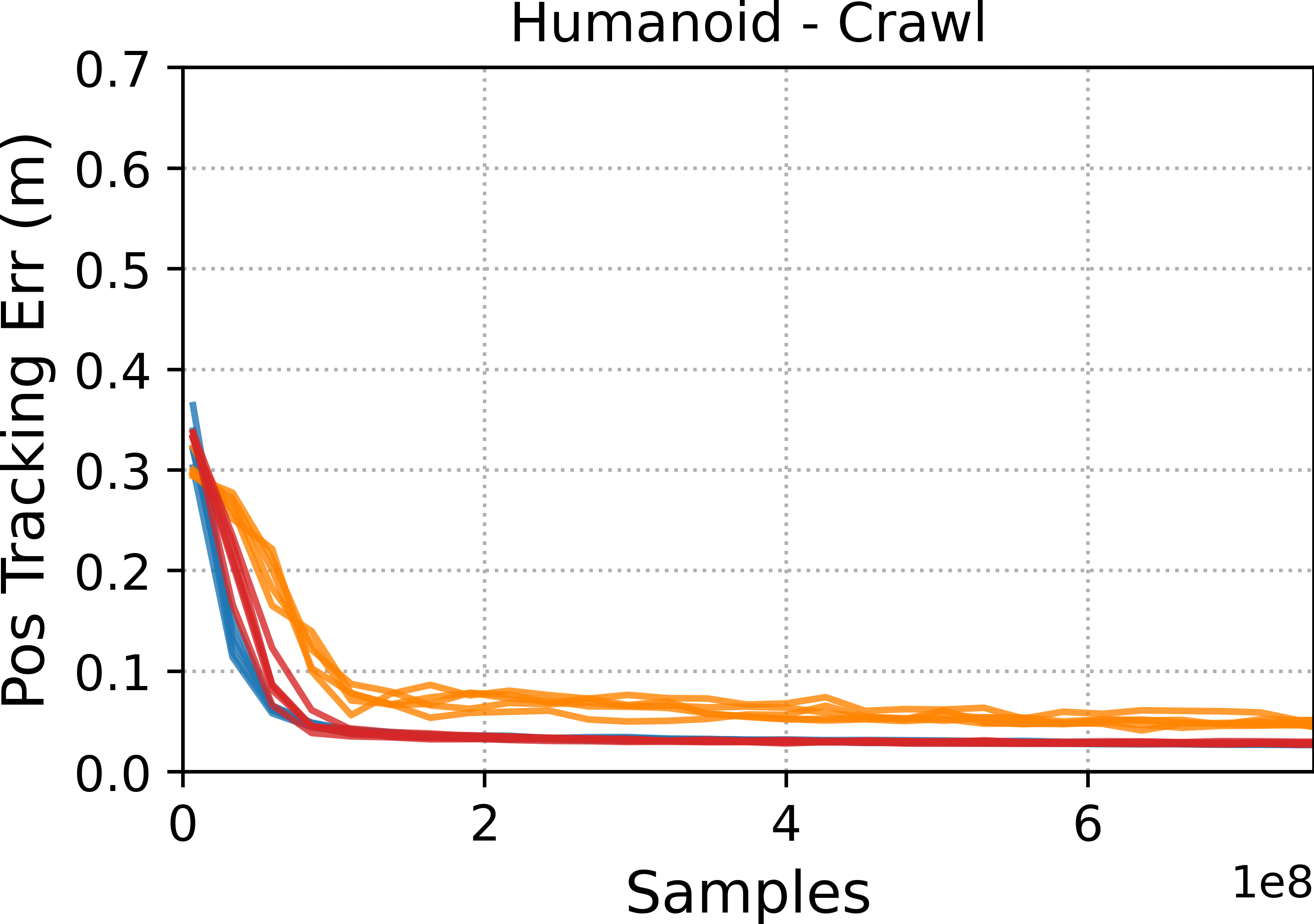}}
    \subfigure{\includegraphics[height=0.16\linewidth]{figures/motion_comparison_learning_curves/cartwheel_learning_curve.png}}
    \subfigure{\includegraphics[height=0.16\linewidth]{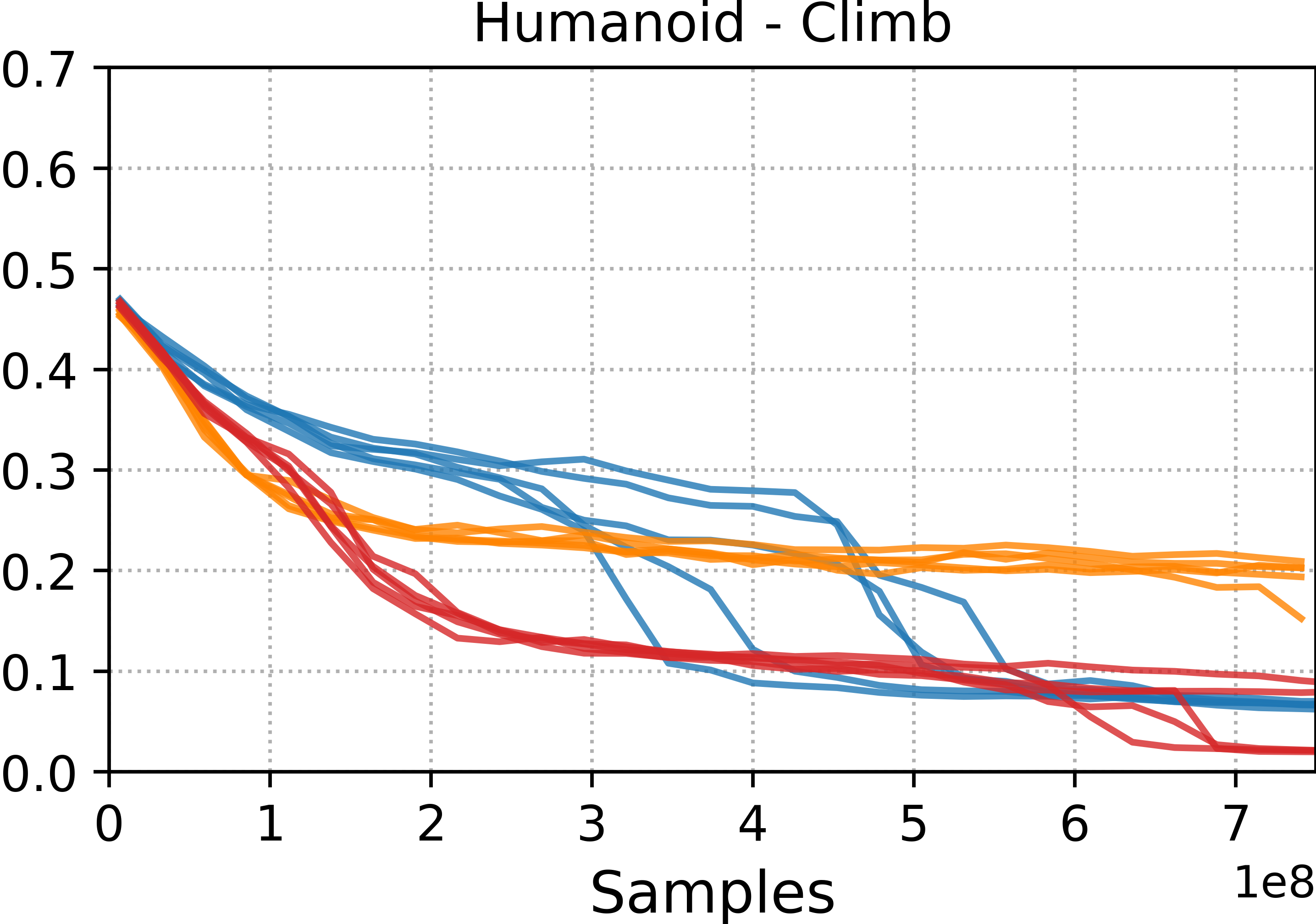}}
    \subfigure{\includegraphics[height=0.16\linewidth]{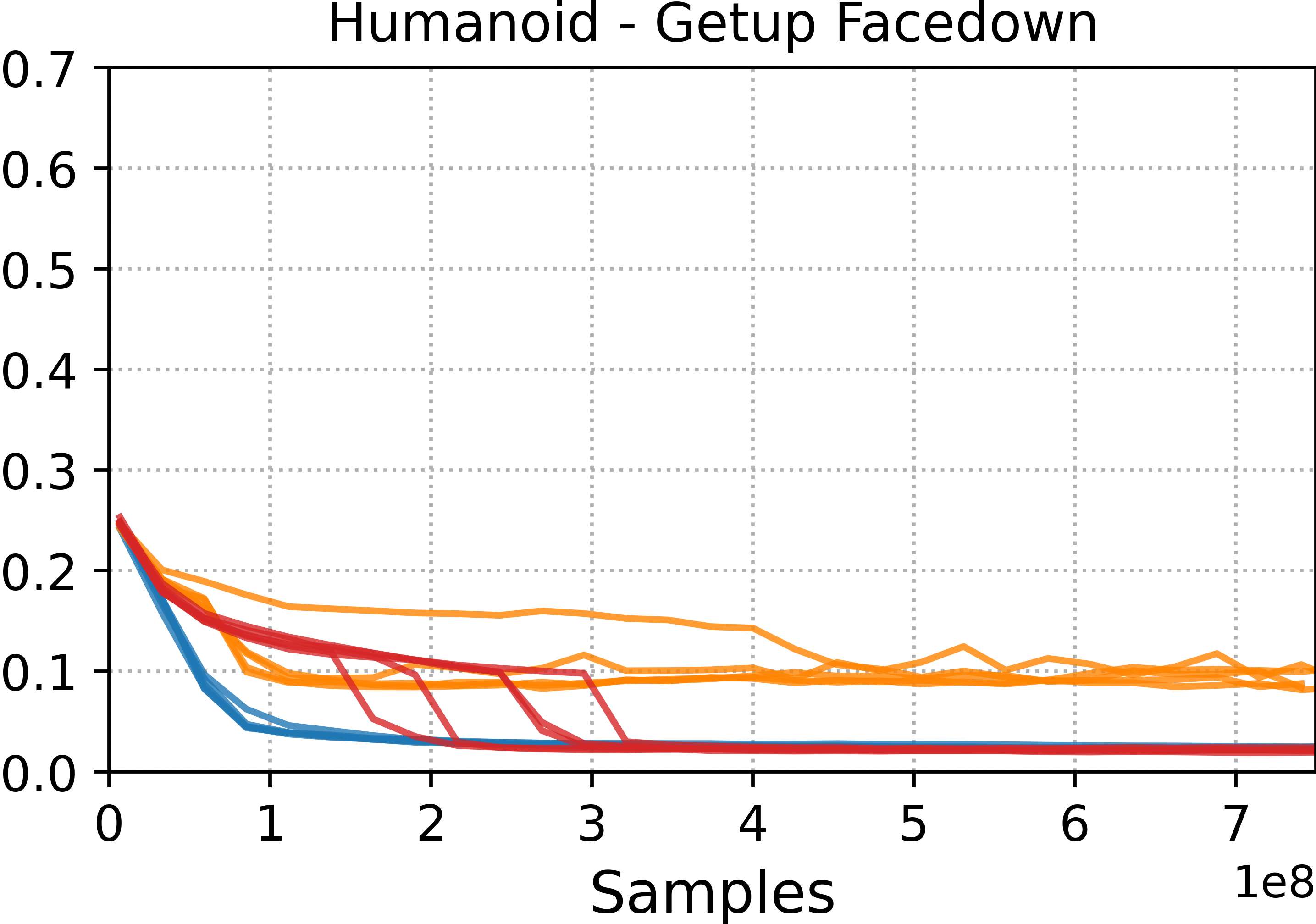}}\\
    \vspace{-0.2cm}
    \subfigure{\includegraphics[height=0.16\linewidth]{figures/motion_comparison_learning_curves/dancedb_learning_curve.png}}
    \subfigure{\includegraphics[height=0.16\linewidth]{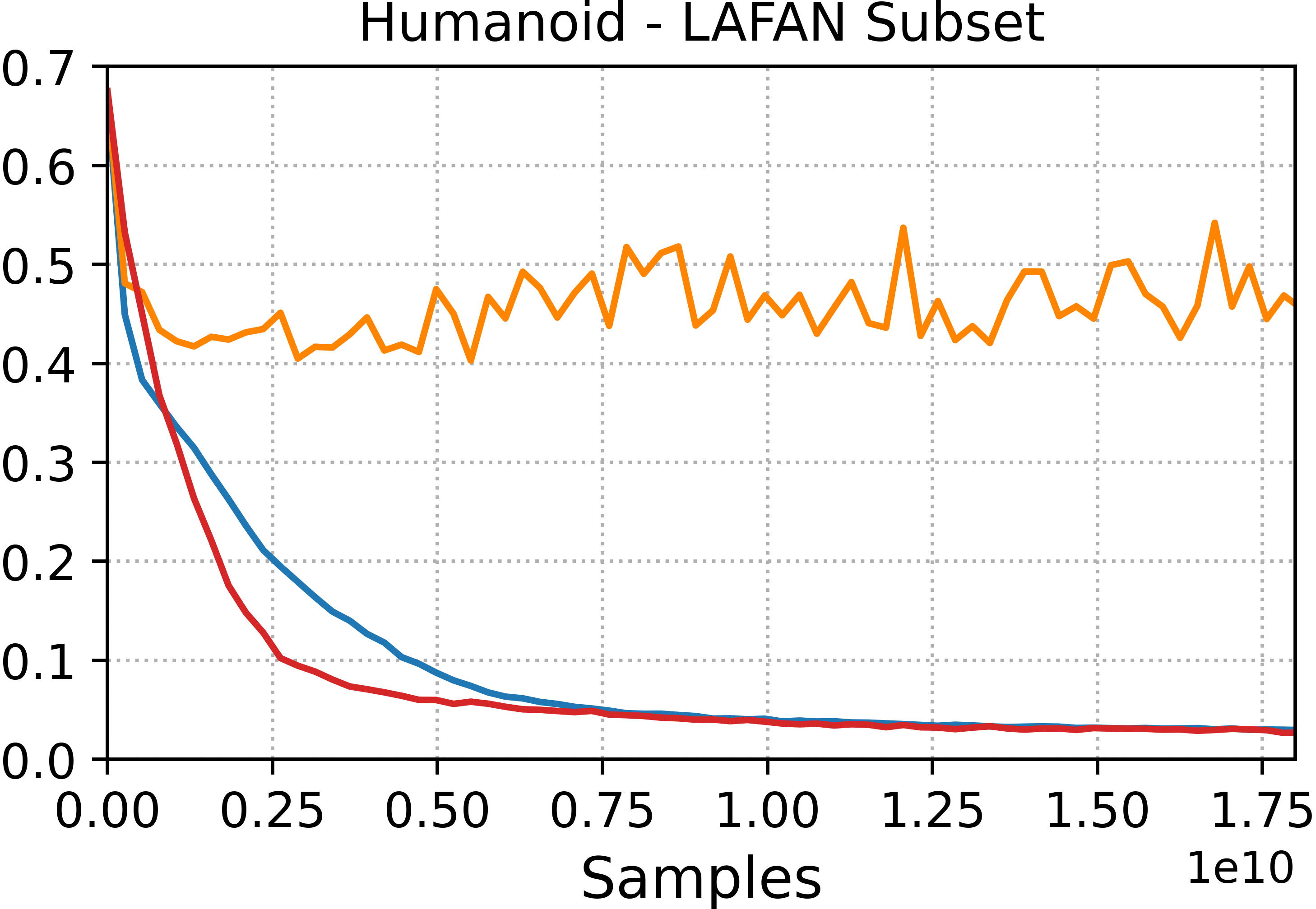}}\\
    \vspace{-0.2cm}
    \subfigure{\includegraphics[width=0.34\linewidth]{figures/motion_comparison_learning_curves/imitation_learning_curve_legend.png}}\\
\caption{Learning curves comparing the tracking performance of simulated humanoid characters trained via AMP \citep{2021-TOG-AMP}, DeepMimic \citep{deep_mimic}, and our method ADD. Statistics are computed over 5 training runs initialized with different random seeds, except for the LaFAN1 subset (1 run due to computational cost). ADD is capable of learning highly agile and acrobatic skills, achieving comparable tracking performance and sample efficiency to DeepMimic, without requiring manual reward engineering. Moreover, ADD exhibits better consistency across seeds, whereas DeepMimic often converges to suboptimal behaviors in some seeds.}
\vspace{10pt}
\label{fig:motion_imitation_learning_curve_full}
\end{figure*}

\subsection{Learning Curves}
\label{supp:motion_imitaion_learning_curves}
Figures \ref{fig:EVAL-Learning Curve} and \ref{fig:motion_imitation_learning_curve_full} present the learning curves of DeepMimic, AMP, and ADD across all skills for the EVAl robot and the humanoid character, respectively.

\subsection{{Steering Task Reward Functions}}
\label{supp:task_reward}
For ADD, task objectives are appended to the differential vector $\Delta_t$,
\[
\Delta_t = \begin{bmatrix}
\Delta_t^{\text{tracking}} \\
v^* - \rvv_t^T \rvd^*_t \\
-\left|\left|\rvv_t - (\rvv_t^{T} \rvd_t^{*}) \rvd_t^{*}\right|\right|
\end{bmatrix},
\] and are amplified by a factor $50$ after normalization.
The reward is then simply $r_t = - \log (1 - D(\Delta_t))$. For AMP and DeepMimic, the reward is changed to
\begin{equation}
    r_t = 0.5 \, r^{\text{tracking}}_t + 0.5 \, r^G_t,
\end{equation}
where
\begin{equation}
r_t^{G} = \exp \left[-2 \big( (v^{*} - \rvv_t^{T} \rvd_t^{*})^{2} + 0.1\left|\left|\rvv_t - (\rvv_t^{T} \rvd_t^{*}) \rvd_t^{*}\right|\right|^2 \big)\right].
\end{equation}

\subsection{Robustness to Perturbations}
\begin{figure}[t]
	\centering
        \captionsetup{skip=0pt}
    \subfigure{\includegraphics[height=0.5\columnwidth]{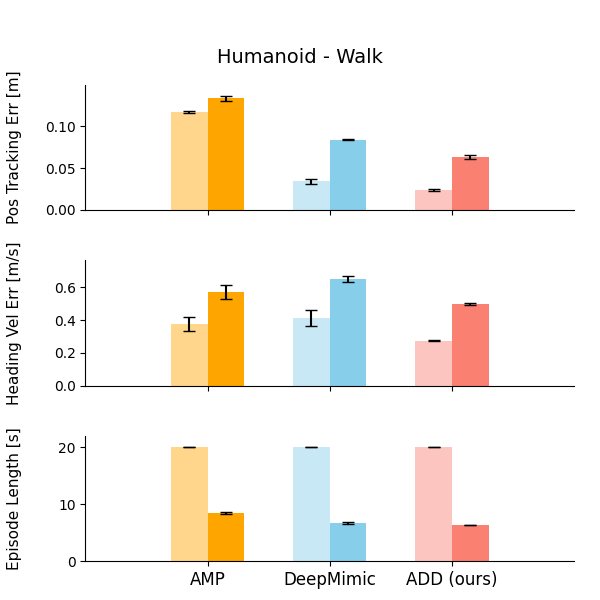}}
    \subfigure{\includegraphics[height=0.5\columnwidth]{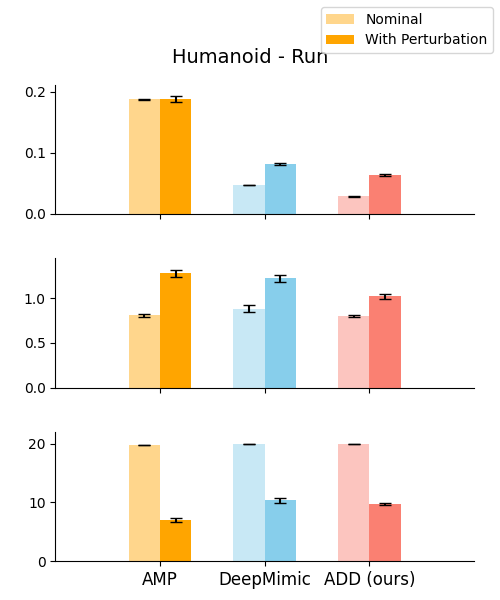}}\\
\caption{{Robustness of locomotion policies in Sec. 6.5 under random external force perturbations. Policies are trained without perturbations and stress-tested by randomly applying forces up to 300 N during inference. Performance is reported in terms of the episode length until loss of balance, as well as position tracking error and target velocity error before falling. Lighter bars denote nominal performance, and darker bars indicate performance under perturbation. ADD exhibits levels of degradation comparable to DeepMimic, reflecting a similar degree of robustness.}}
\label{fig:perturbation_bar_plot}
\end{figure}

To assess the robustness of locomotion policies trained using our approach, we evaluate their performance under random force perturbations and compare the results with prior methods. We take the policies trained in Section~6.5 and randomly apply random forces of up to 300 N on the character’s root. These perturbations are introduced only at test time; the policies are not exposed to such disturbances during training. 
Performance is measured with two criteria: (1) the duration the policy maintains balance before falling, and (2) the accuracy with which the policy follows the reference motion and steering commands while subjected to perturbations, prior to falling. As shown in Figure~\ref{fig:perturbation_bar_plot}, ADD exhibits a performance degradation comparable to DeepMimic, indicating a similar degree of robustness to perturbations. 

\section{Walker Additional Details}
\label{supp:walker_additional_details}
All policies are trained for approximately 60 million samples, which takes a wall-clock time of around \SI{24}{\hour} for the manual rewards and \SI{26}{\hour} for ADD on 2 CPU cores. 

\subsection{Reward Function}
\label{supp:walker_reward_func}
ADD's reward is $r_t = - \log (1 - D(\Delta_t))$, where $\Delta_t$ is simply:
\[
\Delta_t = \begin{bmatrix}
1.2 - h_t \\
1 - u_t \\
8.0 - v_t
\end{bmatrix},
\]
with $h_t$ being the walker's height, $u_t$ the cosine of the torso angle, and $v_t$ the walker's horizontal speed. 1.2, 1, and 8.0 are target values from the problem definition, not tunable hyperparameters. Compared to the reward function in \citet{2018DMControl}, ADD employs a much simpler formulation with significantly fewer manually-tuned hyperparameters. 

The manually-designed reward function used in \citet{2018DMControl} is as follows:
\begin{equation}
\begin{aligned}
r_{\text{manual}} &= r_{\text{stand}} \times \frac{5 \cdot r_{\text{move}} + 1}{6} \\
r_{\text{stand}} &= \frac{3 \cdot \text{tol}\left( h_t; 1.2, \infty, 0.1, 0.6, \text{gaussian} \right) + \frac{1 + u_t}{2}}{4} \\
\quad r_{\text{move}} &= \text{tol}\left( v_t; 8.0, \infty, 0.5, 4.0, \text{linear} \right)
\end{aligned}
\end{equation}

The tolerance function is defined as:
\begin{align}
\text{tol}(x; a, b, v_m, m, \text{sigmoid}) = \notag\\
\begin{cases}
1, & \text{if } a \le x \le b \\
f(d(x; a, b, m); v_m, \text{sigmoid}), & \text{otherwise}
\end{cases},
\end{align}
where
\[
d(x; a, b, m) = \frac{\max(a - x, x - b)}{m}.
\]
Here, $[a,b]$ represents the bounds, $m$ is the margin, $v_m$ is the value at margin, and $\text{sigmoid}$ is the type of sigmoid function.

If $\text{sigmoid} = \text{linear}$:
\begin{equation}
f(d; v_m, \text{linear}) = 
\begin{cases}
1 - (1 - v_m) \cdot d, & \text{if } |(1 - v_m) \cdot d| < 1 \\
0, & \text{otherwise}
\end{cases}.
\end{equation}

If $\text{sigmoid} = \text{gaussian}$:
\begin{equation}
f(d; v_m, \text{gaussian}) = \exp\left( -\ln\left( \frac{1}{v_m} \right) \cdot d^2 \right).
\end{equation}

\subsection{Hyperparameters}
The hyperparameters for ADD in the Walker task are documented in Table \ref{tab:suppADDWalkerParams}.

\begin{table}[t]
    \centering
    \caption{ADD walker experiment hyperparameters.}
    \label{tab:suppADDWalkerParams}
    \begin{tabular}{|l|c|}
        \hline
        \textbf{Parameter} & {\bf Value} \\ \hline
        $\lambda^{\mathrm{GP}}$ Gradient Penalty & $0.001$ \\ \hline
        $K$ Update Minibatch Size &  $512$  \\ \hline
        $\pi$ Policy Stepsize&  $5 \times 10^{-4}$  \\ \hline
        $V$ Value Stepsize &  $5 \times 10^{-4}$  \\ \hline
        $D$ Discriminator Stepsize & $5 \times 10^{-5}$ \\ \hline
        $\mathcal{B}$ Experience Buffer Size &  $4096$  \\ \hline
        $\gamma$ Discount&  $0.99$  \\ \hline
        SGD Momentum &  $0.9$  \\ \hline
        GAE($\lambda$) &  $0.95$  \\ \hline
        TD($\lambda$) &  $0.95$  \\ \hline
        PPO Clip Threshold &  $0.02$  \\ \hline
    \end{tabular}
\end{table}

\section{Go1 Additional Details}
\label{supp:go1_additional_details}
All policies are trained for approximately 1 billion samples, which takes about 3 hours for both ADD and the manually designed reward baseline on an A100 GPU. During training, early termination is applied in both methods when the torso, thigh, or hip of the quadruped makes contact with the ground.

\subsection{Reward Function}
\label{supp:go1_reward_func}
The reward for ADD is given by $r_t = - \log (1 - D(\Delta_t))$, where $\Delta_t$ is:
\begin{equation}
\Delta_t = 
\begin{bmatrix}
0 -  \mathbf{v}_{b,z} \\
0 - \|\boldsymbol{\omega}_{b,xy}\|^2 \\
0 - \|\rvz_{xy}\|^2 \\
0.3 - h_b \\
500 - \max\left(\|\boldsymbol{\tau}_j\|^2, 500\right) \\
100 - \max\left(\|\dot{\mathbf{q}}_j\|^2, 100\right) \\
100000 - \max\left(\|\ddot{\mathbf{q}}_j\|^2, 100000\right) \\
0.5 - \max\left(\|\dot{\mathbf{q}}_j^*\|^2, 0.5\right) \\
\mathbf{v}_{b,x}^* - \mathbf{v}_{b,x} \\
\mathbf{v}_{b,y}^* - \mathbf{v}_{b,y} \\
\boldsymbol{\omega}_{b,z}^* - \boldsymbol{\omega}_{b,z} \\
0.02 - \sum_{f=0}^4 t_{\text{air},f}\\
\end{bmatrix}.
\label{eq:go1_delta_vector}
\end{equation}
The notations follow the definitions in Table \ref{tab:symbol_definitions}. The $z$ axis is aligned with gravity.

\begin{table}[h]
\centering
\caption{Definition of symbols.}
\label{tab:symbol_definitions}
\begin{tabular}{ll}
\toprule
\textbf{} & \textbf{} \\
Joint positions & $\rvq_j$ \\
Joint velocities & $\dot{\rvq}_j$ \\
Joint accelerations & $\ddot{\rvq}_j$ \\
Target joint positions & $\rvq_j^*$ \\
Joint torques & $\boldsymbol{\tau}_j$ \\
Base linear velocity & $\mathbf{v}_b$ \\
Base angular velocity & $\boldsymbol{\omega}_b$ \\
Base height & $h_b$ \\
Commanded base linear velocity & $\mathbf{v}_b^*$ \\
Commanded base angular velocity & $\boldsymbol{\omega}_b^*$ \\
Number of collisions & $n_c$ \\
Feet air time & $\rvt_{\text{air}}$ \\
Environment time step & $dt$ \\
Gravity vector & $\rvz$ \\
\bottomrule
\end{tabular}
\end{table}

The manual reward function we compare to is provided in Table \ref{tab:go1_manual_reward_terms}. The reward terms are taken from what \citet{legged-gym-pmlr-v164-rudin22a} used in their open-source repository. The weights are largely similar to the weights they used for the A1 robot, with some weights tuned to achieve better performance on the Go1 robot. Extensive hyperparameter tuning was performed to ensure the baseline method achieves good performance. ADD has fewer tunable hyperparameters than the manually-designed reward function.

\begin{table}[t]
\centering
\caption{Definition of reward terms used in \citep{legged-gym-pmlr-v164-rudin22a}, with $\phi(x) := \exp\left(-\frac{\|x\|^2}{0.25}\right)$. The $z$ axis is aligned with gravity.}
\label{tab:go1_manual_reward_terms}
\begin{tabular}{lll}
\toprule
\textbf{} & \textbf{definition} & \textbf{weight} \\
\midrule
Linear velocity tracking & $\phi(\mathbf{v}_{b,xy}^* - \mathbf{v}_{b,xy})$ & $1\,dt$ \\
Angular velocity tracking & $\phi(\boldsymbol{\omega}_{b,z}^* - \boldsymbol{\omega}_{b,z})$ & $0.5\,dt$ \\
Linear velocity penalty & $-\rvv_{b,z}^2$ & $2\,dt$ \\
Angular velocity penalty & $-\|\boldsymbol{\omega}_{b,xy}\|^2$ & $0.05\,dt$ \\
Orientation penalty & $-\|\rvz_{xy}\|^2$ & $0\,dt$ \\
Root Height & $-(0.3 - h_b)^2$ & $50\,dt$ \\
Joint velocity & $- \|\dot{\mathbf{q}}_j\|^2$ & $0\,dt$ \\
Joint acceleration & $-\|\ddot{\mathbf{q}}_j\|^2$ & $2.5 \times 10^{-7}\,dt$ \\
Joint torques & $-\|\boldsymbol{\tau}_j\|^2$ & $0.0002\,dt$ \\
Action rate & $-\|\dot{\mathbf{q}}_j^*\|^2$ & $0.01\,dt$ \\
Feet air time & $\sum_{f=0}^4 (\rvt_{\text{air},f} - 0.2)$ & $1\,dt$ \\
\bottomrule
\end{tabular}
\end{table}

\subsection{Hyperparameters}
Training hyperparameters of ADD and the manual reward function are detailed in Table \ref{tab:suppADDGo1Params} and \ref{tab:suppManualGo1Params}, respectively.

\begin{table}[H]
    \centering
    \caption{ADD Go1 quadruped training hyperparameters.}
    \label{tab:suppADDGo1Params}
    \begin{tabular}{|l|c|}
        \hline
        \textbf{Parameter} & {\bf Value} \\ \hline
        $\lambda^{\mathrm{GP}}$ Gradient Penalty & $0.1$ \\ \hline
        $K$ Update Minibatch Size &  $4096 \times 6$  \\ \hline
        $\pi$ Policy Stepsize&  $1 \times 10^{-4}$  \\ \hline
        $V$ Value Stepsize &  $1 \times 10^{-4}$  \\ \hline
        $D$ Discriminator Stepsize & $5 \times 10^{-4}$ \\ \hline
        $\mathcal{B}$ Experience Buffer Size &  $4096 \times 24$  \\ \hline
        $\gamma$ Discount&  $0.99$  \\ \hline
        GAE($\lambda$) &  $0.95$  \\ \hline
        TD($\lambda$) &  $0.95$  \\ \hline
        PPO Clip Threshold &  $0.02$  \\ \hline
    \end{tabular}
\end{table}

\begin{table}[H]
    \centering
    \caption{Manual reward Go1 quadruped training hyperparameters.}
    \label{tab:suppManualGo1Params}
    \begin{tabular}{|l|c|}
        \hline
        \textbf{Parameter} & {\bf Value} \\ \hline
        $K$ Update Minibatch Size &  $4096 \times 6$  \\ \hline
        $\pi$ Policy Stepsize&  $1 \times 10^{-4}$  \\ \hline
        $V$ Value Stepsize &  $1 \times 10^{-4}$  \\ \hline
        $\mathcal{B}$ Experience Buffer Size &  $4096 \times 24$  \\ \hline
        $\gamma$ Discount&  $0.99$  \\ \hline
        GAE($\lambda$) &  $0.95$  \\ \hline
        TD($\lambda$) &  $0.95$  \\ \hline
        PPO Clip Threshold &  $0.02$  \\ \hline
    \end{tabular}
\end{table}

\subsection{Full Learning Curves}
\label{supp:go1_learning_curves}
\begin{figure*}[h]
	\centering
        \captionsetup{skip=0pt}
    \subfigure{\includegraphics[height=0.16\linewidth]{figures/go1/go1_ang_vel_penalty.png}}
    \subfigure{\includegraphics[height=0.16\linewidth]{figures/go1/go1_dof_acc.png}}
    \subfigure{\includegraphics[height=0.16\linewidth]{figures/go1/go1_linear_vel_tracking_err.png}}
    \subfigure{\includegraphics[height=0.16\linewidth]{figures/go1/go1_ang_vel_tracking_err.png}}\\
    \vspace{-0.2cm}
    \subfigure{\includegraphics[height=0.16\linewidth]{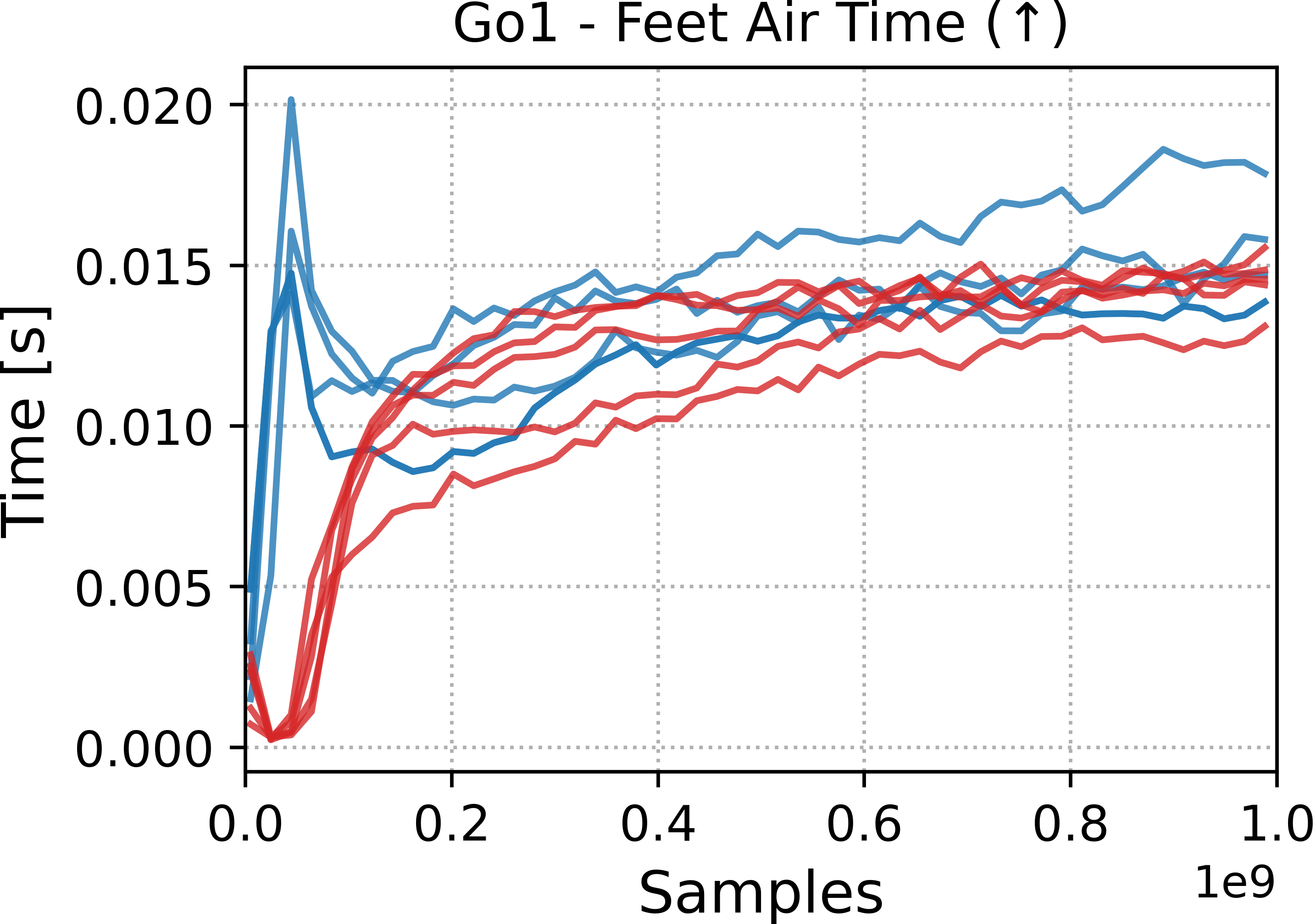}} 
    \subfigure{\includegraphics[height=0.16\linewidth]{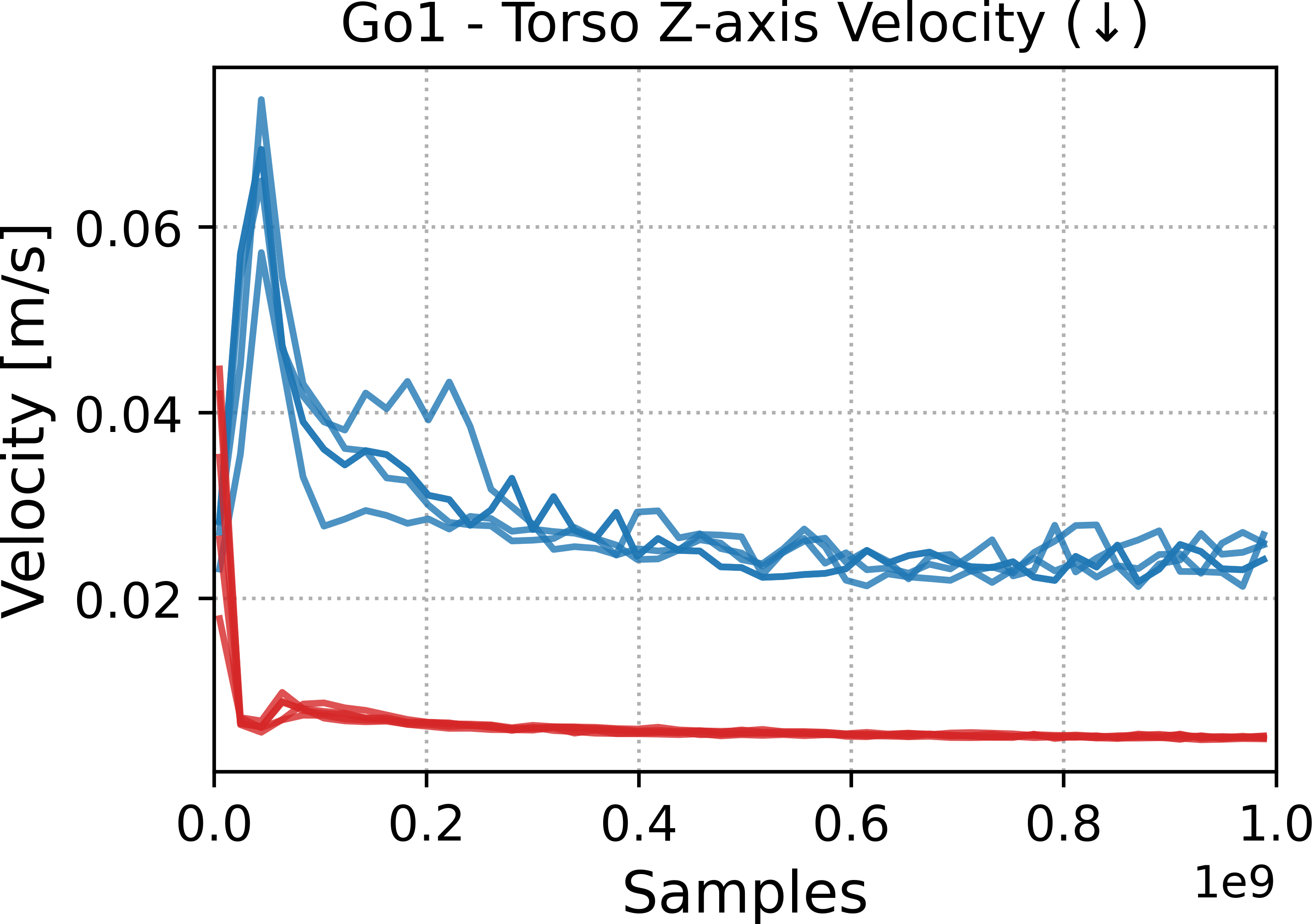}}
    \subfigure{\includegraphics[height=0.16\linewidth]{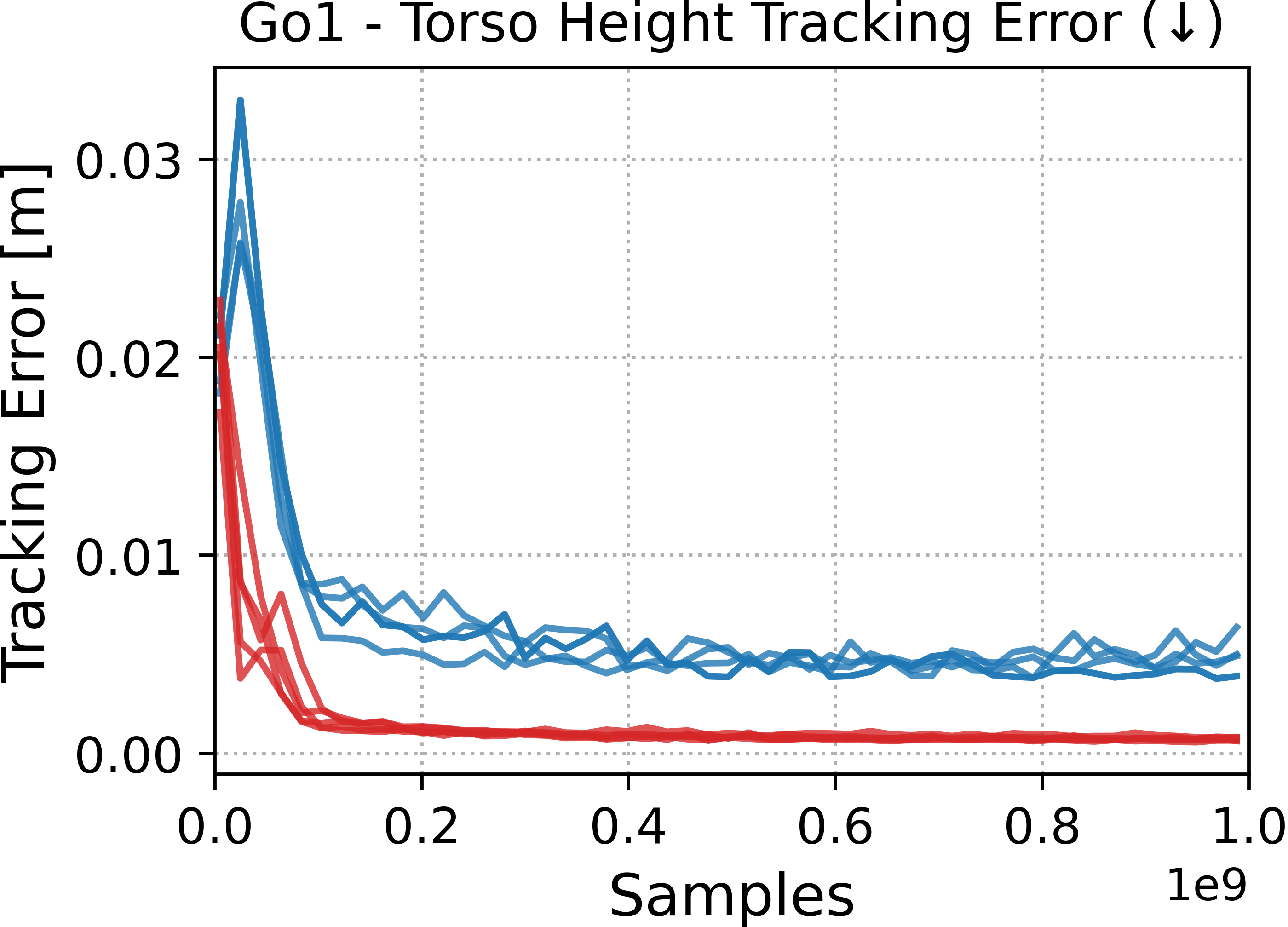}}
    \subfigure{\includegraphics[height=0.16\linewidth]{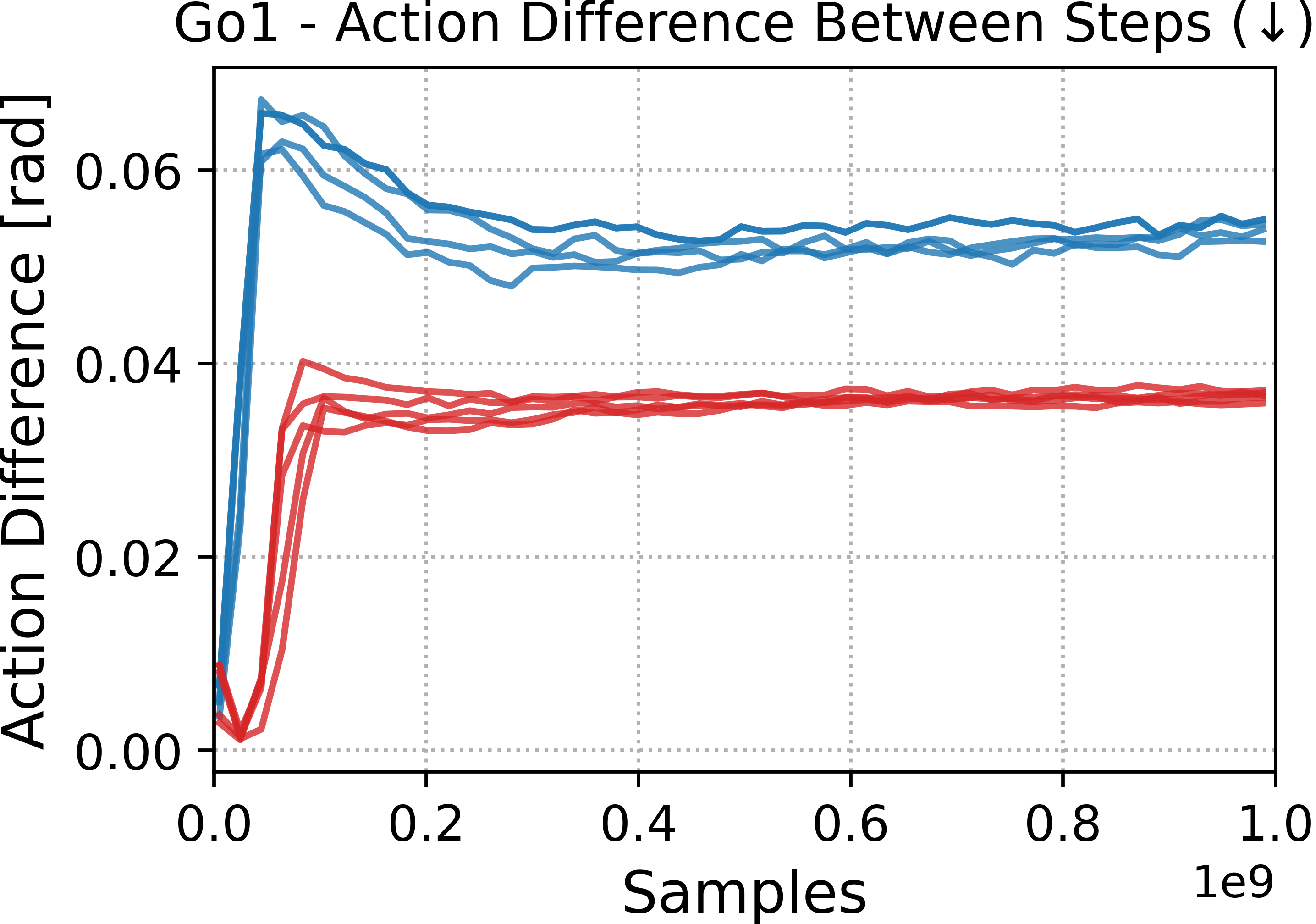}}\\
    \vspace{-0.2cm}
    \subfigure{\includegraphics[height=0.16\linewidth]{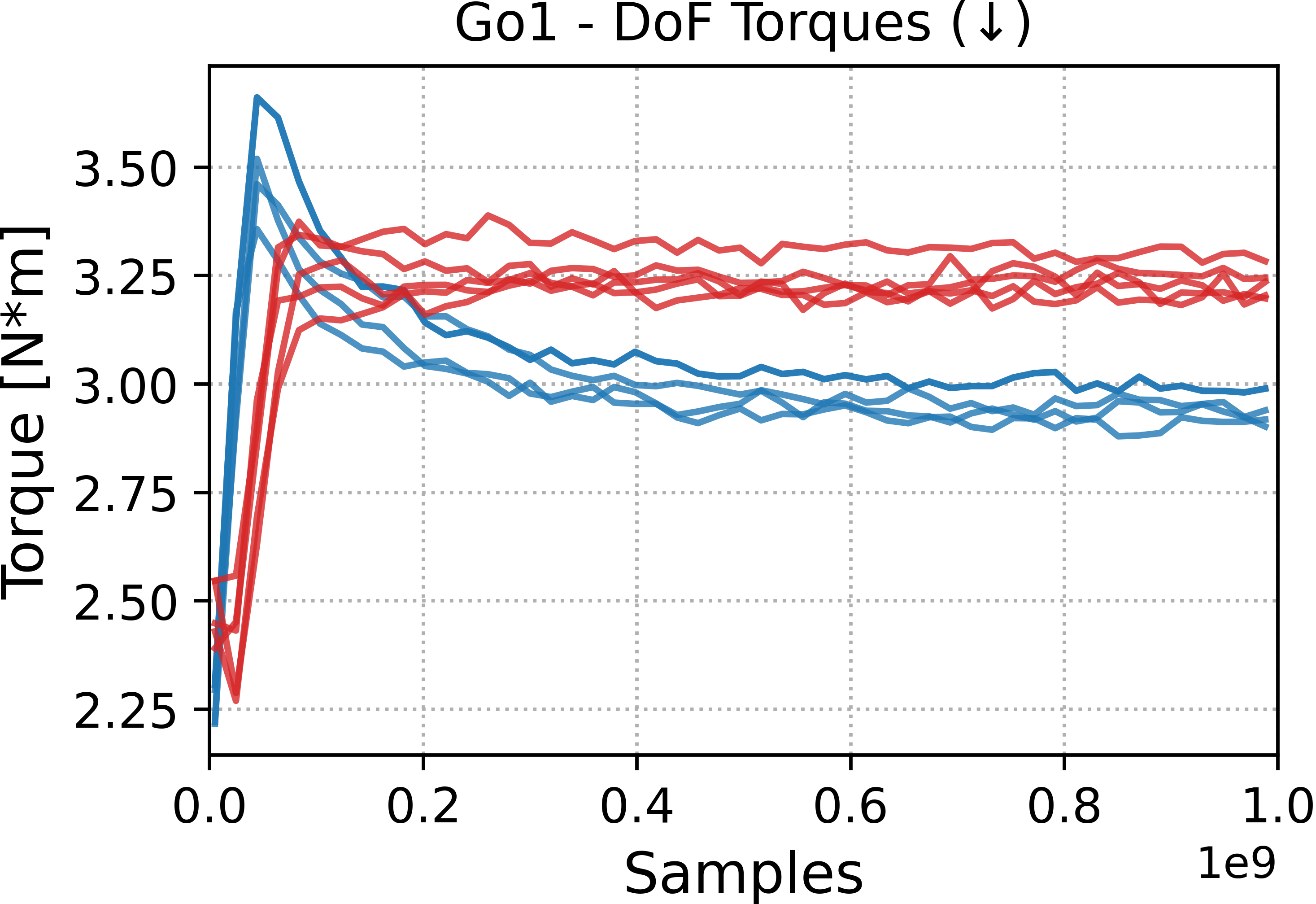}}
    \subfigure{\includegraphics[height=0.16\linewidth]{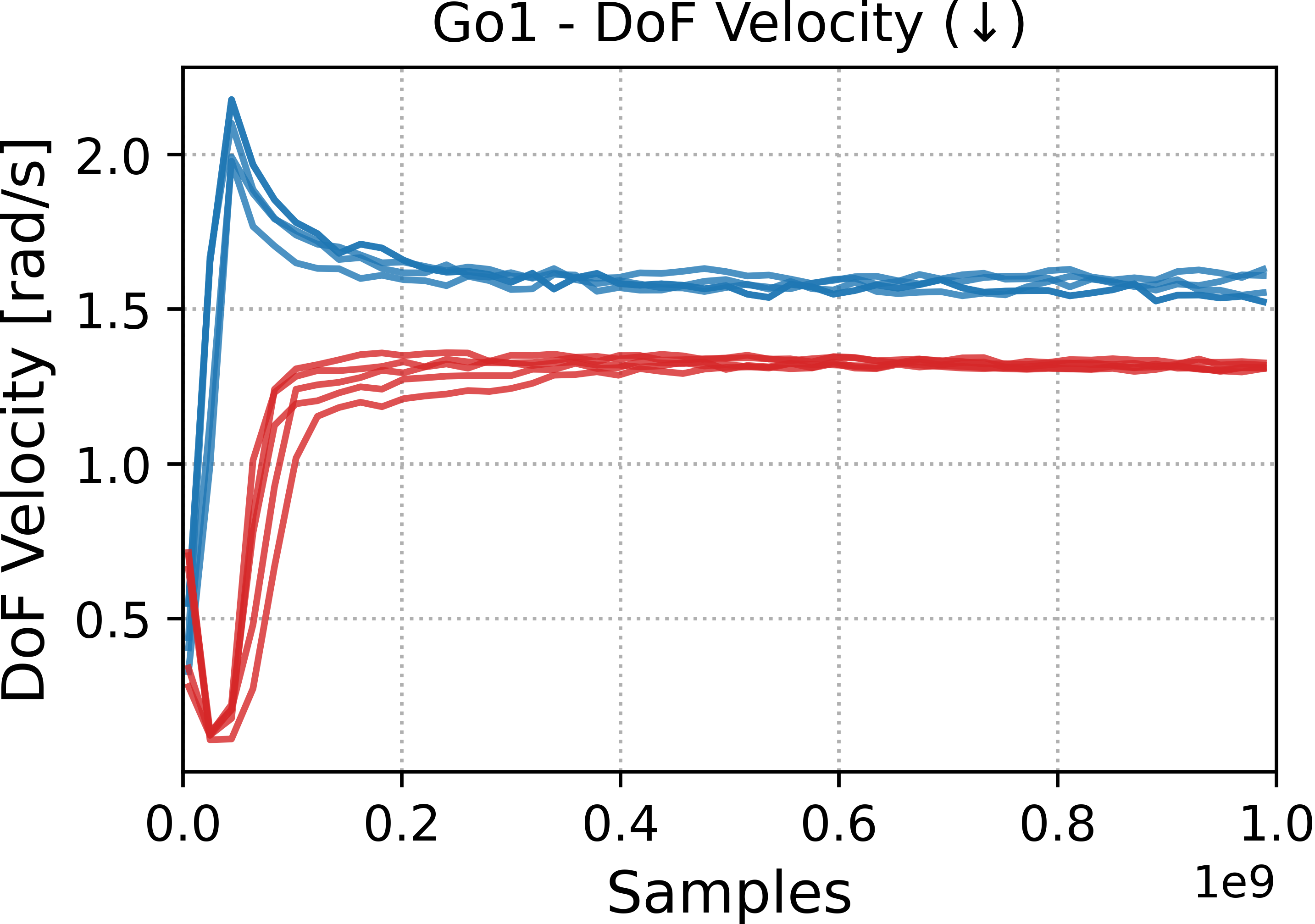}}\\
    \vspace{-0.2cm}
    \subfigure{\includegraphics[width=0.25\linewidth]{figures/walker_learning_curve_legend.png}}\\
\caption{Full learning curves for all training objectives on the quadruped task, with results averaged over five training runs initialized with different random seeds. ADD outperforms the manually designed reward function from \citet{legged-gym-pmlr-v164-rudin22a} on several metrics associated with torso stability and smooth control. Overall, ADD achieves comparable sample efficiency, final performance, and consistency to the manually-designed reward function.}
\label{fig:go1_learning_curve_full}
\end{figure*}

The comprehensive collection of learning curves on the quadrupedal task can be found in Figure~\ref{fig:go1_learning_curve_full}.

\end{document}